\documentclass[prb,aps,reprint,superscriptaddress,amsmath,amssymb,a4paper,floatfix,longbibliography,showkeys]{revtex4-1}
\usepackage{amsmath}
\usepackage{array}
\usepackage{capt-of}
\usepackage{color}
\usepackage{dcolumn}
\usepackage{enumerate}
\usepackage{epsfig}
\usepackage{epstopdf}
\makeatletter
\newcommand\footnoteref[1]{\protected@xdef\@thefnmark{\ref{#1}}\@footnotemark} \makeatother
\usepackage{graphics}
\usepackage[pdftex,colorlinks=true,citecolor=blue]{hyperref}
\usepackage{longtable}
\usepackage{multirow}
\usepackage{tabularx}
\usepackage{upgreek}

\usepackage[pagewise]{lineno}
\usepackage{tikz,xcolor}

\graphicspath{{figures/}}
\definecolor{lime}{HTML}{A6CE39}
\DeclareRobustCommand{\orcidicon}{%
	\begin{tikzpicture}
	\draw[lime, fill=lime] (0,0) 
	circle [radius=0.16] 
	node[white] {{\fontfamily{qag}\selectfont \tiny ID}};
	\draw[white, fill=white] (-0.0625,0.095) 
	circle [radius=0.007];
	\end{tikzpicture}
	\hspace{-2mm}
}

\foreach \x in {A, ..., Z}{%
	\expandafter\xdef\csname orcid\x\endcsname{\noexpand\href{https://orcid.org/\csname orcidauthor\x\endcsname}{\noexpand\orcidicon}}
}

\begin{document}
%
%
%%%%%%%%%%%%%%%%%%%%%%%%%%%%%%%%%%%%%%%%%%%%%%%%%
%%%% Document information and authors %%%%
%%%%%%%%%%%%%%%%%%%%%%%%%%%%%%%%%%%%%%%%%%%%%%%%%
%
%%%%%%%%%
% Title %
%%%%%%%%%
%
\title{Binding and Electronic Level Alignment of $\boldsymbol{\pi}$-Conjugated Systems on Metals}
%
%%%%%%%%%%%
% Authors %
%%%%%%%%%%%
%
\author{Antoni Franco-Ca\~{n}ellas\orcidA{}}

\affiliation{Institut f{\"u}r Angewandte Physik, Universit{\"a}t T{\"u}bingen, Auf der Morgenstelle 10, 72076
  T{\"u}bingen, Germany
%ORICD of the author https://orcid.org/0000-0001-7767-9611
}
\author{Steffen Duhm\orcidB{}}

\affiliation{Institute of Functional Nano \& Soft Materials (FUNSOM), Jiangsu Key Laboratory for Carbon-Based
  Functional Materials \& Devices and Joint International Research Laboratory of Carbon-Based Functional
  Materials and Devices, Soochow University, Suzhou 215123, People's Republic of China
%ORICD of the author https://orcid.org/0000-0002-5099-5929
}
\author{Alexander Gerlach\orcidC{}}

\affiliation{Institut f{\"u}r Angewandte Physik, Universit{\"a}t T{\"u}bingen, Auf der Morgenstelle 10, 72076
  T{\"u}bingen, Germany
%ORICD of the author https://orcid.org/0000-0003-1787-1868
}
\author{Frank Schreiber\orcidD{}} \email{frank.schreiber@uni-tuebingen.de} \affiliation{Institut f{\"u}r Angewandte
  Physik, Universit{\"a}t T{\"u}bingen, Auf der Morgenstelle 10, 72076 T{\"u}bingen, Germany
%ORICD of the author https://orcid.org/0000-0003-3659-6718
}
\date{\today}
%
%%%%%%%%%%%%%%%%%%%%%%%%%%%%%%%%%%%%%%%%%%%%%%%%%
%%%% Abstract %%%%
%%%%%%%%%%%%%%%%%%%%%%%%%%%%%%%%%%%%%%%%%%%%%%%%%
%
\begin{abstract}
  We review the binding and energy level alignment of $\pi$-conjugated systems on metals, a field which
  during the last two decades has seen tremendous progress both in terms of experimental characterization as
  well as in the depth of theoretical understanding.
  Precise measurements of vertical adsorption distances and the electronic structure together with
  \textit{ab-initio} calculations have shown that most of the molecular systems have to be considered as
  intermediate cases between weak physisorption and strong chemisorption. In this regime, the subtle interplay
  of different effects such as covalent bonding, charge transfer, electrostatic and van der Waals interactions
  yields a complex situation with different adsorption mechanisms.
  In order to establish a better understanding of the binding and the electronic level alignment of
  $\pi$-conjugated molecules on metals, we provide an up-to-date overview of the literature, explain the
  fundamental concepts as well as the experimental techniques and discuss typical case studies. Thereby, we
  relate the geometric with the electronic structure in a consistent picture and cover the entire range from
  weak to strong coupling.
\end{abstract}


\maketitle
%
%%%%%%%%%%%%%%%%%%%%%%%%%%%%%%%%%%%%%%%%%%%%%%%%%
%%%% Main text %%%%
%%%%%%%%%%%%%%%%%%%%%%%%%%%%%%%%%%%%%%%%%%%%%%%%%
%
\tableofcontents

\section{Introduction} \label{sec:intro}
The interface between $\pi$-conjugated organic semiconductor molecules and metals is at the heart of a
number of important scientific questions, both from a fundamental as well as from an applied
perspective. It is a key issue for the different energy-level alignment (ELA) schemes as well as for charge
carrier injection/extraction efficiencies and related issues in organic (opto)electronics
devices~\cite{Ishii_1999_AdvMater, Kahn_2003_JPolymSciBPolymPhys, Koch_2008_JPhysCondensMatter,
  Widdascheck_2019_AdvFunctMater, Fahlman_2019_NatRevMater}. At the same time, already the question of the
interaction and binding is non-trivial, in particular, for systems which are between the limiting cases of
(clearly weak) physisorption and (clearly strong) chemisorption. Importantly, there is a subtle interplay
between geometric and electronic structure, with a frequently substantial (but not necessarily dominating)
contribution of dispersion interactions, which makes predictions of the metal-organic interface rather
challenging, if only ``simple rules'' are employed. Rather advanced theoretical methods, developed in the
last decade, have enabled substantial progress~\cite{Tkatchenko_2010_MRSBulletin, Berland_2015_RepProgPhys,
  Liu_2015_PhysRevLett, Maurer_2016_ProgSurfSci, Hermann_2017_ChemRev, Liu_2017_JChemPhys}. In parallel
with that, a satisfactory understanding of these systems requires the experimental determination of both
the exact adsorption (i.e., binding) geometry as well as the resulting electronic structure including
possible charge transfer, interface dipoles, and shifts of the electronic levels. Fortunately, the last
years have also seen tremendous progress in experimental results, so that we are now looking at a
reasonably large and representative set of experimental data on a number of systems, which allow a more
comprehensive discussion. This is the main goal of the present review.

We shall first emphasize the importance of the structural properties. In line with the motivation above, it
has become clear that the precise knowledge of the molecular arrangement on the surface is necessary to assess
and interpret the electronic properties and eventually the ELA. The nature of the interaction of (aromatic)
$\pi$-systems with metals is less obvious than, say, CO on Ni(111) or other chemisorbing
systems~\cite{Norsko_1990_RepProgPhys}, which can be safely assumed to exhibit a well-established chemical
bond on the one hand, and, say, noble gases, which are obviously physisorption, i.e.\,dispersion-interaction
dominated on the other hand~\cite{Diehl_2004_JPhysCondensMatter}. The interaction and interface for the
intermediate case has been subject to intense research with two largely complementary approaches:
\begin{enumerate}[I.]
\item Experimental high-precision determination of adsorption distances, mostly using the X-ray standing wave
  (XSW) technique. Remarkably, while XSW had been developed in the 1960s for the localization of interstitial
  dopants in the bulk~\cite{Zegenhagen_1993_SurfSciRep} and thereafter used also for simple adsorbates on
  surfaces~\cite{Cowan_1980_PhysRevLett, Bedzyk_1985_PhysRevB}, the first investigations of larger aromatic
  compounds were published only in 2005~\cite{Gerlach_2005_PhysRevB, Hauschild_2005_PhysRevLett}. Since then,
  numerous studies using the XSW technique have revealed that $\pi$-conjugated molecules on metals show a
  surprisingly rich phenomenology, e.g., with significant distortions of the molecules on noble metal surfaces.
  Of course, also other techniques such as photoelectron diffraction (PhD)~\cite{Hofmann_1994_SurfSci},
  rod-scans in X-ray diffraction~\cite{Krause_2003_JChemPhys} or LEED I-V~\cite{Stellwag_1995_SurfSci,
    Zheleva_2012_JPhysChemC, Sirtl_2013_PhysChemChemPhys}, which are used for structural investigations on
  surfaces, have their merits but do not exhibit the same precision and/or element specificity as the XSW
  technique.
\item Quantum theoretical calculations that managed to include long-range dispersion forces in density
  functional theory (DFT) codes~\cite{Grimme_2006_JComputChem, Rydberg_2003_PhysRevLett,
    Dion_2004_PhysRevLett, Tkatchenko_2009_PhysRevLett, Berland_2019_PhysRevB}, which became more popular
  than previous attempts involving, e.g., Hartree-Fock self-consistent field wave
  functions~\cite{Bagus_2002_PhysRevLett} or M{\o}ller-Plesset perturbation theory
  (MP2)~\cite{Abbasi_2009_JPhysChemC}. Different schemes going beyond standard DFT were developed to tackle
  the fundamental problem, how to treat exchange-correlation effects. While those approximations with
  dispersion corrections involve increased computational costs, they have become more and more accurate for
  calculating the adsorption geometry of organic molecules on metals.  In this context, the reader seeking
  more information is referred to reviews of vdW-corrected DFT~\cite{Klimevs_2012_JChemPhys,
    Berland_2015_RepProgPhys, Hermann_2017_ChemRev} and to Ref.~\onlinecite{Maurer_2016_ProgSurfSci} for its
  application in the context of metal-organic interfaces.
\end{enumerate}

Regarding the electronic properties of such systems, it has already been recognized in the 1970s that the
electronic structure of molecular solids is considerably different to that in the gas
phase~\cite{Seki_1974_BullChemSocJpn, Salaneck_1978_PhysRevLett}. However, it took twenty more years until
``energy-level alignment'' and ``interface dipoles'' for $\pi$-systems at interfaces came into the focus of
research~\cite{Ohno_1991_PhysRevB, Narioka_1995_ApplPhysLett, Lee_1998_ApplPhysLett} and the seminal review
by Ishii et al.\ has been published~\cite{Ishii_1999_AdvMater}. In the last decade a systematic
understanding and phenomenology has been established~\cite{Kahn_2003_JPolymSciBPolymPhys,
  Koch_2007_ChemPhysChem, Braun_2009_AdvMater, Hwang_2009_MaterSciEngR, Greiner_2012_NatMater,
  Opitz_2017_JPhysCondensMatter, Cinchetti_2017_NatMater, Akaike_2018_JpnJApplPhys}
\textbf{\cite{Zojer_2019_AdvMaterInterfaces}}. In particular, ELA is now relatively well studied for
multilayers on \emph{inert} substrates, i.e., if substrate-adsorbate interactions can be
neglected~\cite{Ley_2013_AdvFunctMater, Oehzelt_2014_NatCommun, Yang_2017_JPhysDApplPhys}. However, this is
frequently not the case on metal substrates, and it is clear that the complete binding scenario including
possible distortions of the adsorbate is required for a thorough understanding of
ELA~\cite{Koch_2008_JPhysCondensMatter, Monti_2012_JPhysChemLett, Klappenberger_2014_ProgSurfSci,
  Goiri_2016_AdvMater, Otero_2017_SurfSciRep}.
\begin{figure} \centering
  \includegraphics[width=\columnwidth]{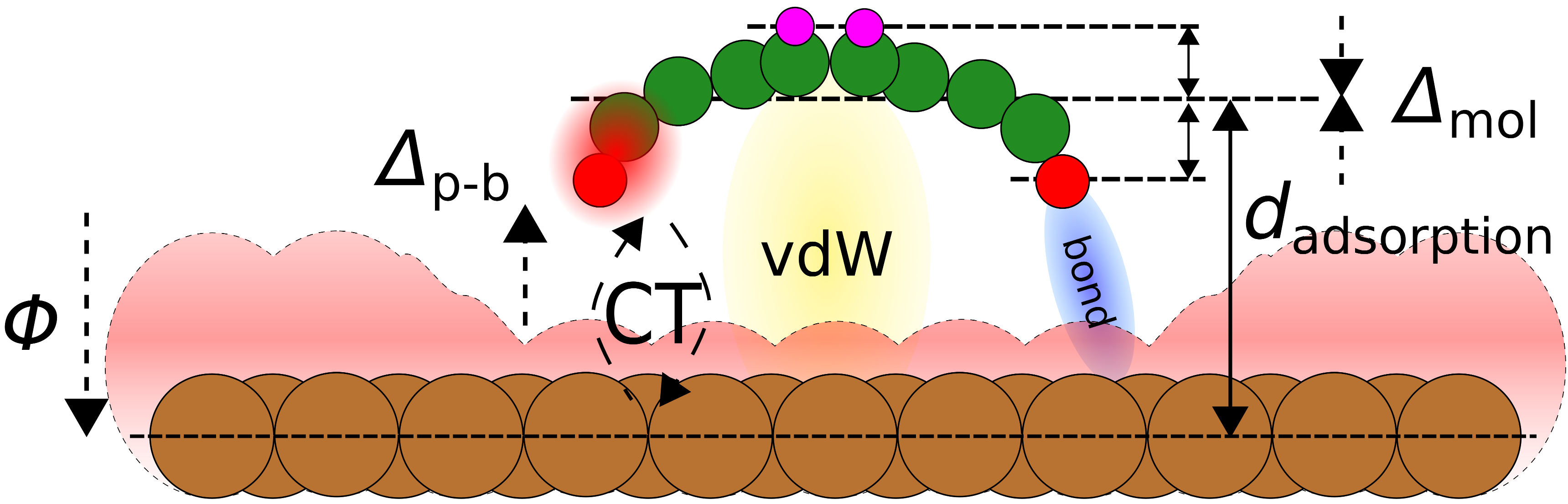}
  \caption{Sketch of the fundamental quantities and phenomena central to this review: $\phi$ is the
    substrate work function, $\Delta_\mathrm{p-b}$ the change of surface dipole due to the push-back
    effect, CT the charge transfer effects between adsorbate and substrate, vdW are van der Waals forces,
    $\Delta_\mathrm{mol}$ the intramolecular dipoles intrinsic as well as adsorption induced,
    $d_\mathrm{adsorption}$ the average adsorption distance of the molecule and ``bond'' refers to the
    possible formation of chemical bonds between the molecule and the substrate. The magnitude of the
    energy shifts (dashed arrows) is intimately related to the adsorption geometry (solid arrows) as will
    be discussed in the main text.}
  \label{fig:intro_summary}
\end{figure}

As mentioned above, it is by now accepted that ELA at organic-metal interfaces is of utmost importance for
the performance of organic (opto)electronic devices~\cite{Fung_2016_AdvMater, Lussem_2016_ChemRev,
  Nakano_2017_AdvMater, Opitz_2017_JPhysCondensMatter, Akaike_2018_JpnJApplPhys,
  Bao_2019_AdvMaterInterfaces, Rockson_2019_ACSApplMaterInterfaces, Gurney_2019_RepProgPhys}. Moreover,
energy-levels and thus charge injection barriers can be tuned by engineering the substrate work
function~\cite{Koch_2005_PhysRevLett, Li_2009_ChemMater, Zhou_2012_Science, Vilan_2017_ChemRev,
  Widdascheck_2019_AdvFunctMater}. This can be done by pre-covering a metal electrode with a monolayer of
an electron acceptor (donor) for increasing (decreasing) the effective substrate work function and thus
lowering the hole (electron) injection barrier into subsequently deposited organic
layers~\cite{Koch_2005_PhysRevLett, Broker_2008_ApplPhysLett, Rana_2012_PhysStatSolA,
  Gao_2013_ApplPhysLett, Timpel_2018_AdvFunctMat}. The contact formation at such strongly coupled
interfaces goes usually along with a complex electronic scenario involving donation and back-donation of
charges (see Figure~\ref{fig:intro_summary})~\cite{Romaner_2007_PhysRevLett, Tseng_2010_NatChem,
  Vitali_2010_NatMater, Haming_2012_PhysRevB, Hofmann_2013_NewJPhys, Stadtmuller_2014_NatComms,
  Zamborlini_2017_NatCommun, Chen_2019_JPhysCondensMatter}. Furthermore, the adsorption distances including
a possible intramolecular distortion impacting the molecular dipole are essential. This is why in-depth
discussion of the electronic structure usually requires a precise determination of the geometric structure
(Figure~\ref{fig:intro_summary}), and why XSW results have a key role in this context.
Several original research articles (e.g.\ Refs.~\onlinecite{Romaner_2007_PhysRevLett,
  Koch_2008_JAmChemSoc, Duhm_2008_OrgElec, Stadler_2009_NaturePhysics, Kroger_2010_NewJPhys,
  Heimel_2013_NatureChem, Stadtmuller_2014_NatComms, Baby_2017_ACSNano,
  Franco-Canellas_2017_PhysRevMaterials, Klein_2019_JPhysChemC, Chen_2019_JPhysCondensMatter}) and, more
recently, some review articles and book contributions (Refs.~\onlinecite{Gerlach_2013_book,
  Stadtmuller_2015_JElectronSpectroscRelatPhenom, Duhm_2015_chapter, Willenbockel_2014_PhysChemChemPhys,
  Maurer_2016_ProgSurfSci, Otero_2017_SurfSciRep, Kera_2018_JPhysSocJpn}) have demonstrated that
correlating electronic structure and vertical adsorption heights gives new insights.

\begin{figure}
  \centering
  \includegraphics[width=\columnwidth]{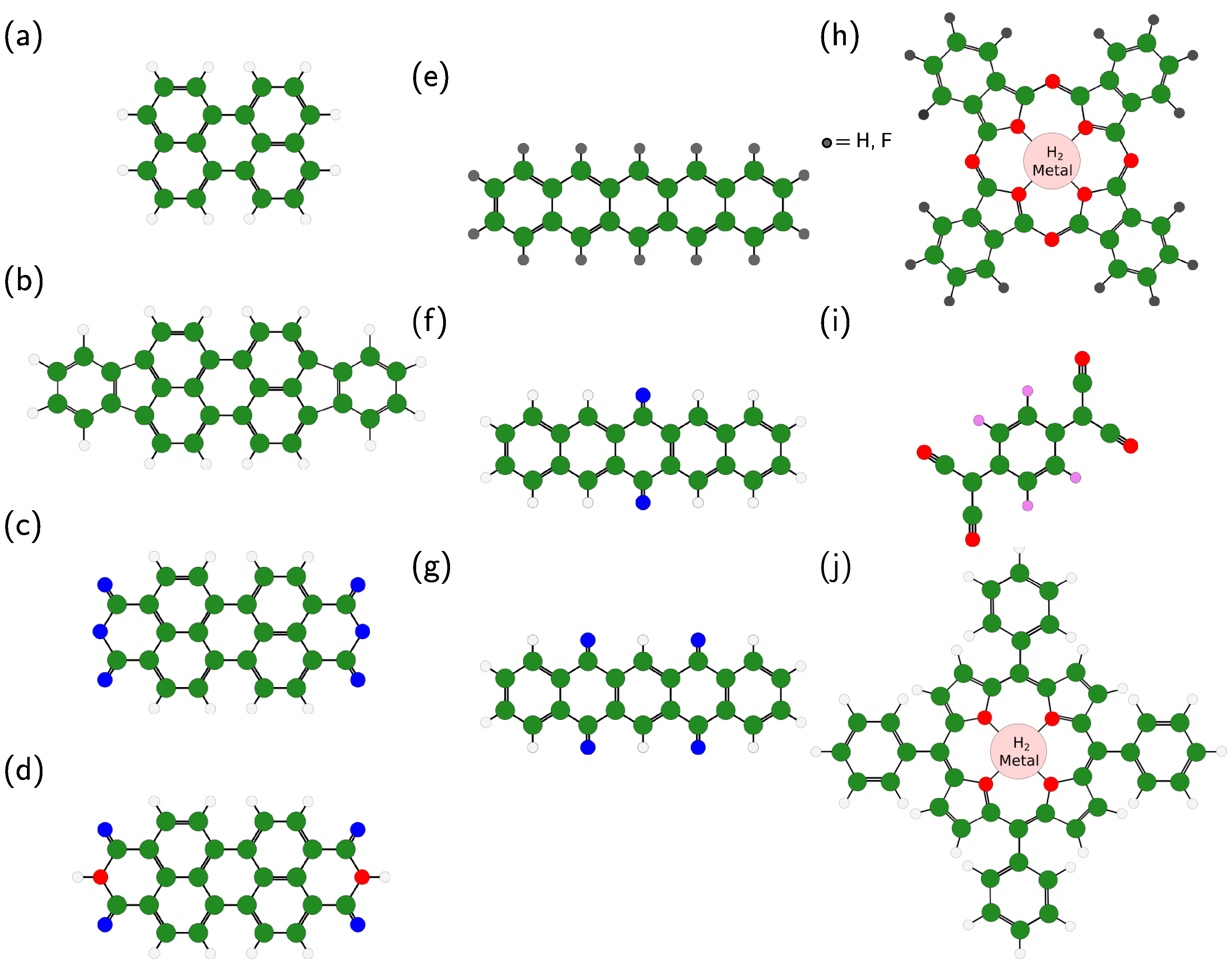}
  \caption{Chemical structure of the main molecules reviewed here (green for carbon, blue for oxygen, red
    for nitrogen, pink for fluorine and white for hydrogen). (a) Perylene. (b) Diindenoperylene (DIP). (c)
    Perylene-3,4,9,10-tetracarboxylic dianhydride (PTCDA). (d)
    Perylene-3,4,9,10-tetracarboxylic-3,4,9,10-diimide (PTCDI). (e) Pentacene (PEN) if the peripheral
    atoms are hydrogen or perfluoropentacene (PFP) if they are fluorine. (f) 6,13-pentacenequinone
    (P2O). (g) 5,7,12,14-pentacenetetrone (P4O). (h) (Metal) phthalocyanines (MePc) with or without
    perfluorination. (i) 2,3,5,6-tetrafluoro-7,7,8,8 tetracyanoquinodimethane (F$_4$TCNQ). (j) (Metal)
    tetraphenylporphyrin (MeTPP).  }
  \label{fig:molecules}
\end{figure}

This review is organized as follows: Initially, we explain the basic concepts of organic-metal contact
formation (Sec.~\ref{sec:overview}), followed by some experimental considerations related to XSW,
photoelectron spectroscopy and complementary techniques (Sec.~\ref{sec:experimental}). After providing a
comprehensive list of XSW data obtained for conjugated organic molecules (COMs, representative chemical
structures are shown in Figure~\ref{fig:molecules}) on metals, which may serve as reference and general
overview, we discuss several typical adsorbate systems (Sec.~\ref{sec:case_studies}). In each case, we explore
the geometric and electronic structure of these systems as well as how these properties are related for
different adsorption scenarios. Finally, we shall summarize the important findings (Sec.~\ref{sec:summary}).

\section{General considerations and fundamentals} \label{sec:overview}
First, we shall introduce the basic quantities, concepts and phenomena that describe and govern the
metal-organic interface, in particular with respect to the different effects influencing the ELA in the
monolayer regime.

\subsection{Interface energetics} \label{ssec:energetics}
\begin{figure}
  \centering
  \includegraphics[width=\columnwidth]{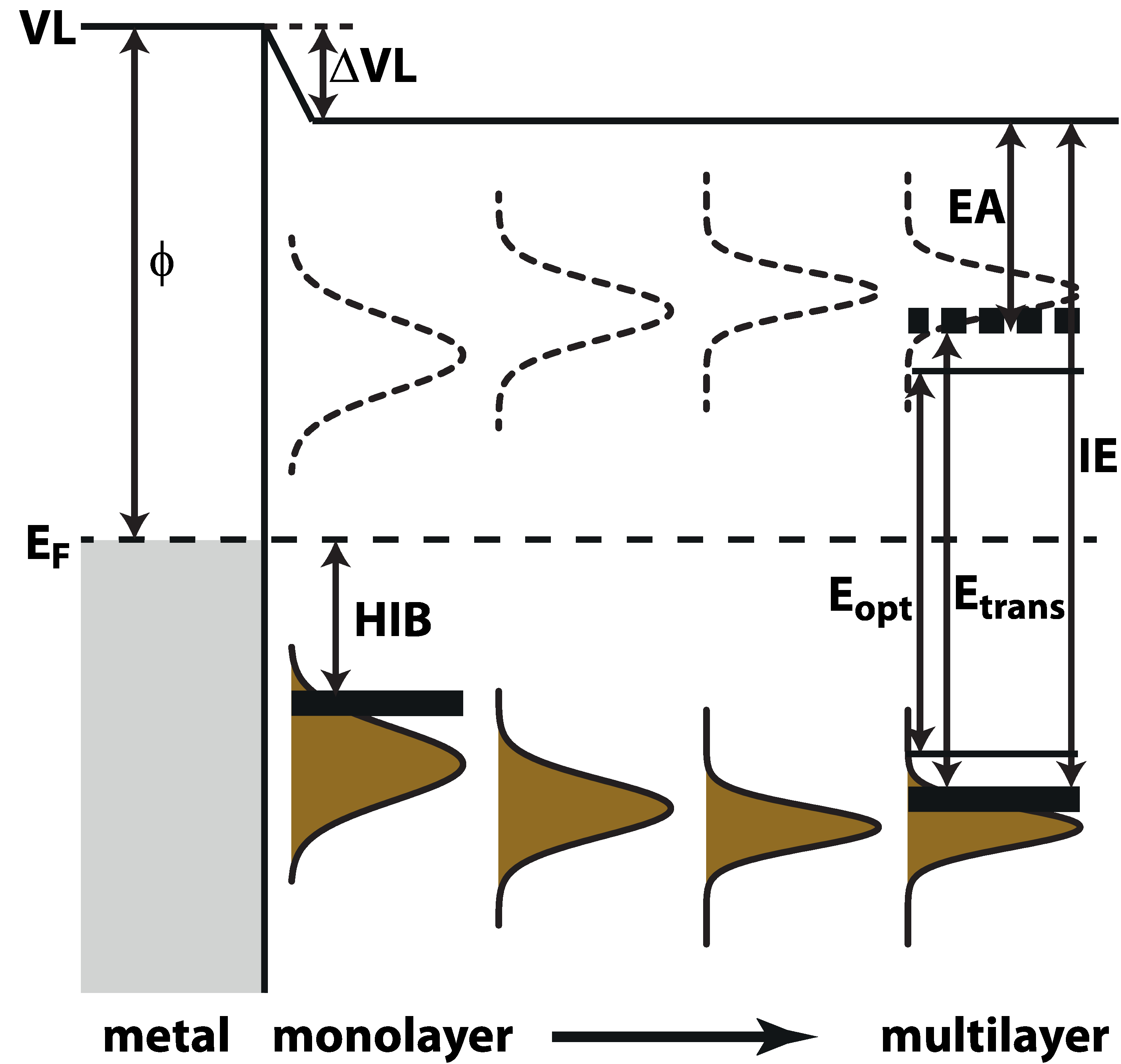}
  \caption{Schematic energy-level diagram of a weakly interacting organic-metal interface. The metal is
    characterized by its work function $\phi$, which is the energetic difference between the vacuum level (VL)
    and the Fermi level $E_F$. In the shown limiting case of physisorption, the vacuum level shift $\Delta$VL
    is due to the push-back effect. For the organic adsorbate the density of states of the frontier molecular
    orbitals (HOMO and LUMO) are approximated as Gaussian peaks and are shown from monolayer to multilayer
    coverage. The energetic difference between $E_F$ and the onset of the HOMO level defines the hole
    injection barrier (HIB). The ionization energy (IE), the electron affinity (EA), the transport gap
    $E_\mathrm{trans}$ and the optical gap $E_\mathrm{opt}$ are usually taken from multilayer measurements.}
  \label{fig:ela_schematics}
\end{figure}

The most relevant energy-levels at an organic-metal interface in the limiting case of physisorption are shown
in Figure~\ref{fig:ela_schematics}. A metal has electrons occupying energy-levels up to the Fermi level
$E_F$. The energy to bring them to the vacuum level (VL) corresponds to the metal work function $\phi$. In the
COM the most important energy-levels are those of the highest occupied molecular orbital (HOMO) and the lowest
unoccupied molecular orbital (LUMO), which are also referred to as the frontier molecular orbitals. The energy
difference between the HOMO and the LUMO defines the transport gap $E_\mathrm{trans}$.  Because typical
exciton binding energies of COM thin films are in the range of several hundred
meV~\cite{Friend_2012_FaradayDiscuss, Forrest_2015_PhilTansRSocA, Djurovich_2009_OrgElec}, i.e.\ much higher
than for most inorganic semiconductors, the optical gap $E_\mathrm{opt}$ is considerably smaller than
$E_\mathrm{trans}$~\cite{Gao_2001_ApplPhysLett, Zahn_2007_ChemRev, Bredas_2014_MaterHoriz}, with the latter
being the relevant parameter for ELA and the charge-transport characteristics of the thin film. We note that
the ionization energy (IE), which is defined as the energy necessary to move an electron from the HOMO to the
vacuum, and the electron affinity (EA), which is the energy necessary to bring an electron from the vacuum to
the LUMO, cannot be considered as materials parameters: The collective impact of intramolecular dipole
moments, which depend on the molecular orientation within the thin film, influences the IE as well as the
EA~\cite{Han_2013_ApplPhysLett, Heimel_2011_ChemMater, Duhm_2008_NatMater,
  DAvino_2016_JPhysCondensMatter}. Therefore, one has to determine these values for each specific thin film
structure.

For the IE and EA often the \textit{onsets} of experimentally determined HOMO and LUMO levels are used
\cite{Kahn_2003_JPolymSciBPolymPhys, Zahn_2006_ChemPhys}, (\textit{cf.}~Figure~\ref{fig:ela_schematics})
because the onsets govern the transport properties \cite{Horowitz_2015_JApplPhys}. However, the onset of a
peak measured by (inverse) photoemission depends, naturally, on the experimental resolution. Furthermore,
the signal-to-noise-ratio can also play a significant role for the onset, especially if the peak shape is
not simply Gaussian and/or gap states are involved \cite{Zu2019JPhysChemLett,
  Yang_2017_JPhysDApplPhys}. Whether the use of onsets or peak maxima is more beneficial depends on the
specific adsorbate/substrate system and the scientific question. Unfortunately, no general convention has
been established yet and, consequently, great care has to be taken when comparing values from different
publications or when comparing experiment and theory.

For COM thin films polarization leads to a rearrangement of energy-levels in the solid state compared to the
gas phase~\cite{Hill_2000_ChemPhysLett, Schwoerer_2007_book, DAvino_2016_JPhysCondensMatter,
  Liu_2017_JChemPhys, Li_2018_PhysRevB}. The polarizability of metals is, in general, much higher than that of
organic thin films. The image-charge effect (often called screening) leads thus to a further narrowing of the
transport gap in monolayers on a metal substrate compared to multilayers~\cite{Hill_2000_JApplPhys,
  Koch_2007_ChemPhysChem} as shown in Figure~\ref{fig:ela_schematics}. Moreover, even for physisorption the
vicinity of a metal leads to broadening of the energy-levels through electronic, quantum-mechanical
interaction of the localized molecular states with the continuum of metal states~\cite{Anderson_1961_PhysRev,
  Heimel_2016_chapter, Willenbockel_2014_PhysChemChemPhys}.

\begin{figure}
  \centering
  \includegraphics[width=\columnwidth]{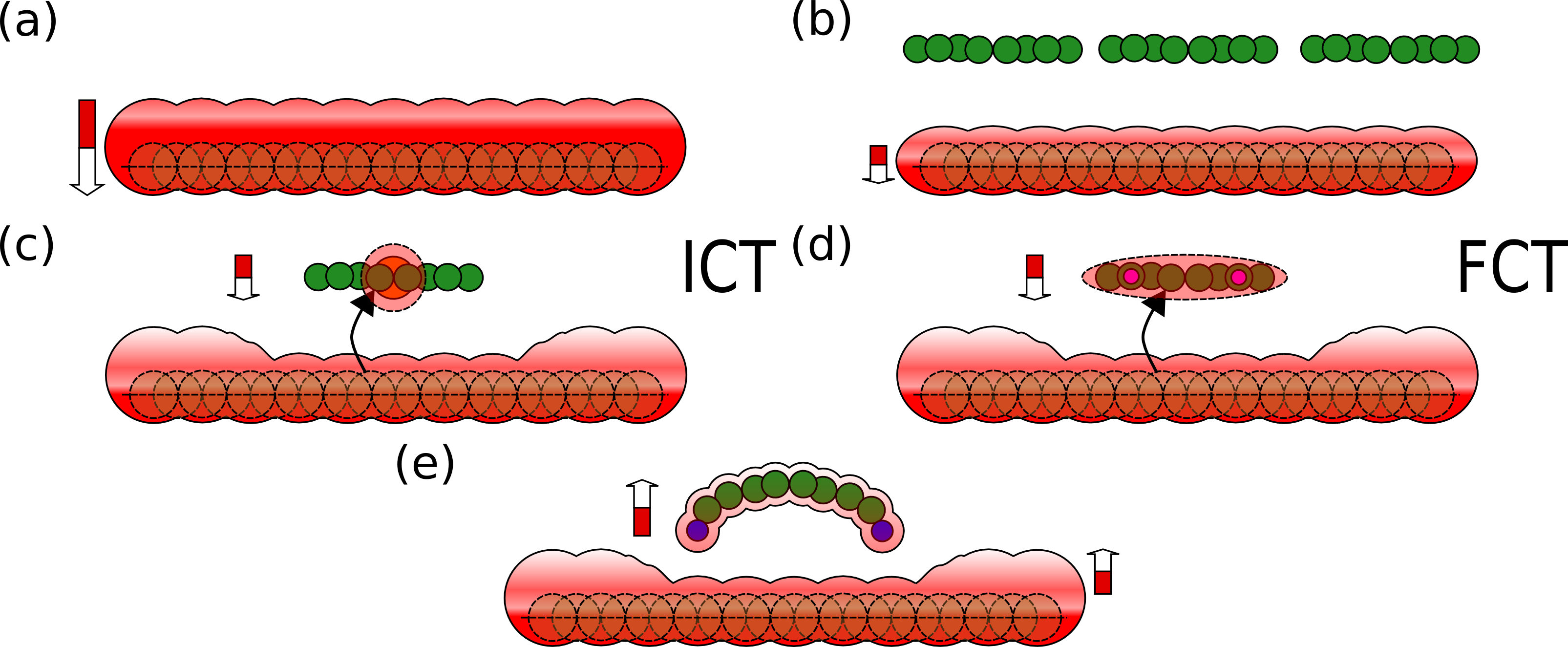}
  \caption{Dipoles at organic-metal interfaces. (a) The electron density spilling out of a clean metal surface
    gives rise to a surface dipole. (b) This surface dipole is weakened by the adsorption of COMs (push-back
    effect due to Pauli repulsion). (c), (d) Charge transfer yields an additional interface dipole. The charge
    transfer can be integer (ICT) or fractional (FCT). (e) Adsorption induced distortions or the adsorption of
    polar COMs can lead to additional dipoles.}
  \label{fig:interface_dipole}
\end{figure}

Upon contact formation of a COM and a metal, vacuum level alignment is rather the exception than the
rule~\cite{Li_2017_AdvMater, Koch_2008_JPhysCondensMatter, Kahn_2003_JPolymSciBPolymPhys, Braun_2009_AdvMater,
  Ishii_1999_AdvMater, Egger_2015_NanoLett, Akaike_2018_JpnJApplPhys}. There are various reasons for vacuum
level shifts $\Delta$VL upon contact formation, which are not restricted to metal substrates, but may also
take place when the molecules are adsorbed on inorganic semiconductors and
insulators~\cite{Greiner_2012_NatMater, Hewlett_2016_AdvMater, Fu_2018_ACSApplMaterInterfaces,
  Erker_2019_JPhysChemLett, Futscher_2019_JPhysCondensMatter}. The magnitude of interface dipoles is often
related to vertical adsorption distances and the most relevant possible contributions as they are illustrated
in Figure~\ref{fig:interface_dipole} are:

\begin{enumerate}[I.]
\item \emph{Push-back effect} $\Delta_\mathrm{p-b}$ caused by the Pauli repulsion between the electrons of
  the adsorbate and the metal
\item \emph{Charge transfer} between adsorbate and substrate
\item \emph{Chemical bond formation} between adsorbate and substrate
\item \emph{Molecular dipole moment} $\Delta_\mathrm{mol}$, which can be intrinsic (polar COMs) or due to
  adsorption induced distortions
\end{enumerate}

\begin{figure}
  \centering
  \includegraphics[width=0.8\columnwidth]{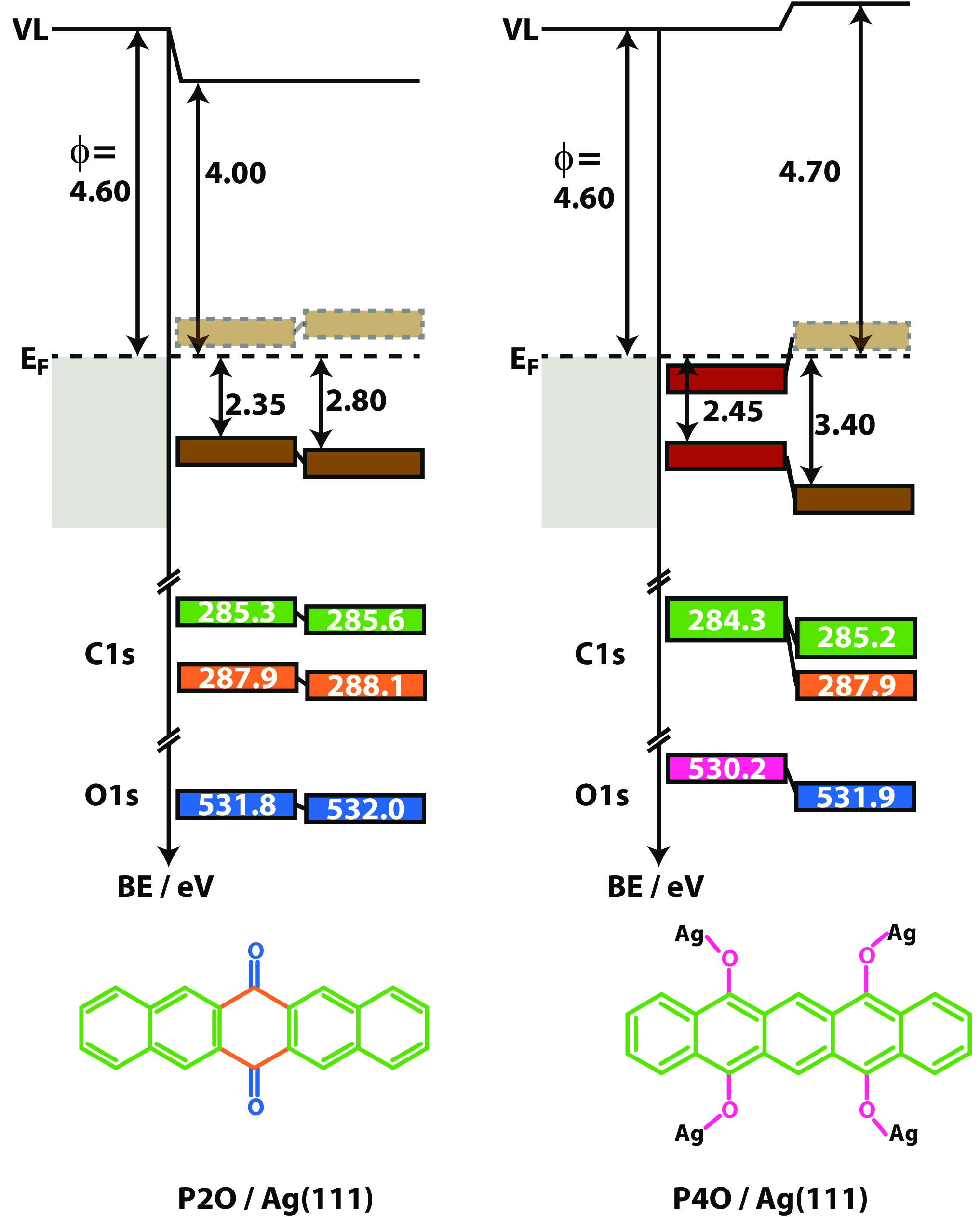}
  \caption{Schematic energy-level diagrams for P2O and P4O on Ag(111). On the left the Ag substrate with its
    work function $\phi$ and Fermi level $E_F$ is displayed. The middle panels correspond to a P2O monolayer,
    with the position of the vacuum level (VL), the position of the (former) LUMO, the HOMO position and the
    energetic position of C\,1s and O\,1s core-levels. In the right panels the corresponding values for
    multilayer coverage are displayed. All binding energy values are given in eV, energy axes are not to
    scale. The molecular structures on the bottom show possible resonance structures in the monolayer. The
    energy-level diagrams are drawn using UPS and XPS data published in
    Refs.~\onlinecite{Heimel_2013_NatureChem, Wang_2018_JPhysChemC}.}
  \label{fig:ELA_concept}
\end{figure}

For the physisorbed system shown in Figure~\ref{fig:ela_schematics} only the push-back effect is
considered. For systems with stronger interactions the impact of the interfacial coupling on $\Delta$VL has to
be taken into account. In general, whether an adsorbate is physisorbed or chemisorbed on a substrate is
clearly defined by adsorption energies~\cite{Atkins_2006_book} and can be accessed
theoretically~\cite{Liu_2013_NatCommun, Jakobs_2015_NanoLett, Maurer_2016_ProgSurfSci, Hollerer_2017_ACSNano,
  Sarkar_2018_ChemMater, Yang_2018_ACSApplMaterInterfaces}. However, the adsorption type is not directly
accessible by standard experimental techniques. To overcome this issue we use a simple definition based on
peak shifts between mono- and multilayer in photoemission data~\cite{Wang_2018_JPhysChemC}: For rigid shifts
of valence electron features (typically the HOMO-derived peak) and core-levels we assume physisorption and
chemisorption in all other cases. Within this definition it becomes apparent that the pentacene oxo-derivative
P2O is physisorbed on Ag(111) and P4O is chemisorbed on the same substrate. Thus, we use schematic
energy-level diagrams based on photoemission data~\cite{Heimel_2013_NatureChem, Wang_2018_JPhysChemC} of P2O
and P4O on Ag(111) (Figure~\ref{fig:ELA_concept}) to illustrate the impact of organic-metal coupling strength
on interface dipoles and ELA.

The push-back effect, which leads to $\Delta_\mathrm{p-b}$, is related to the electron density spilling
out into vacuum at clean metal surfaces~\cite{Wandelt_1997_ApplSurfSci, Smoluchowski_1941_PhysRev,
  Witte_2005_ApplPhysLett}. Push-back takes place at virtually all organic-metal interfaces as the surface
dipole part of the metal work function will be decreased by the mere presence of the molecular monolayer
\cite{Bagus_2002_PhysRevLett, DeRenzi_2005_PhysRevLett, Koch_2007_ChemPhysChem}. There is a clear
correlation between $\Delta_\mathrm{p-b}$ and adsorption distances~\cite{Rusu_2010_PhysRevB,
  Egger_2015_NanoLett, Toyoda_2010_JChemPhys, Ferri_2017_PhysRevMaterials}. For physisorbed systems the
push-back effect is often the main contribution to $\Delta$VL and can be held responsible for most of the
0.60\,eV shift at the P2O/Ag(111) interface.

For the discussion of interfacial charge transfer it is helpful to distinguish between \emph{integer} and
\emph{fractional} charge transfer (Figure~\ref{fig:interface_dipole}c and~d)~\cite{Winkler_2015_MaterHoriz,
  Hollerer_2017_ACSNano, Hofmann_2015_ACSNano}. While the latter is usually related to chemical bond
formation, the former can also occur for weakly interacting systems and is then a result of \emph{Fermi level
  pinning}~\cite{Braun_2009_AdvMater, Khoshkhoo_2017_OrgElec}. This happens for high (or low) substrate work
functions for which a vacuum level controlled ELA would lead to a situation with the HOMO (LUMO) being above
(below) the Fermi level. In such cases thermodynamic equilibrium is maintained by an interfacial charge
transfer. Thus, notably, also in the absence of any specific interfacial interaction charge transfer across
an organic-inorganic interface can take place. Interestingly, the HOMO- (LUMO-) levels are typically pinned
several hundred meV below (above) $E_F$~\cite{Braun_2009_AdvMater, Ley_2013_AdvFunctMater,
  Oehzelt_2014_NatCommun, Yang_2017_OrgElec, Horowitz_2015_JApplPhys, Fukagawa_2007_AdvMater}. This is due to
a certain degree of disorder in molecular thin films leading to a broadening of HOMO and LUMO
density-of-states (DOS)~\cite{Oehzelt_2014_NatCommun, Yang_2017_JPhysDApplPhys}. The relationship between DOS
shape and ELA has been addressed in several publications~\cite{Zuo_2016_PhysRevB, Horowitz_2015_JApplPhys,
  Kalb_2010_PhysRevB, Oehzelt_2014_NatCommun, Yang_2017_JPhysDApplPhys} and is beyond the scope of this
review. Likewise, for ELA at organic-organic interfaces, the reader is referred to
Refs.~\onlinecite{Tang_2007_JApplPhys, Braun_2009_AdvMater, Poelking_2014_NatMater, Chen_2011_AdvFunctMater,
  Opitz_2017_JPhysCondensMatter, Li_2017_AdvMater, Oehzelt_2015_SciAdv}.

\begin{figure}
  \centering
  \includegraphics[width=0.95\columnwidth]{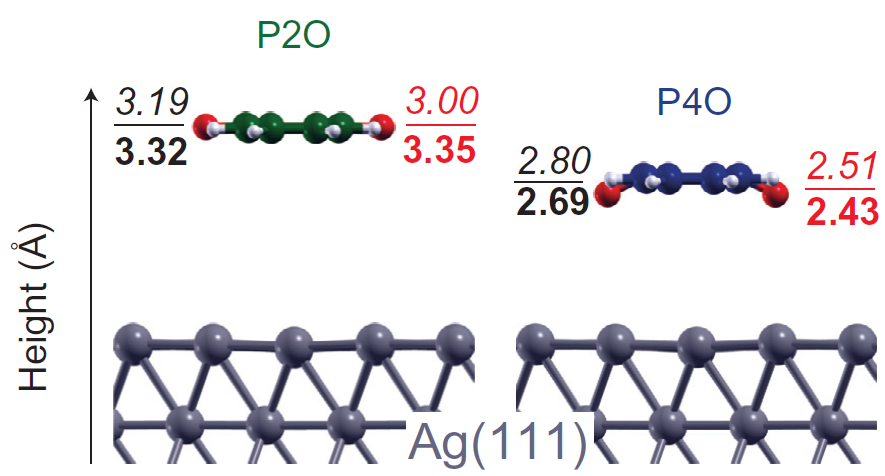}
  \caption{Vertical adsorption heights of P2O and P4O on Ag(111). Bold numbers refer to experimental and
    italic numbers to theoretical results, black numbers to carbon and red numbers to oxygen. For better
    visibility, the molecular distortions are not drawn to scale. Taken from
    Ref.~\onlinecite{Heimel_2013_NatureChem}.}
  \label{fig:dH_PxO_Ag111}
\end{figure}

The above mentioned screening effect leads to rigid energy-level shifts (typically several hundred meV) of
valence and core-levels to higher binding energies between monolayer and multilayer coverage of organic thin
films~\cite{Hill_2000_JApplPhys, Koch_2007_ChemPhysChem}. This is the case for physisorbed P2O on Ag(111)
(Figure~\ref{fig:ELA_concept}). For chemisorbed systems, the expected shifts due to screening can be
overcompensated by the strong chemical coupling at the organic-metal interface. This becomes apparent for P4O
on Ag(111); in this particular case, chemisorption goes along with a filling of the former LUMO. The charge
transfer counteracts the VL decrease by push back leading to a constant VL upon contact formation. The
apparent vacuum level alignment is, however, most likely coincidental. For related systems also a pronounced
\emph{increase} in the effective metal work function has been observed~\cite{Duhm_2006_JPhysChemB,
  Kawabe_2008_OrgElec, Koch_2005_PhysRevLett}. Such systems will be discussed in more detail in
Sec.~\ref{sec:CTC}. The relatively strong chemisorption of P4O on Ag(111) leads to a rehybridization of the
molecules in the monolayer (a possible resonance structure is shown in the bottom of
Figure~\ref{fig:ELA_concept}). This is in line with the experimentally determined vertical adsorption
distances (Figure~\ref{fig:dH_PxO_Ag111}), which show a pronounced distortion of P4O upon adsorption on
Ag(111)~\cite{Heimel_2013_NatureChem}.

Overall, the PxO/Ag(111) systems (Figure~\ref{fig:ELA_concept}) demonstrate some potential pitfalls in
interpreting energy-level diagrams: For P2O/Ag(111) an apparent interface dipole mimics strong interaction,
whereas the charge transfer at the P4O/Ag(111) interface leads to apparent vacuum level alignment. Thus,
additional information is necessary to fully understand and describe organic-metal interface energetics. In
particular, a precise knowledge of the vertical adsorption distance is necessary for a proper description of
the adsorption behavior.

\subsection{Role of the substrate} \label{ssec:concepts_substrate}
As discussed above, one can distinguish two limiting cases within the domain of metal-organic interactions,
namely, physisorption and chemisorption. The adsorption distances are therefore expected to range between
the sum of the van der Waals radii $\sum r_\mathrm{vdW}$ for pure physisorptive bonding and the sum of the
covalent radii $\sum r_\mathrm{cov}$ for pure chemisorptive bonding. The corresponding values for carbon
atoms interacting with the three noble metals are given in Tab.~\ref{tab:substrate_parameters}. Obviously,
the differences between van der Waals and covalent bonding for a given substrate material (being
$1.1-1.3$\,\AA) are much larger than the differences related to the choice of the substrate, i.e.\ Cu, Ag
or Au.  XSW experiments, however, consistently show that typical adsorption distances on these substrates
are not similar and that the different \emph{reactivity} of those materials is a key factor. For that
purpose, the electronic properties of the substrates have to be discussed in some
detail~\cite{Brivio_1999_RevModPhys}.

\begin{table*}
  \begin{ruledtabular}
    \caption{Selected substrate parameters: Atomic number $Z$; sum of van der Waals radii $\sum
      r_\mathrm{vdW}$ for carbon and noble metal atoms~\cite{Bondi_1964_JPhysChem}; sum of covalent radii
      $\sum r_\mathrm{cov}$ for carbon and noble metal atoms~\cite{Cordero_2008_DaltonTrans}; lattice plane
      spacing $d_0$ for the (111) Bragg reflection; corresponding photon energy $E_\mathrm{Bragg} = hc/2d_0$
      in back-reflection ($\theta_\mathrm{Bragg} \approx 90^\circ$); work function $\phi_\mathrm{subs}$ of the
      bare substrates~\cite{Derry_2015_JVacSciTechnolA}.  Note that the small surface relaxations of Cu(111)
      \cite{Lindgren_1984_PhysRevB} and Ag(111)~\cite{Statiris_1994_PhysRevLett} are often neglected for the
      determination of the adsorption distances, whereas the reconstruction of Au(111)
      ~\cite{Takeuchi_1991_PhysRevB, Sandy_1991_PhysRevB} should be taken into account.}
    \label{tab:substrate_parameters}
    \begin{tabular}{ l| c c c c c c c}
      & $Z$ & $\sum r_\mathrm{vdw}$\,(\AA) & $\sum r_\mathrm{cov}$\,(\AA) & $d_0$\,(\AA) & $E_\mathrm{Bragg}$\,(keV) & $\phi_\mathrm{subs}$\,(eV) & (111) surface\\
      \hline
      Cu    & 29 & 3.17	& 2.08 & 2.086	& 2.972	 & $\sim$4.9 & small relaxation~\cite{Lindgren_1984_PhysRevB} \\
      \hline
      Ag    & 47 & 3.49	& 2.21 & 2.357	& 2.630	 & $\sim$4.6 & small relaxation~\cite{Statiris_1994_PhysRevLett} \\
      \hline
      Au    & 79 & 3.43	& 2.12 & 2.353	& 2.634	 & $\sim$5.3 & $(22 \times \sqrt{3})$ herringbone  reconstruction~\cite{Takeuchi_1991_PhysRevB,Sandy_1991_PhysRevB} \\
    \end{tabular}
  \end{ruledtabular}
\end{table*}
		
\begin{figure}
  \centering
  \includegraphics[width=\columnwidth]{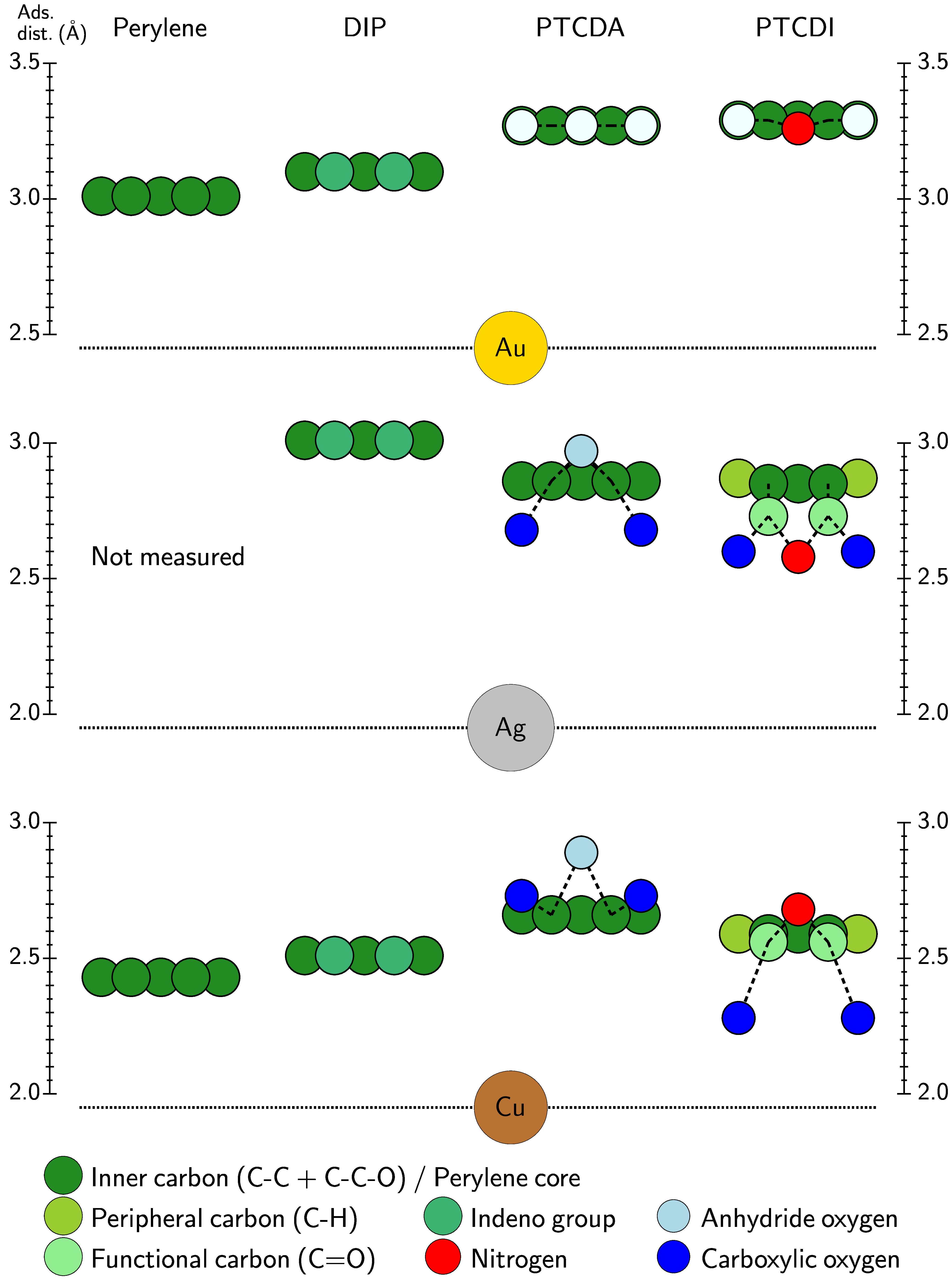}
  \caption{Experimentally determined vertical adsorption distances of perylene (derivatives) on the
    (111)-surfaces of the noble metals. Adapted from Ref.~\onlinecite{Franco-Canellas_2017_PhysRevMaterials}
    with permission. The data is taken from these references: Perylene and PTCDI from
    Ref.~\onlinecite{Franco-Canellas_2017_PhysRevMaterials}, PTCDA on Au(111) from
    Ref.~\onlinecite{Henze_2007_SurfSci}, PTCDA on Ag(111) from Ref.~\onlinecite{Hauschild_2010_PhysRevB},
    PTCDA on Cu(111) from Ref.~\onlinecite{Gerlach_2007_PhysRevB}, DIP from
    Ref.~\onlinecite{Burker_2013_PhysRevB}.}
  \label{fig:perylene_derivatives}
\end{figure}

In metals, narrow $d$- and broad $sp$-bands form the valence-band states, where the latter are more likely to
interact with a given adsorbate. At a certain distance, the molecular orbitals will start to overlap with
those of the surface atoms. Initially, the adsorbate orbitals will broaden and shift in energy (see
Figure~\ref{fig:ela_schematics}) as a consequence of the interactions with the rather delocalized
$sp$-electrons and only if the $d$-orbitals are involved will the adsorbate levels split into bonding and
antibonding states, generally one being below and the other above the metal band. In this context, one can
relate the interaction strength and the degree of chemisorption to the different orbitals involved. For
instance, the term \emph{weak chemisorption} is used for the case where only $sp$-orbitals are involved. When
$d$-orbitals are also at play, the filling of the bonding and antibonding states influences the interaction
strength as well. Thus, a strong bond is associated to the filling of only bonding states. Conversely, the
partial or total filling of antibonding states induces a repulsive interaction that counteracts the attractive
forces exerted by the $sp$-electrons. The degree of filling is related to the relative position of the
$d$-states with respect to the Fermi level. Also, the broadening of these states is responsible for the degree
of repulsion with the adsorbate states. Indeed, a broader state increases the overlap with the adsorbate
orbitals and subsequently the cost of orthogonalizing the wave functions to avoid Pauli repulsion. In light of
this, moving from left to right in the periodic table, i.e.\ from transition to coinage metals, the outmost
$d$-states shift down in energy away from the Fermi level~\cite{Hammer_1995_Nature}, thus explaining the
decreasing reactivity within this series. The broadening of the band, on the other hand, increases when moving
down the column or from right to left in the periodic table, which explains why Cu is said to be more reactive
than Au. This trend is also reflected in the averaged vertical adsorption distances $d_H$ of the carbon atoms
in the molecular backbone of adsorbates on such surfaces. Figure~\ref{fig:perylene_derivatives} shows that
$d_H$ decreases for each perylene derivative on the (111)-surfaces of noble metals in the order
Au--Ag--Cu. This finding can be considered as a qualitative trend for most COMs on these surfaces, but precise
quantitative predictions can only be done if the nature of the adsorbate is taken into account.

For the interaction with a given adsorbate, not only the chemical composition of the bulk crystal is
important, but also its surface structure and termination. Both the transfer of charge across the interface
and the formation of bonds often need some energy barriers to be overcome.  For all metal substrates the work
function $\phi$ decreases with increasing ``openness'' of the surface being considered. Thus, closed-packed
surface structures, i.e.\ fcc(111), bcc(110), and hcp(001), show the highest $\phi$ and the lowest
reactivity. Likewise, defects, step edges and kinks act as interaction centers for adsorbates, which in some
cases migrate across the flat terraces until they find a suitable location. In all these cases, the
electronic and/or chemical interaction, with the extreme case of adsorbate dissociation, is favored by the
lower energy barriers caused by elements that disrupt the surface potential landscape due to dangling bonds
or excess/defect of charges, which may be recovered by the interaction with the adsorbate.
		
Of particular importance for CT effects is the presence of surface states, which form as a consequence of the
reduced coordination of the topmost atoms compared to those in the bulk~\cite{Shockley_1939_PhysRev,
  Reinert_2001_PhysRevB}. Due to the termination of the crystal and the change of the electronic band
structure new states confined to a region very close to the surface may exist. While these states appear even
on perfect surfaces, the presence of defects, impurities or even adsorbates may create new \emph{interface
  states} localized around them. Similar to the doping in semiconductors, surface/interface states may act as
a center for charge exchange or reaction when a certain adsorbate is present.

In the context of this review, a large fraction of the studies in the literature have focused on the
(111)-surfaces of Au, Ag and Cu. These are relatively inert and less prone to reacting with aromatic
adsorbates. Also, for the noble metals they are the ones with the lowest energy, meaning that they are
preferred in evaporation processes, giving them a slightly higher practical relevance than, e.g., (110) and
(100).  Recently, also other orientations of the noble metals~\cite{Mercurio_2013_PhysRevB,
  Weiss_2017_PhysRevB, Felter_2019_Nanoscale}, alloys~\cite{Bauer_2016_PhysRevB,
  Stadtmueller_2019_JPhysCondensMatter} as well as ZnO~\cite{Hewlett_2016_AdvMater,
  Erker_2019_JPhysChemLett, Niederhausen_2020_PhysRevMaterials} have been investigated, see the list in
Sec.~\ref{sec:case_studies}. For more details on the substrate surface without adsorbates, we refer to
Ref.~\onlinecite{Woodruff_2010_JPhysCondensMatter}

\subsection{Role of the molecule} \label{ssec:concepts_molecule}
The description of organic molecules is largely based on the concept of \emph{localized
  bonds}~\cite{Bader_1991_ChemRev}. On metal surfaces, however, this approach might be questioned and is
scrutinized, e.g., by specific chemical modifications of the $\pi$-conjugated systems being investigated. It
is well known and understood how functional groups impact gas phase properties of
COMs~\cite{Liang_2013_AngewChemIntEd, Li_2012_AccChemRes, Bunz_2015_AccChemRes,
  Pecher_2019_WIREsComputMolSci}. The particular nature of those functional groups may stabilize the COM or
modify the HOMO-LUMO gap and other energy-levels. For instance, electronegative side-groups like fluorine
generally increase the EA and render the COM thus more n-type~\cite{Sakamoto_2004_JAmChemSoc}.

For molecules in contact with the metal substrate functionalization can lead to additional effects like
fostering or hindering intermolecular interactions and thereby increasing or decreasing the interaction
strength. That way, e.g.\ perfluorination of pentacene reinforces the repulsion with metal substrates and
can change the interaction from chemisorption to physisorption~\cite{Koch_2008_JAmChemSoc}. Moreover, in
the contact layer the desired functionalization effect can even be nullified as shown in the bottom of
Figure~\ref{fig:ELA_concept} for P4O: The possible resonance structure of weakly interacting P2O molecules
in the contact layer to Ag(111) are identical to the gas phase structure. Importantly, the conjugation
does not extend over the pentacene backbone but is broken by the keto-groups. For chemisorbed P4O on
Ag(111), however, by re-hybridization on the surface the conjugation can extend over the entire backbone
of the molecule and thereby resemble PEN molecules~\cite{Heimel_2013_NatureChem}. For perylene
derivatives, substitution can lead to significant differences of the adsorption distances and adsorption
induced distortions~\cite{Franco-Canellas_2017_PhysRevMaterials}, which are especially pronounced on the
relatively reactive Ag(111) and Cu(111) surfaces (Figure~\ref{fig:perylene_derivatives}).

Notably, all \emph{site-specific interactions} affect also the electronic structure and can therefore be
used to tailor interface energetics~\cite{Peisert_2015_JElectronSpectroscRelatPhenom,
  Romaner_2007_PhysRevLett, Yamane_2010_PhysRevLett}.  In general, a competition of adsorbate-substrate
interaction between the $\pi$-system of the COM and the functional groups can take place. For example, a
submonolayer of the acceptor molecule HATCN (1,4,5,8,9,11-hexaazatriphenylenehexacarbonitrile,
C$_{18}$N$_{12}$) is lying flat on Ag(111) to maximize the interaction of the $\pi$-system and the
substrate. Increasing the coverage to a full monolayer, however, induces a re-orientation of the HATCN to
an edge-on geometry due to the efficient interaction of the cyano-groups with the
substrate~\cite{Broker_2010_PhysRevLett}.  Moreover, the flexibility of the COM plays also an important
role.  While peripheral substitution of the molecules often leads to large adsorption induced molecular
distortions~\cite{Romaner_2007_PhysRevLett, Franco-Canellas_2018_PhysRevMaterials}, functional groups
belonging to a central part of the conjugated molecular backbone induce no or only negligible
distortions~\cite{Yang_2016_PhysRevB} -- even at strongly coupled organic-metal interfaces.

\subsection{Role of in-plane interactions} \label{ssec:concepts_intermolecular}
While the focus of this review is on the vertical interactions, we may briefly comment on the impact of
lateral forces. Obviously, the influence of the surrounding molecules dominates the purely organic
environment of the multilayer regime, most prominently through the $\pi$--$\pi$ interactions of adjacent
molecules~\cite{Salzmann_2012_ACSNano, Hinderhofer_2012_ChemPhysChem}. For a monolayer on a metal, though,
the molecule-molecule (i.e.\ in-plane) interactions are usually much weaker than those between the molecules
and the substrate. Thus, lateral interactions are often only a small correction, and the substrate largely
controls the interface properties and the ELA. 
There are two notable exceptions, though. One is for molecules with a large intrinsic molecular
dipole. For these, the electrostatic interaction of nearby molecules, which can be experimentally tuned
via the molecular coverage, influences the alignment of the molecular dipoles and is directly responsible
for the overall interface dipole, which in turn induces work-function changes of the
substrate~\cite{Fukagawa_2006_PhysRevBa}. The other is for heteromolecular monolayers adsorbed on metal
substrates~\cite{Stadtmuller_2015_JElectronSpectroscRelatPhenom, Goiri_2016_AdvMater,
  Bouju_ChemRev_2017}. In this case, combining pairs of donor-acceptor molecules has been proven to be an
effective method to tune the metal work function~\cite{El-Sayed_2013_ACSNano}.

From a more fundamental perspective, it is known that an increased intermolecular interaction can weaken the
molecule-substrate coupling and vice versa~\cite{Kilian_2008_PhysRevLett}, as evidenced by changes in the
adsorption distance and the frontier orbitals of the molecule. In this regard, for homomolecular systems, the
balance favoring one or the other may be tuned by changing the temperature~\cite{Kilian_2008_PhysRevLett,
  Scholl_2010_Science, Kroger_2010_NewJPhys, Kroger_2011_PhysRevB}, the
coverage~\cite{Stadler_2009_NaturePhysics, Duhm_2013_ACSApplMaterInterfaces} as well as the nature of the
substrate~\cite{Wiener_2013_PhysRevB, Lu_2016_JPhysCondensMatter, Franco-Canellas_2017_PhysRevMaterials}. A
nice example of this is found in Ref.~\onlinecite{Kilian_2008_PhysRevLett} with STM and XSW measurements of
PTCDA taken at RT, which show the well-known herringbone structure, and at LT, where the first layer becomes
disordered. The decrease of the intermolecular interactions at LT lowers the average adsorption distance of
PTCDA and increases the bending of the oxygen atoms towards the surface and goes along with an increased
filling of the former LUMO level~\cite{Kilian_2008_PhysRevLett}, all pointing towards an enhanced coupling
with the substrate.  For a detailed discussion of the in-plane arrangement of molecules and their epitaxy
with the substrate, the reader is referred to Refs.~\onlinecite{Mannsfeld_2006_ModPhysLettB,
  Hooks_2001_AdvMater, Forker_2017_SoftMatter}.

\section{Experimental methods} \label{sec:experimental}
Pivotal to this review are studies performed with the XSW technique and photoelectron spectroscopy. In this
section we will give a general overview of the fundamentals and the experimental challenges of these, mainly
within the context of organic-metal interfaces. Some other complementary techniques in this context will also
be mentioned without going into much detail or claiming to be exhaustive.

\subsection{The X-ray standing wave technique} \label{ssec:experimental_xsw}
\begin{figure*} \centering
  \includegraphics[width=0.9\textwidth]{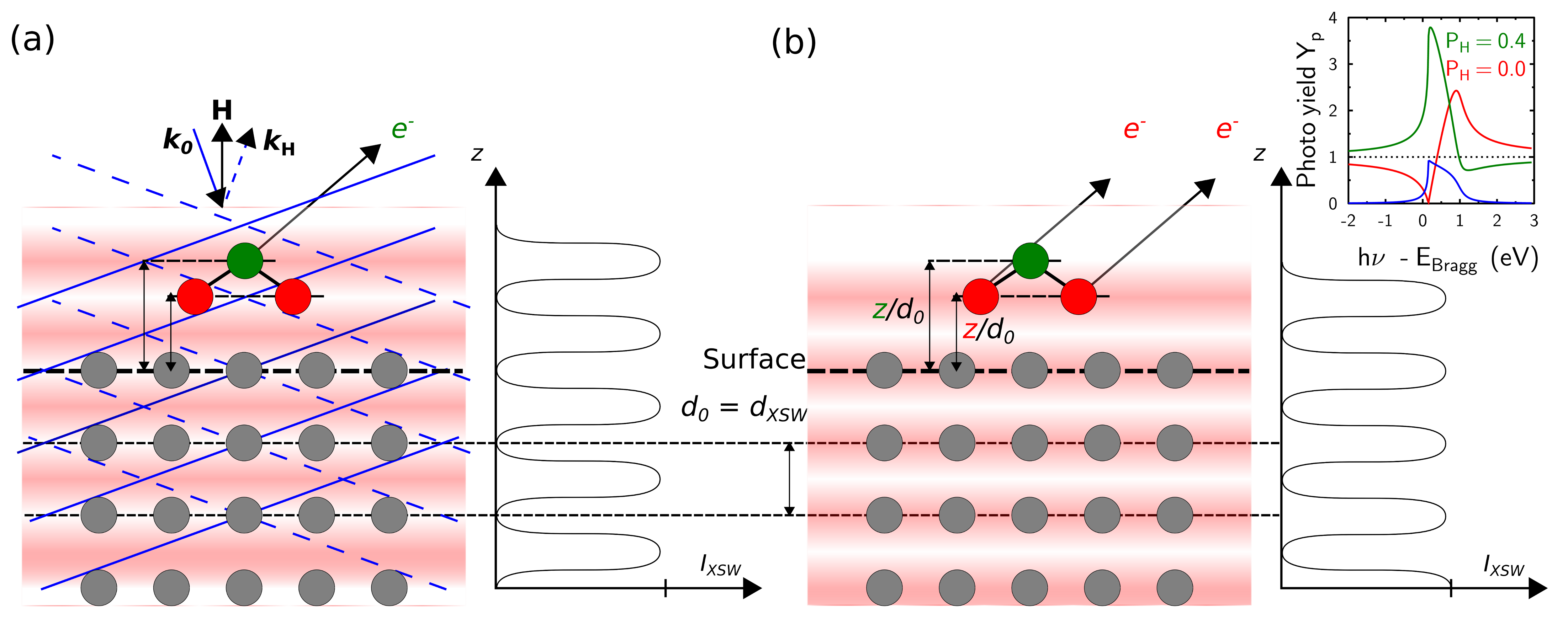}
  \caption{Schematics of the XSW-field formation. (a) An incoming X-ray plane wave with wave vector
    $\mathbf{k}_0$ interferes with the Bragg-reflected wave with the corresponding wave vector
    $\mathbf{k}_\mathrm{H}$ and creates an interferences field inside as well as above the crystal
    surface. Here, the periodicity between maxima (or minima) of intensity $d_\mathrm{XSW}$ equals the
    lattice plane spacing $d_0$ along the diffraction direction $\mathbf{H}$. (b) For a given scattering
    geometry the interference field is stationary, but if one scans the incident angle or photon energy
    around the Bragg condition, the intensity profile $I_\mathrm{XSW}$ can be shifted by $d_0/2$. Since the
    absorption of X-rays around the Bragg conditions, which depends on the relative position of the atoms
    ($z/d_0$) within the field, determines the number of emitted photoelectrons, the photo yield variation
    $Y_p$ reveals the corresponding adsorption distance. Importantly, different elements within the adsorbed
    molecule, as symbolized by the green and red color of the atoms, can be distiguished using their specific
    core-level signal.
    In the inset, two simulated XSW scans are shown, with the normalized reflectivity (blue line) and the
    photoelectron yields as a function of the beam energy relative to the Bragg condition (green and red
    line). The coherent positions $P_H$ of the two species, as introduced in Eq.~(\ref{eq:yieldF}), can be
    easily converted to the adsorption distance. The strong dependence of the photoelectron yield with
    respect to the position of the emitter within the XSW field, which is the origin of the precision of
    this technique, can be readily seen by the characteristic modulation of these curves.}
  \label{fig:xsw_schematics}
\end{figure*}

The XSW technique is an interferometric method that exploits the standing wave field $I_\mathrm{XSW}$ created
by Bragg diffraction of the incoming X-ray beam. By measuring characteristic photoemission signals, that are
related to the local field strength at the position of the excited atomic species, one can derive
high-precision and chemically sensitive adsorption distances of molecules on single crystals (see
Figure~\ref{fig:xsw_schematics} for a schematic).

\subsubsection{Concept of XSW measurements}
In 1964 Boris W.\ Batterman first demonstrated that the fluorescence intensity emitted from a single
crystal, illuminated with X-rays, changed characteristically when rocking the crystal around the Bragg
condition due to the presence of an X-ray standing wave field~\cite{Batterman_1964_PhysRev}. By matching the
maxima and minima of the fluorescence signal with the reflected X-ray intensity around the Bragg angle, he
correlated them with the relative position of the atomic planes of the crystal. A few years later, he used
this concept to locate, within a silicon single crystal, the position of arsenic atoms, used as dopants,
relative to the silicon atomic planes~\cite{Batterman_1969_PhysRevLett}. Soon thereafter, it was exploited
that the XSW field extends outside the crystal surface~\cite{Andersen_1976_PhysRevLett}, opening up the door
to not only the mapping of dopants within a crystal structure~\cite{Zegenhagen_1993_SurfSciRep,
  Zegenhagen_2019_JpnJApplPhys} and the study of buried interfaces~\cite{Bedzyk_1988_Science,
  Lee_1996_PhysicaB, Schneck_2015_CurrOpinColloidInterfaceSci}, but also the location of adsorbates on the
crystal surface~\cite{Cowan_1980_PhysRevLett}. For the latter, initially, adsorbed atoms were
studied~\cite{Cowan_1980_PhysRevLett, Bedzyk_1985_PhysRevB}, Langmuir-Blodgett
films~\cite{Nakagiri_1985_ThinSolidFilms} and atomic layers followed~\cite{Bedzyk_1989_PhysRevLett}. Some
smaller molecules deposited on different surfaces started to be studied in the
1990s~\cite{Sugiyama_1995_PhysRevB, Fenter_1998_SurfSci, Shuttleworth_2002_ChemPhysLett,
  Mulligan_2003_SurfSci}, while, the first measurements of conjugated organic molecules on metal surfaces
came a few years later~\cite{Kilian_2002_PhysRevB, Stanzel_2004_SurfSci, Hauschild_2005_PhysRevLett,
  Gerlach_2005_PhysRevB}.

Without entering into the exact mathematical derivation of the XSW field, which is based on dynamical
diffraction theory~\cite{Authier_2003_book}, one can explain the basic principle of the XSW technique using
the fundamental equation~\cite{Zegenhagen_1993_SurfSciRep, Woodruff_1998_ProgSurfSci,
  Vartanyants_2001_RepProgPhys, Gerlach_2013_book,Burker_2014_book, Zegenhagen_2019_JpnJApplPhys}
\begin{equation}
  \label{eq:yieldF} Y_\mathit{p}(h\nu) = 1 + R + 2 \sqrt{R} \, f_\mathrm{H} \cos(\nu -2 \pi P_\mathrm{H})
\end{equation}
which relates the normalized photo yield $Y_p$ from a given chemical species, the intrinsic reflectivity $R$
of the crystal and the relative phase $\nu$ between the incoming and the reflected wave with the two
structural parameters $f_\mathrm{H}$ and $P_\mathrm{H}$. The so-called coherent position $P_\mathrm{H}$, which
takes values between 0 and 1 (being both geometrically equivalent), is directly related to the (mean) position
of the species being considered via
\begin{equation}
  \label{eq:d_H} d_\mathrm{H} = (n+P_\mathrm{H}) d_0 \quad n=0,1,2,\dots
\end{equation}
where $n$ introduces an ambiguity that stems from the periodicity of the XSW field of period $d_0$
(\textit{cf.}~Table~\ref{tab:substrate_parameters}). In most cases, this ambiguity can be removed with common
sense and the physical constraints of the system. The index $H$ in Eq.~(\ref{eq:yieldF}) and (\ref{eq:d_H})
refers to the reciprocal lattice vector of the Bragg reflection employed in the experiment. The coherent
fraction $f_\mathrm{H}$ is related to the vertical ordering of the species contributing to a given
$d_\mathrm{H}$. It assumes values between 0 and 1, with 0 as the outcome of randomly distributed emitters
around $d_\mathrm{H}$ and 1 the case where all are adsorbing at $d_\mathrm{H}$. Different effects contribute
to the decrement of $f_\mathrm{H}$, for instance, thermal vibrations and static disorder. We note that,
besides the obvious reason that the adsorption distance only makes sense within the first adsorbed layer of
molecules, the significance of the experimental $P_\mathrm{H}$ values is limited by $f_\mathrm{H}$. In other
words, for a highly disordered layer ($f_\mathrm{H} \approx 0$), it is pointless to associate any adsorption
distance. This has two direct consequences: first, $f_\mathrm{H}$ can be used as confidence parameter for the
obtained adsorption distance and secondly, coverages below or equal to a full monolayer are desirable to avoid
artificially decreasing $f_\mathrm{H}$.

We note that for practical purposes Eq.(\ref{eq:yieldF}) has to be refined to account, e.g., for the
broadening of the reflectivity curve due to the monochromator and the imperfections of the crystals. Also,
non-dipole effects in the photoemission process, which affect the angular distribution of the
photoelectrons~\cite{Schreiber_2001_SurfSci, Straaten_2018_JElectronSpectroscRelatPhenom}, have to be
considered in the data analysis.

\subsubsection{Experimental considerations}
Generally, datasets for two experimental quantities are required to model the XSW field and subsequently
extract the position of a given species relative to the lattice planes of the crystal, i.e.\ the reflectivity
$R = R(E)$ and photo yield $Y_p(E)$ when scanning around the Bragg condition $E = E_\mathrm{Bragg}$.  The
reflectivity can be measured with a camera directed at a fluorescence screen conveniently located in the
chamber and the photo yield is extracted from fluorescence, Auger or photoelectron spectroscopy data from the
species of interest: Here, we restrict our discussion to photoelectrons, which are measured through XPS scans
performed with different photon energies around $E_\mathrm{Bragg}$
(\textit{cf.}~Table~\ref{tab:substrate_parameters}).

The use of the XSW technique is constrained by rather demanding experimental requirements. Certainly, the
first major challenge is the indispensable crystal quality of the substrate, both at the surface as well as in
the bulk, which is responsible for the coherence of the standing wave field.  In addition to the high photon
flux required for these experiments the need to tune the X-ray energy limits the usage of the technique to
synchrotron facilities~\cite{Materlik_1984_PhysLettA}.  Here, beamlines with insertion devices, crystal
monochromators and complex X-ray optics can provide a stable and highly brilliant X-ray beam, that can be
(de-)focused to avoid beam damage on the samples.  The reader is referred to
Refs.~\onlinecite{Zegenhagen_2010_JElectronSpectroscRelatPhenom, Zegenhagen_2013_book,
  Lee_2018_SynchrotronRadiatNews} for a more detailed explanation of the beamline requirements.

The experimental geometry, namely, the relative direction of the incoming beam with respect to the sample
and the electron analyzer is essential. It can be shown that when creating the interference field in
back-reflection (see Figure~\ref{fig:xsw_schematics}), i.e.\ the incoming and the Bragg-reflected beam
being almost perpendicular to the diffracting crystal planes (diffraction angles $\theta_\mathrm{Bragg}$
close to $90^\circ$), the intrinsic angular width of the reflectivity curve is largest. Thereby, the need
for nearly perfect crystallinity of the substrates is relaxed~\cite{Woodruff_1988_SurfSci}. XSW experiments
performed under these conditions are referred to as normal-incidence (NI)XSW and have become standard for
measuring adsorption distances of larger molecules on metals.

Recently, it has been demonstrated that dedicated beamlines such as I09 at Diamond Light Source
(UK)~\cite{Lee_2018_SynchrotronRadiatNews}, which is operational since 2013, can implement significant
improvements in performance and usability compared to previous installations. Due to the optimized
experimental setup and data-acquisition methods the signal-to-background and signal-to-noise ratio of the
photoelectron spectra could be improved without risking extensive beam damage even for molecular systems.  If
the electron analyzer is positioned at an angle of 90$^\circ$ with respect to the incident X-ray beam (as
realized at I09), the substrate background in the spectra is strongly suppressed and also the non-dipole
contributions to the photoelectron yield are minimized. Overall, the challenges associated with XSW
measurements have to some degree shifted away from the technical side and more towards the sample preparation
and data analysis. Indeed, by using a proper core-level model to account for the different contributions to
the photoelectron yield one can extract adsorption distances for the chemically inequivalent species within a
molecule.  Hence, the systematic combination of XSW experiments with high-resolution XPS allows to resolve
intramolecular distortions that were not accessible before and thereby extend the significance of XSW results
beyond average adsorption distances~\cite{Mercurio_2013_PhysRevB, Blowey_2017_FaradayDiscuss,
  Franco-Canellas_2017_PhysRevMaterials}. For that matter, accurate~\cite{Smith_2016_Carbon} and preferably
theory-backed core-level models~\cite{Diller_2017_JChemPhys, Baby_2015_BeilsteinJNanotechnol} are necessary.

Over the years, different software packages have been used for handling (NI)XSW data. For fitting the
core-level spectra the commercial \textsc{CasaXPS}~\cite{CASAXPS} has become very popular. For the analysis of
the resulting photoelectron yield data, on the other hand, there are various specialized tools
available. Recently, Bocquet et al.\ discussed the general formalism and contributed a new open source program
with graphical user interface (\textsc{Torricelli}) that facilitates the fitting of XSW
data.~\cite{Bocquet_2019_ComputPhysCommun} As also pointed out in
Ref.~\onlinecite{Straaten_2018_JElectronSpectroscRelatPhenom}, the analysis can be non-trivial, if the large
angular aperture of the analyzer and the finite tilt of the sample are considered.  As a concluding remark, we
also note that via off-normal XSW measurements, i.e.\ using a Bragg reflection with a finite in-plane
component of \textbf{H}, it is in principle possible to triangulate the position of adsorption
sites~\cite{Cheng_2003_PhysRevLett}. Since for large adsorbate molecules this can be
difficult~\cite{Weiss_2017_PhysRevB}, our focus is on the vertical structure along the surface normal.

\subsection{Photoelectron spectroscopy} \label{ssec:PES_issues}

\begin{figure}
  \centering
  \includegraphics[width=\columnwidth]{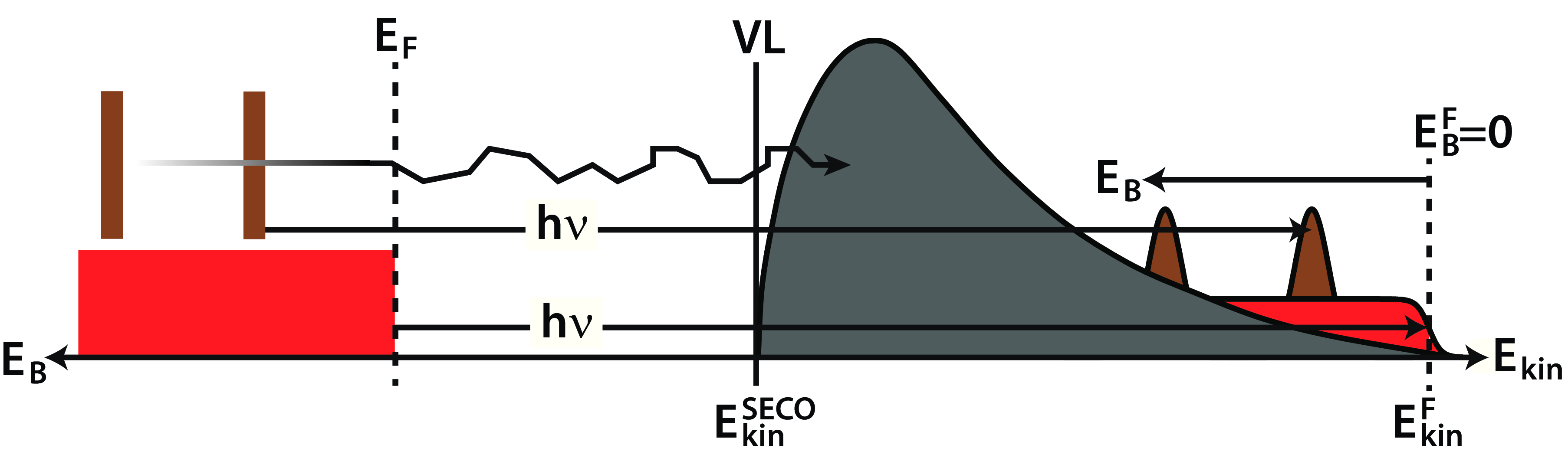}
  \caption{Schematic valence energy-levels and UP spectrum of a COM monolayer on a metal substrate. The
    sample is illuminated by photons with energy $h\nu$. Photoelectrons from the metal Fermi level $E_F$ have
    the highest kinetic energy ($E_\mathrm{kin}^F$). Usually, the Fermi level is used as energy reference to
    define the binding energy $E_\mathrm{B}$ of valence electron features. Inelastically scattered secondary
    electrons lost the information about their initial state and the position of the secondary electron
    cutoff $E_\mathrm{kin}^\mathrm{SECO}$ allows to determine the vacuum level (VL) of the sample. The sketch
    is strongly simplified, in particular the measurement process itself, i.e., the impact of the
    spectrometer on measured kinetic energies, is neglected. More detailed sketches can be found, e.g., in
    Refs.~\onlinecite{Ishii_1999_AdvMater, Cahen_2003_AdvMater}.}
  \label{fig:PES_principle}
\end{figure}

Photoelectron spectroscopy (PES) is a well-known and established technique to determine the electronic
structure of solids and is described in detail in various books and review articles
\cite{Egelhoff_1987_SurfSciRep, Cardona_1978_book, Nefedov_1988_book, Hufner_2003_book, Schattke_2008_book,
  Fadley_2010_JElectronSpectroscRelatPhenom, Bagus_2013_SurfSciRep, Suga_2014_book,
  Moser_2017_JElectronSpectroscRelatPhenom, Dil_2019_ElectronStruct}. In this section we will, thus, deal with
issues specific to PES on organic thin films~\cite{Rao_1979_ApplSpectroscRev, Seki_1986_ChemPhys,
  Netzer_1992_CritRevSolidStateMaterSci, Cahen_2003_AdvMater, Crispin_2003_JPolymSciB,
  Crispin_2004_JAmChemSoc, Rocco_2008_JChemPhys, Ueno_2008_ProgSurfSci, Puschnig_2009_Science,
  Bussolotti_2015_JElectronSpectroscRelatPhenom, Schultz_2017_AdvMaterInterfaces,
  Munoz_2018_MacromlRapidCommun, Kirchhuebel_2019_PhysChemChemPhys} including the main pitfalls and obstacles.

\subsubsection{Ultraviolet photoelectron spectroscopy}
First, we will focus on ultraviolet photoelectron spectroscopy (UPS). Figure~\ref{fig:PES_principle} displays
on the left side two energy-levels (HOMO and HOMO-1) of an organic thin film on a metal substrate, for which
the continuous occupied DOS is shown. The sample is irradiated with monochromatic UV-light with photon energy
$h\nu$ and the resulting photoemission intensity is shown on the right side of
Figure~\ref{fig:PES_principle}. An electron analyzer measures the kinetic energy $E_\mathrm{kin}$ and
intensity of photoelectrons. The resulting spectra are usually plotted as function of binding energy
$E_\mathrm{B}$ with the Fermi level serving as energy reference ($E_\mathrm{B}^F = 0$\,eV). The information
depth is limited by the inelastic mean free path of photoelectrons. The so-called ``universal curve'' gives a
value of $\sim$7\,\AA{} for electrons with a kinetic energy of 15\,eV (typical for measurements with He\,I)
in organic materials~\cite{Seah_1979_SurfInterfaceAnal}. Consequently, for (sub)monolayer coverages of a flat
lying COM film on a metal, molecular features and substrate features appear concomitantly.

\begin{figure*}
  \centering
  \includegraphics[width=0.85\textwidth]{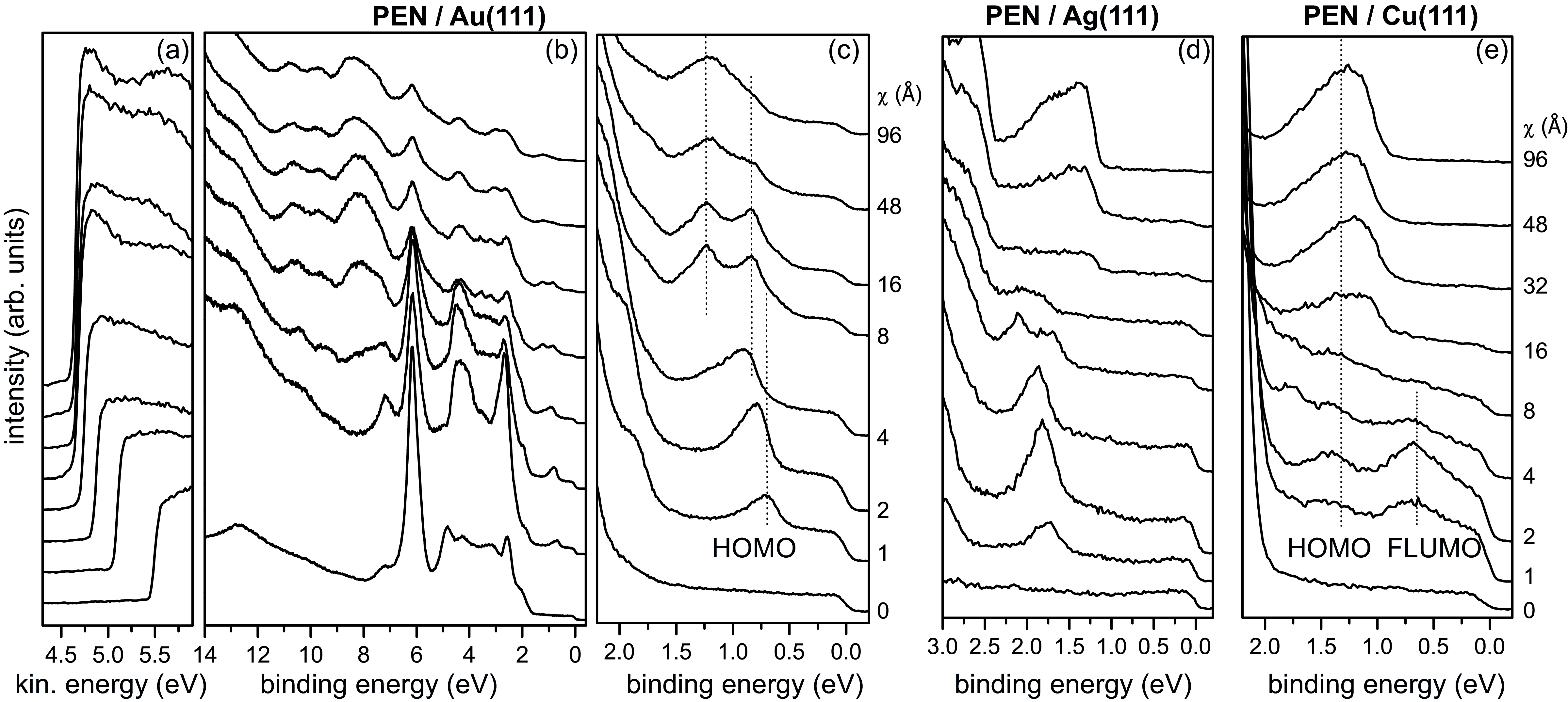}
  \caption{UP spectra of pentacene deposited on the (111)-surfaces of noble metals measured with the He\,I
    excitation line. $\chi$ denotes the nominal pentacene thickness. For PEN on Au(111) the SECO region
    (a), a valence survey spectrum (b), and a zoom into the region close to the Fermi level (c) are
    shown. For PEN on Ag(111) (d) and on Cu(111) (e) only the zooms are shown. All valence electron
    spectra are measured with an emission angle of 45$^\circ$. The lines are guides for the eye and mark
    the evolution of HOMO and FLUMO (former LUMO) features of PEN on Au(111) and Cu(111),
    respectively. The energy scale of the plots of the secondary electron spectra is corrected by the
    analyzer work function and the applied bias voltage. Thus, the SECO position corresponds to the VL
    position above the Fermi level. Adapted from Ref.~\onlinecite{Lu_2016_JPhysCondensMatter}.}
  \label{fig:PES_PEN_X111}
\end{figure*}

In addition to valence electron features, also secondary electrons (gray in Figure~\ref{fig:PES_principle})
contribute to the spectrum. These electrons have been inelastically scattered in the sample and thus lost the
information about their initial state. However, they can be used to determine the VL of the sample, since at a
certain kinetic energy ($E_\mathrm{kin}^\mathrm{SECO}$) the energy of secondary electrons is not sufficient to
overcome the surface potential of the sample. At this energy their intensity is dropping to zero, which is
often called secondary-electron cutoff (SECO). The VL of the sample (w.r.t.\,$E_F$) is given by the difference
of the photon energy and the whole width of the spectrum, i.e.:
\begin{equation}
  \label{eq:VL} \mathrm{VL} = h\nu -(E_\mathrm{kin}^F - E_\mathrm{kin}^\mathrm{SECO}).
\end{equation}

This rather simplified description is sufficient to determine the VL of samples with a homogenous surface
potential. However, as mentioned above, the adsorption of COMs usually modifies the work function of clean
metal surfaces. For submonolayer coverages, or in the case of island growth, the sample features local
surface potentials~\cite{Wandelt_1997_ApplSurfSci, Koller_2007_OrgElec,
  Duhm_2009_JElectronSpectroscRelatPhenom, Schultz_2017_AdvMaterInterfaces, Schultz_2019_PhysStatusSolidiB,
  Wang_2016_SciRep, Tada_2011_NatMater, Siles_2018_Small}. Depending on the lateral dimensions of these
inhomogeneities, either two separate SECOs can be observed (for large island sizes) or the SECO position is
determined by the area-weighted mean of the local surface potentials. A detailed description and guidelines
on how to analyze SECOs are given in Ref.~\onlinecite{Schultz_2019_PhysStatusSolidiB}. This publication
describes, furthermore, how to determine IEs of organic thin films, which are defined by the SECO and the
onset of the HOMO-derived peak~\cite{Cahen_2003_AdvMater, Kahn_2016_MaterHoriz,
  Koch_2008_JPhysCondensMatter, Kahn_2003_JPolymSciBPolymPhys, Ishii_1999_AdvMater,
  Schultz_2019_PhysStatusSolidiB}.

In general, for discussing interfacial interactions often ``monolayer'' and ``multilayer'' energy-levels are
compared (\textit{cf.}~Figure~\ref{fig:ELA_concept}). The thickness of vacuum-sublimed thin films in organic
molecular beam deposition (OMBD) is usually measured by a quartz-crystal micro balance and corresponds, thus,
to a nominal mass thickness. In that process, layer-by-layer growth is rather the exception than the rule and
island or Stranski-Krastanov (island on wetting layer) growth dominates~\cite{Venables_1984_RepProgPhys,
  Yang_2015_ChemRev, Winkler_2016_SurfSci, Kowarik_2017JPhysCondensMatter,
  Schreiber_2004_PhysStatSolA}. Thus, the first step in interpreting photoemission data is to identify the
spectrum which is most dominated by monolayer contributions. A first hint gives the evolution of the SECO as
adsorption induced charge rearrangements often saturate upon monolayer formation.  For PEN, which may be
regarded as the ``fruit fly'' of organic surface science~\cite{Kuroda_1961_CanJChem, Lee_1977_ChemPhysLett,
  Sebastian_1981_ChemPhys, Moerner_1999_Science, Kang_2003_ApplPhysLett, Eremtchenko_2005_PhysRevB,
  Fukagawa_2006_PhysRevB,Kafer_2007_PhysRevB, Koch_2007_AdvMater,
  Zheng_2007_Langmuir,Dougherty2008JPCC,Koch_2008_JAmChemSoc, Yamane_2009_JElectronSpectroscRelatPhenom,
  Puschnig_2009_Science,Toyoda_2010_JChemPhys, Han_2013_ApplPhysLett_PEN, Duhm_2013_ACSApplMaterInterfaces,
  Park_2016_AnalChem, Ji_2017_CanJChem, Zhang_2017_AdvElectronMater, Kera_2018_JPhysSocJpn,
  Choi_2018_NatMater, Hoffmann-Vogel_2018_RepProgPhys, Klues_2018_CrystEngComm,
  Doring_2019_JPhysCondensMatter, Nakayama_2019_JPhysChemLett, Tadano_2019_PhysRevApplied},
thickness-dependent UP spectra on Au(111), Ag(111) and Cu(111)~\cite{Lu_2016_JPhysCondensMatter} are shown in
Figure~\ref{fig:PES_PEN_X111} as a typical example of UPS at organic-metal interfaces.  Indeed, for PEN on
Au(111) the VL (as deduced from the SECO position in Figure~\ref{fig:PES_PEN_X111}a) decreases rapidly up to
a nominal PEN thickness of 4\,\AA. However, this does not mean that a nominal thickness of 4\,\AA{}
corresponds to a closed monolayer. It simply tells that from this thickness on, subsequently deposited
molecules grow predominantly in multilayers.

The suppression of substrate features, e.g., the Au $d$-bands in a BE range from 2 to 8\,eV in
Figure~\ref{fig:PES_PEN_X111}b or the Fermi-edge in Figure~\ref{fig:PES_PEN_X111}c, with increasing coverage
can be used to estimate the growth mode of the adsorbate. However, the applicability of the universal curve
to organic thin films has been questioned~\cite{Bussolotti_2013_PhysRevLett, Graber_2011_SurfSci,
  Ozawa_2014_JElectronSpectroscRelatPhenom} and only qualitative statements are straightforward. For
PEN/Au(111) the substrate Fermi-edge is still visible for a nominal coverage of 96\,\AA , which corresponded
to more than twenty layers of flat lying PEN. This clearly shows that the growth mode is not
layer-by-layer. For the spectra with a nominal thickness of 96\,\AA{} on Ag(111) and Cu(111), on the other
hand, the Fermi-edge is (almost) invisible, pointing to less pronounced island growth.

The shape of HOMO-derived UPS peaks is often not simply Gaussian. For well ordered monolayers and sufficient
experimental resolution, hole-phonon coupling~\cite{Koch_2007_AdvMater, Kera_2009_ProgSurfSci,
  Kera_2015_JElectronSpectroscRelatPhenom} becomes evident in UP spectra as can be seen by the high-BE
shoulder of the HOMO-derived peak in the spectra for a nominal coverage of 2\,\AA{} on Au(111) and Ag(111)
(Figures~\ref{fig:PES_PEN_X111}c and~d). Furthermore, factors like the measurement geometry and the photon
energy impact photoemission intensities. For example, the emission from the HOMO of flat-lying
$\pi$-conjugated molecules has typically a maximum for an emission angle of 45$^\circ$ and a minimum for
normal emission~\cite{Hollerer_2017_ACSNano, Puschnig_2009_Science, Liu_2014_JElectronSpectroscRelatPhenom,
  Yagishita_2015_JElectronSpectroscRelatPhenom}.

The spectra in Figure~\ref{fig:PES_PEN_X111} are measured with an hemispherical analyzer and
angle-integrated over $\pm 12^\circ$ along $k_x$, which is a typical measurement geometry. Also such
angle-integrated spectra can reflect energy dispersing features, in particular if rotational domains
related to the substrate symmetry are involved. This explains the HOMO-shape of PEN in multilayers on
Ag(111), in which PEN adopts a herringbone arrangement~\cite{Kafer_2007_ChemPhysLett} and exhibits a band
dispersion~\cite{Yoshida_2008_PhysRevB}. Notably, also former LUMO derived energy-levels of organic
monolayers on metals can show intermolecular energy dispersion \cite{Yamane_2013_PhysRevLett,
  Ules_2014_PhysRevB}. For PEN on Au(111) the multilayer growth mode is still under
debate~\cite{Kafer_2007_PhysRevB, Kang_2003_ApplPhysLett} and, hence, the multilayer HOMO features have
not been unambiguously assigned~\cite{Lu_2016_JPhysCondensMatter}. Overall, great care has to be taken
when comparing measurements obtained in different experimental setups. For example, in an early
publication of PEN on Cu(111) the former LUMO-derived peak just below the Fermi level
(Figure~\ref{fig:PES_PEN_X111}e) has been overlooked~\cite{Koch_2008_JAmChemSoc}. Moreover, small
differences in, e.g., temperature, substrate cleanness, evaporation rate or impurities, can have a
significant impact on organic thin film growth and, consequently, the electronic
structure~\cite{Lee_1977_ChemPhysLett, Schreiber_2004_PhysStatSolA, Salzmann_2007_ApplPhysLett,
  Witte_2008_PhysStatSolA, Breuer_2011_CrystGrowthDes, Ji_2017_CanJChem, Jones_2016_AdvFunctMater,
  Cocchi_2018_PhysChemChemPhys}.

\subsubsection{X-ray photoelectron spectroscopy}
\begin{figure}
  \centering
  \includegraphics[width=0.9\columnwidth]{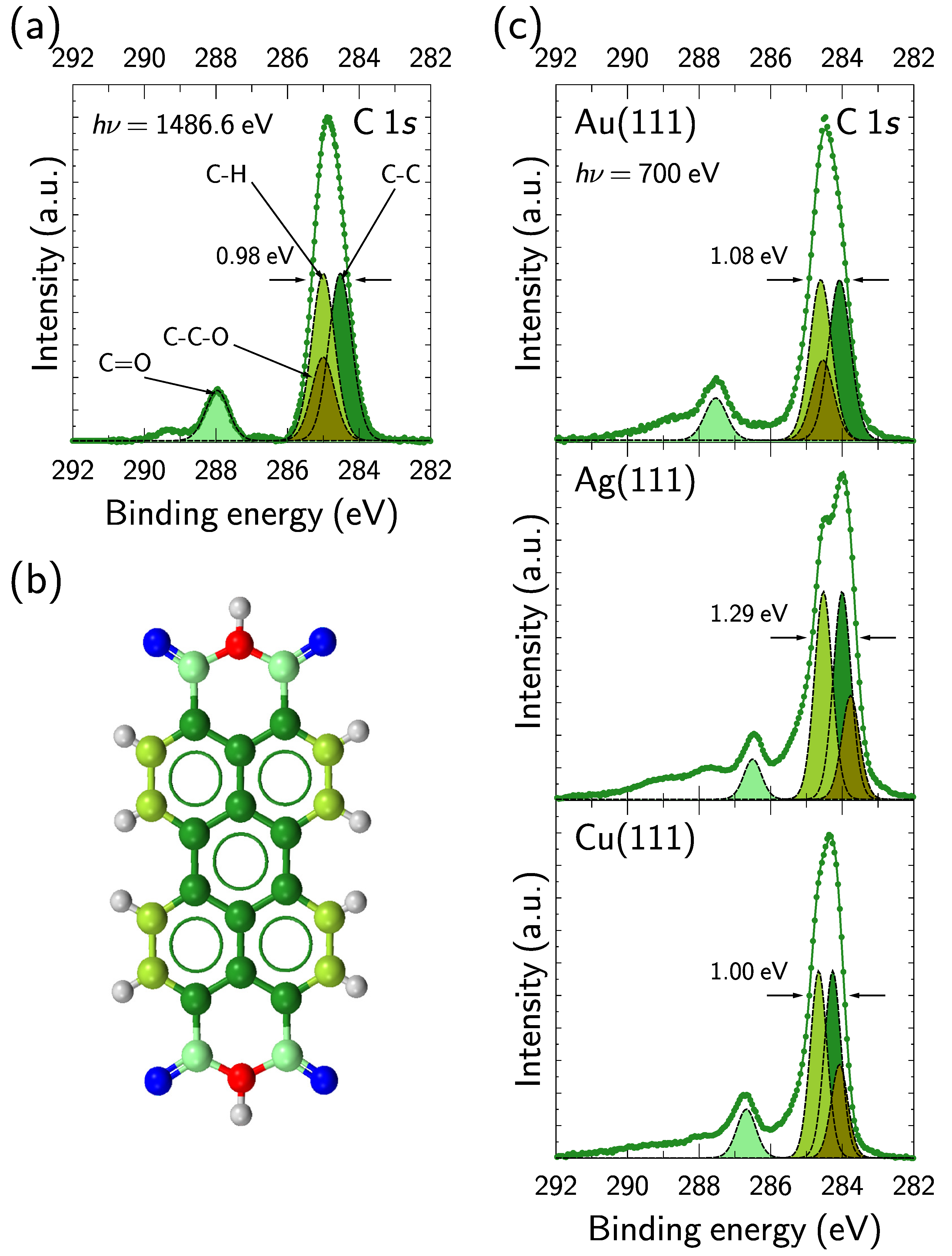}
  \caption{Influence of the environment on the core-level signals. Multilayer spectrum (a) of the C\,1s signal
    of PTCDI (b) compared to (sub)-monolayer coverages of the same molecule adsorbed on the noble metals (c).
    Figure adapted from Ref.~\onlinecite{Franco-Canellas_2017_PhysRevMaterials}. }
  \label{fig:PTCDI_XPS}
\end{figure}

For XPS the electronic structure of the sample is probed with X-rays, whose higher photon energy make
core-levels accessible~\cite{Nefedov_1988_book, Moulder_1995_book, Fadley_2010_JElectronSpectroscRelatPhenom,
  Crist_2019_JESRP}. Core-levels provide information about the local chemical environment of the atoms, which
gives rise to so-called chemical shifts in XP spectra~\cite{Chadwick_1981_JESRP,
  Travnikova_2012_JElectronSpectroscopRelatPhenom}. Figure~\ref{fig:PTCDI_XPS}a shows the C\,1s spectrum of a
PTCDI multilayer on Au(111)~\cite{Franco-Canellas_2017_PhysRevMaterials} illustrating the strong chemical
shift between carbon atoms in the functional groups (C=O) and in the perylene backbone. The chemically
inequivalent carbon atoms within the molecule can be precisely resolved if the shift in energy is large
enough, which is often the case for carbon bound to electronegative atoms (e.g.\ O and F). The binding-energy
(or core-level) shifts associated with the chemical structure can be further modified by the molecular
environment, for instance if there are strong intermolecular interactions, and/or by the proximity of the
substrate. Intrinsic or extrinsic peak broadening is another complication when describing core-level
signals. Its origin is manifold and a proper description often requires electronic structure calculations.

The fine structure of the core-levels can only be resolved using a sufficiently high energy resolution, which
is feasible when measuring at synchrotron radiation facilities or with monochromatized lab-sources. Generally,
monolayer spectra may include fingerprints of (chemical) interactions with the
substrate~\cite{Freund_1996_SurfSciRep, Crispin_2003_JPolymSciB, Scholl_2004_JChemPhys,
  Lanzilotto_2018_ChemEurJ}. For PTCDI on Au(111) the interaction is
weak~\cite{Franco-Canellas_2017_PhysRevMaterials} and therefore the multilayer (Figure~\ref{fig:PTCDI_XPS}a)
and monolayer (Figure~\ref{fig:PTCDI_XPS}c) spectra are -- except for a rigid shift due to screening -- almost
identical.  PTCDI monolayers on Ag(111) and Cu(111), however, are
chemisorbed~\cite{Franco-Canellas_2017_PhysRevMaterials}, which is reflected in non-rigid shifts of the
components attributed to different chemical environments of the multilayer versus the monolayer spectra; in
particular, the energetic spacing between the C=O-derived peak and the main peak decreases for the PTCDI
monolayer (Figure~\ref{fig:PTCDI_XPS}c). In addition to the insight in the chemistry of adsorbates, chemical
shifts in XPS facilitate the experimental determination of individual vertical binding distances for each
carbon species using the XSW technique (Figure~\ref{fig:perylene_derivatives}).

\subsection{Complementary techniques} \label{ssec:subsection_other_techniques}
The XSW technique is mostly applied to the measurement of adsorption distances and molecular distortions
perpendicular to the surface. Alternatively, photoelectron diffraction (PhD)~\cite{Woodruff_2007_SurfSciRep}
provides a full, local 3D positioning of a given species with respect to the surface
atoms~\cite{Lorenzo_2011_SurfSci, Duncan_2012_JPhysChemC, White_2018_SurfSci}. However, while rather
successful for certain (preferably simple) systems, the data analysis and interpretation are not
straightforward since inequivalent positions may be difficult to decouple, thus challenging its application
to larger adsorbates~\cite{Duncan_2010_SurfSci, Salomon_2015_JPhysChemC}. Similarly, LEED I-V may provide a
3D picture of an adsorbate~\cite{Stellwag_1995_SurfSci, Zheleva_2012_JPhysChemC,
  Sirtl_2013_PhysChemChemPhys}, but the analysis of the data is computationally expensive and requires some
initial guess of the adsorbate position. Both, PhD and LEED I-V require a certain degree of
registry/commensurability between the adsorbate and the substrate, which limits their use to systems with
in-plane order. Surface X-ray diffraction~\cite{Feidenhansl_1989_SurfSciRep, Robinson_1992_RepProgPhys,
  Meyerheim_1998_ApplPhysA, Krause_2003_JChemPhys}, and specifically the so-called rod scans (along $q_z$)
for the vertical structure, are slightly less demanding to model (thanks to the applicability of the
kinematic, i.e.\ single-scattering, approximation), but the sensitivity to light elements is limited because
of low X-rays scattering cross-sections. Importantly, it is very difficult to obtain element-specific
positions needed to determine possible distortions/bending of the adsorbates. In contrast, relative
positions, such as tilt angles, can be inferred with NEXAFS by exploiting the geometry-dependent absorption
of X-rays without the need of long-range order~\cite{Stoehr_1992_book, Diller_2012_JChemPhys,
  Breuer_2015_JElectronSpectroscRelatPhenom}. We note that other popular techniques in the study of inorganic
surfaces and interfaces such as reflection high-energy electron diffraction (RHEED), scanning-electron
microscopy (SEM) or ion scattering-based techniques are not very common in our context because of the
probable beam damage induced by the high energy of incoming particles. Some notable exceptions can be found
though~\cite{SchmitzHubsch_1997_PhysRevB}.

Although not the focus of this review, a full characterization of the interface geometry involves the
in-plane structure of the adsorbed layers and their registry with the surface atoms. Scanning-tunneling
microscopy (STM)~\cite{Rosei_2003_ProgSurfSci, Barth_2007_AnnuRevPhysChem, Han_2012_SurfSciRep,
  Klappenberger_2014_ProgSurfSci, Wagner_2015_ProgSurfSci, Gottfried_2015_SurfSciRep,
  Dong_2016_ProgSurfSci, Wakayama_2016_JpnJApplPhys, Bouju_ChemRev_2017, He_2017_ChemPhysChem,
  Goronzy_2018_ACSNano} and low-energy electron diffraction (LEED)~\cite{Mannsfeld_2006_ModPhysLettB,
  Hooks_2001_AdvMater, Forker_2017_SoftMatter} are the most popular techniques in this context in addition
to grazing-incidence X-ray diffraction~\cite{Feidenhansl_1989_SurfSciRep}, which provides the highest
resolution. STM offers real-space images with atomic resolution, which can be combined with local
spectroscopy measurements. However, any quantitative determination of vertical adsorption structures with
STM is still challenging \cite{Woodruff_2019JpnJApplPhys}. Notably, with atomic force microscopy (AFM)
under UHV conditions one can estimate vertical bonding distances, although input from DFT-modeling is
necessary to extract absolute numbers~\cite{Schuler_2013_PhysRevLett}. LEED offers reciprocal-space
information that is averaged over a large sample area. With more elaborate versions such as LEED I-V,
mentioned above, and spot-profile analysis LEED (SPA-LEED) a precise description of the adsorbate unit cell
can be achieved~\cite{Kilian_2004_SurfSci}. Finally, low-energy electron microscopy (LEEM) provides in
particular real-time information of the in-plane arrangement and morphology during
growth~\cite{Levesque_2016_JPhysChemC,Henneke_2017_NatMater, Felter_2019_Nanoscale}.

In Sec.~\ref{ssec:PES_issues} photoemission spectroscopy was introduced as a tool to study the
energy-level alignment and interfacial coupling by using the energy information of photoelectrons. Beyond
that, i.e.\ by exploiting also the momentum information of the electrons~\cite{Seki_1986_ChemPhys,
  Ueno_2008_ProgSurfSci, Offenbacher_2015_JElectronSpectroscRelatPhenom,
  Udhardt_2017_JPhysChemC,Hollerer_2017_ACSNano, Yagishita_2015_JElectronSpectroscRelatPhenom,
  Liu_2014_JElectronSpectroscRelatPhenom, Puschnig_2009_Science, Bussolotti_2017_NatCommun}, new
possibilities arise, which --~although not within the scope of this article~-- shall be briefly summarized
here. By angle-resolved and photon-energy dependent UPS measurement possible
in-plane~\cite{Ueno_2008_ProgSurfSci, Yamane_2017_JPhysChemLett, Koller_2007_Science,
  Kakuta_2007_PhysRevLett} and out-of-plane~\cite{Ueno_2008_ProgSurfSci, Aghdassi_2018_Nanotechnology,
  Yamane_2013_PhysRevLett} band dispersions of organic thin films can be accessed. Moreover, even for
largely angle-integrated measurements the photoelectron angular distribution (PAD) provides insight into,
e.g., the orientation of COMs on the surface \cite{Hollerer_2017_ACSNano, Puschnig_2009_Science,
  Liu_2014_JElectronSpectroscRelatPhenom, Yagishita_2015_JElectronSpectroscRelatPhenom}. The vibrational
fine structure of HOMO-peaks allows to assess charge reorganization energies and thus to estimate hopping
mobilities by a `first-principle' experiment~\cite{Coropceanu_2002_PhysRevLett, Kera_2009_ProgSurfSci,
  Perroni_2014_Electronics, He_2016_JPhysChemC, Graus_2016_PhysRevLett}. Moreover, the development of
instrumentation over the last decades has made it possible to measure photoelectron reciprocal-space maps,
often termed ``orbital tomography'', which can be used to reconstruct molecular orbitals in real space
and/or to precisely assign photoemission intensities to a particular molecular orbital
\cite{Ules_2014_PhysRevB, Yang_2019_JPhysChemLett} \cite{Dauth_2014_NewJPhys, Graus_2018_JPhysSocJpn,
  Hollerer_2017_ACSNano, Puschnig_2009_Science, Egger_2019_NewJPhys}.

Two-photon photoemission (2PPE) spectroscopy provides insight into electron dynamics of interface
states~\cite{Ueba_2013_JPhysChemC, Marks_2014_JElectronSpectroscRelatPhenom, Gerbert_2017_JPhysChemC,
  Yamada_2018_ProgSurfSci, Lerch_2018_JPhysCondensMatter}. In conjunction with real-space information, e.g.,
by LEED or STM, detailed insight in organic-metal coupling is possible~\cite{Schwalb_2008_PhysRevLett,
  Zaitsev_2016_PhysRevB, Lerch_2017_JPhysChemC}. Accessing the unoccupied density of states by inverse
photoemission (IPES) is demanding, as cross-sections and overall energy resolution are notoriously low and,
most importantly, beam damage can be a problem for organic thin films~\cite{Kanai_2009_ApplPhysA,
  Zahn_2006_ChemPhys,Krause_2013_OrgElec}. Some of these issues can be overcome by low-energy inverse
photoemission spectroscopy (LEIPS)~\cite{Yoshida_2015_JElectronSpectroscRelatPhenom,
  Han_2013_ApplPhysLett_PEN}. Scanning tunneling spectroscopy (STS), as a local probe, accesses unoccupied
states as well as occupied states close to $E_F$ and can furthermore identify site-specific interactions at
organic-metal interfaces~\cite{Song_2017_ACSNano, Klappenberger_2014_ProgSurfSci,
  Kraft_2006_PhysRevB,Rohlfing_2007_PhysRevB, Gonzalez-Lakunza_2009_ChemPhysChem,
  FernandezTorrente_2008_JPhysCondensMatter, Nazin_2003_Science}, also measurable with high-resolution
electron energy-loss spectroscopy (HREELS)~\cite{Eremtchenko_2003_Nature, Lu_2009_ChemPhysLett,
  Navarro_2015_JPhysChemC, Ruckerl_2017_BeilsteinJNanotechnol}. With \textit{in situ} optical differential
reflectance spectroscopy (DRS) optical properties can be measured during
deposition~\cite{Forker_2009_PhysChemChemPhys, Forker_2012_AnnuRepProgChemSectCPhysChem,
  Navarro-Quezada_2015_RevSciInstrum, Nickel_2018_JPhysChemC, Meisel_2018_JPhysChemC}.

Temperature-programmed desorption (TPD), with a proper modelling of the data, is used to measure the
adsorption energy of a particular adsorbate on a given
substrate~\cite{Redhead_1962_Vacuum,Lord_1974_SurfSci,Chan_1978_ApplicationsofSurfaceScience,
  Chan_1978_ApplicationsofSurfaceSciencea, Habenschaden_1984_SurfSci, Fichthorn_2002_PhysRevLett,
  Winkler_2016_SurfSci, Maass_2018_JChemPhys}. This parameter is important for a precise and quantitative
distinction between adsorption regimes and has become, together with the adsorption distance, a benchmark
parameter for state-of-the-art DFT calculations~\cite{Liu_2015_PhysRevLett,Maurer_2016_ProgSurfSci}. Other
uses include the study of thermally-activated on-surface reactions~\cite{Haq_2011_JAmChemSoc,
  Rockert_2014_ChemEurJ, Chen_2014_JPhysChemC, Ferrighi_2015_ChemEurJ}, the assessment of the thin-film
growth, desorption kinetics and the thermal stability of a given system~\cite{Lukas_2001_JChemPhys,
  Kafer_2009_ApplPhysAMaterSciProcess, Frank_2009_ChemPhysLett, Gotzen_2011_Langmuir, Breuer_2013_JChemPhys,
  Winkler_2016_SurfSci, Thussing_2017_JPhysChemC}. Finally, \textit{in-situ} IR spectroscopy provides insight
into the vibrational modes and changes thereof upon adsorption. It is also useful for identifying unknown
sample compositions and may give information on changes in the adsorbate
charge~\cite{Broker_2010_PhysRevLett}. Of course, there are also many other spectroscopic techniques
including, e.g., Raman and photoluminescence, that can be applied to study some of the isssues discussed
here, but rather indirectly and outside the scope of this review.

\vfill
\begin{minipage}{\columnwidth}
  \centering
  \noindent\rule{\columnwidth}{0.4pt}
  \captionof{table}{(On the following pages) -- List of experimental and element-resolved adsorption
    distances $d_H$ determined with the XSW technique for COMs on (111) noble metal surfaces.}
  \label{tab:xsw-data}
\end{minipage}

\newpage

% Include the XSW TABLE (GERLACH TABLE)

\begingroup
\squeezetable
\begin{longtable}[c]{p{18mm}p{22mm}|p{22mm} D{.}{.}{-1}}
  \toprule
  \hline
% \multicolumn{4}{>{\footnotesize}c}{\rule[-3mm]{0mm}{8mm}{\textbf{XSW results of COMs}}} \\
% \hline
% \textbf{Molecule}  & \textbf{Comment}  & \textbf{Signal}  & \boldsymbol{d_H \,}\textbf{(\AA)} \\ 
   \textbf{Molecule}  & \textbf{Comment}  & \textbf{Signal}  & \multicolumn{1}{c}{$\boldsymbol{d_H \,}$\textbf{(\AA)}} \\ 
 \hline
  \endhead
  \multicolumn{4}{c}{\rule[-3mm]{0mm}{8mm}{\textbf{Pentacene (PEN) derivatives on Cu(111)}}} \\
  \hline
      \multirow{6}{*}{F$_4$PEN~\cite{Franco-Canellas_2018_PhysRevMaterials}} & 	& C\,1\textit{s} -- total 		 & 2.37(4) \\
      \cline{3-4}
      &					& \mbox{C\,1\textit{s} -- PEN backbone}  		        & 2.36(2)\\
      \cline{3-4}
      &					& C\,1\textit{s} -- C(1,2)				& 2.24(8) \\
      \cline{3-4}
      &					& C\,1\textit{s} -- C(3,4)				& 2.42(2) \\
      \cline{3-4}
      &					& C\,1\textit{s} -- C--F				& 3.15(5)\\
      \cline{3-4}
      & 					& F\,1\textit{s} 					& 3.40 \\
      \hline
      \multirow{2}{*}{P2O~\cite{Heimel_2013_NatureChem}} & 	& C\,1\textit{s} 					& 2.34 \\
      \cline{3-4}
      & 					& O\,1\textit{s} 					& 2.02 \\
      \hline
      \multirow{2}{*}{P4O~\cite{Heimel_2013_NatureChem}} & 	& C\,1\textit{s} 					& 2.25 \\
      \cline{3-4}
      & 					& O\,1\textit{s}					& 1.98 \\
      \hline
      PEN~\cite{Koch_2008_JAmChemSoc}  & 		& C\,1\textit{s}					& 2.34(2) \\
      \hline
      \multirow{2}{*}{PFP\cite{Koch_2008_JAmChemSoc}} 	& \multirow{2}{*}{}	& C\,1\textit{s}		 & 2.98(7) \\
      \cline{3-4}
      & 					& F\,1\textit{s} 					& 3.08(4) \\
      % \cline{2-4}
      % & \multirow{2}{*}{\cite{Goiri_2014_PhysRevLett} Heteromol. layer 1:1 PFP--CuPc} & C\,1\textit{s} -- C--F & 3.33(7) \\
      % \cline{3-4}
      % &					& F\,1\textit{s}					& 3.37(5)	 \\
      \hline
      \multicolumn{4}{c}{\rule[-3mm]{0mm}{8mm}\textbf{Perylene derivatives on Cu(111)}} \\
      \hline
      DIP~\cite{Burker_2013_PhysRevB} & 0.6\,ML 		& C\,1\textit{s}				& 2.51(3) \\
      \hline
      \multirow{5}{*}{DPDI~\cite{Matena_2014_PhysRevB}}  &\multirow{3}{*}{Mobile phase}& C\,1\textit{s}		 & 2.68(6) \\
      \cline{3-4}
      &   					& N\,1\textit{s} -- NH$_2$				& 2.40(11) \\
      \cline{3-4}
      &   					& N\,1\textit{s} -- NH					& 2.20(3) \\
      \cline{2-4}
      & \multirow{2}{*}{Porous phase}		& C\,1\textit{s}					& 3.00(4)  \\
      \cline{3-4}
      &  					& N\,1\textit{s} -- N--Cu 				& 2.83(3) \\
      \hline
      \multirow{2}{*}{Perylene~\cite{Franco-Canellas_2017_PhysRevMaterials}} & RT & C\,1\textit{s}	 & 2.43(2) \\
      \cline{2-4}
      & 200\,K				& C\,1\textit{s}					& 2.38(2) \\
      \hline
      \multirow{3}{*}{PTCDA~\cite{Gerlach_2007_PhysRevB}} & 	& C\,1\textit{s} 					& 2.61 	\\
      \cline{3-4}
      & 					& O\,1\textit{s} -- carb.	 			& 2.73 	 \\
      \cline{3-4}
      & 					& O\,1\textit{s} -- anh.				& 2.89 	 \\
      \hline
      \multirow{6}{*}{PTCDI~\cite{Franco-Canellas_2017_PhysRevMaterials}} & \multirow{2}{*}{} & \mbox{C\,1\textit{s} -- perylene core}	& 2.61(2) \\
      \cline{3-4}
      & 					& C\,1\textit{s} -- C=O					& 2.56(4)  \\
      \cline{3-4}
      & 					& C\,1\textit{s} -- C--H				& 2.59(4) \\
      \cline{3-4}	
      & 					& \mbox{C\,1\textit{s} -- C--C+C--C--O}			& 2.59(5) \\
      \cline{3-4}
      &					& N\,1\textit{s} 					& 2.68(2)\\
      \cline{3-4}	
      &					& O\,1\textit{s} 					& 2.28(2) \\
      \hline
      \multirow{2}{*}{TAT~\cite{Yang_2016_PhysRevB}} & \multirow{2}{*}{} & C\,1\textit{s} & 2.48(4) \\
      \cline{3-4}
      &					& N\,1\textit{s} 					& 2.44(6) \\
      \hline
      \multicolumn{4}{c}{\rule[-3mm]{0mm}{8mm}\textbf{Phthalocyanine (Pc) derivatives on Cu(111)}} \\
      \hline
      % \multirow{3}{*}{CoPc} &\multirow{3}{*}{} & C\,1\textit{s} & 2.55 \\
      % \cline{3-4}
      % & & N\,1\textit{s} & 2.60 \\
      % \cline{3-4}
      % & & Co\,2\textit{s} & 2.50 \\
      % \hline
      % \multirow{4}{*}{F$_{16}$CoPc} &\multirow{4}{*}{} & C\,1\textit{s} & 2.69 \\
      % \cline{3-4}
      % & & N\,1\textit{s} & 2.79 \\
      % \cline{3-4}
      % & & F\,1\textit{s} & 3.21 \\
      % \cline{3-4}
      % & & Co\,2\textit{s} & 2.71 \\
      % \hline
      \multirow{14}{*}{CuPc} &\multirow{2}{*}{\cite{Kroger_2011_PhysRevB} 0.4\,ML, 300\,K} & C\,1\textit{s}		& 2.64(7) \\
      \cline{3-4}
      & 					& N\,1\textit{s}					&  2.54(7) \\
      \cline{2-4}
      & \multirow{2}{*}{\cite{Kroger_2011_PhysRevB} 0.6\,ML, 300\,K} & C\,1\textit{s} 		& 2.57(7) \\
      \cline{3-4}
      & 					& N\,1\textit{s} 					& 2.48(7) \\
      \cline{2-4}
      & \multirow{2}{*}{\cite{Kroger_2011_PhysRevB} 0.9\,ML, 300\,K} & C\,1\textit{s}			& 2.79(7)\\
      \cline{3-4}
      & 					& N\,1\textit{s} 					& 2.69(7) \\
      \cline{2-4}
      & \multirow{2}{*}{\cite{Kroger_2011_PhysRevB} 0.4\,ML, 183\,K} & C\,1\textit{s} 		& 2.62(7) \\
      \cline{3-4}
      & 					& N\,1\textit{s} 					& 2.56(7) \\
      \cline{2-4}
      & \multirow{2}{*}{\cite{Kroger_2011_PhysRevB} 0.6\,ML, 183\,K} & C\,1\textit{s} 		& 2.53(7) \\
      \cline{3-4}
      & 					& N\,1\textit{s} 					& 2.55(7)\\
      \cline{2-4}
      &\multirow{2}{*}{\cite{Kroger_2011_PhysRevB} 0.9\,ML, 183\,K} & C\,1\textit{s} 			& 2.82(7) \\
      \cline{3-4}
      & 					& N\,1\textit{s} 					& 2.73(7) \\
      % \cline{2-4}
      % & \multirow{2}{*}{\cite{Goiri_2014_PhysRevLett} Heteromol. layer 1:1 PFP--CuPc} & C\,1\textit{s} -- C--C 	& 2.72(5) \\
      % \cline{3-4}
      % &					& N\,1\textit{s}					& 2.66(5)	 \\
      \hline
      \multirow{5}{*}{F$_{16}$CuPc} & \multirow{3}{*}{\cite{Gerlach_2005_PhysRevB}} & C\,1\textit{s}     & 2.61 \\
      \cline{3-4}
      & 					& N\,1\textit{s} 					& 2.70 \\
      \cline{3-4}
      & 					& F\,1\textit{s} 					& 2.88 \\
      \cline{2-4}
      & \multirow{2}{*}{\cite{Oteyza_2010_JChemPhys}} & C\,1\textit{s} 				        & 2.68 \\
      \cline{3-4}
      & 					& F\,1\textit{s} 					& 3.21 \\
      \hline
      \multirow{2}{*}{H$_{2}$Pc} &\multirow{2}{*}{0.7\,ML} 	& C\,1\textit{s} 			& 2.45(7) \\
      \cline{3-4}
      & 					& N\,1\textit{s} 					& 2.39(6) \\
      \hline
      \multirow{4}{*}{GaClPc~\cite{Gerlach_2011_PhysRevLett}} & \multirow{4}{*}{0.8\,ML, Cl down}       & C\,1\textit{s} 	& 4.44(7) \\
      \cline{3-4}
      & 					& N\,1\textit{s} 					& 4.71(3) \\
      \cline{3-4}
      & 					& Ga\,1\textit{s}					& 4.21(5) \\
      \cline{3-4}
      & 					& Cl\,KLL 						& 1.88(3) \\
      \hline
\newpage
      \multirow{4}{*}{VOPc\cite{Blowey_2019_JPhysChemC}} &   \multirow{4}{*}{\shortstack[l]{0.5 ML, RT \\ O up/down coex.}} 	& C\,1\textit{s} 					&  \\
      \cline{3-4}
      & 					& N\,1\textit{s} 					&  \\
      \cline{3-4}
      & 					& V\,2\textit{p} 					&  \\
      \cline{3-4}
      & 					& O\,1\textit{s} 					&  \\
      \hline
      \multirow{3}{*}{ZnPc~\cite{Yamane_2010_PhysRevLett}} &\multirow{3}{*}{0.7 ML} & C 1\textit{s}     & 2.49(3)   \\
      \cline{3-4}
      &  					& N\,1\textit{s} 					& 2.55(2)  \\
      \cline{3-4}
      & 					& Zn\,2\textit{p}$_{3/2}$ 				& 2.25(5)  \\
      \hline
      \multirow{4}{*}{F$_{16}$ZnPc~\cite{Yamane_2010_PhysRevLett}} &\multirow{4}{*}{} & C 1\textit{s}    & 2.66(10)\\
      \cline{3-4}
      &  					& N\,1\textit{s} 					& 2.85(2)  \\
      \cline{3-4}
      &  					& F\,1\textit{s} 					& 3.15(9) \\
      \cline{3-4}
      &  					& Zn\,2\textit{p}$_{3/2}$ 				& 2.58(5) \\
      \hline 
      \multicolumn{4}{c}{\rule[-3mm]{0mm}{8mm}\textbf{Porphyrin derivatives on Cu(111)}} \\
      \hline
      \multirow{6}{*}{2HTPP~\cite{Burker_2014_JPhysChemC}} 	&\multirow{3}{*}{0.8\,ML, 294\,K} & C\,1\textit{s}	& 2.40(3) \\
      \cline{3-4}
      & 					& N\,1\textit{s} -- aminic 			& 2.23(5)  \\
      \cline{3-4}
      &					& N\,1\textit{s} -- iminic 				& 2.02(8) \\
      \cline{2-4}
      &\multirow{3}{*}{0.8\,ML, 146\,K}	& C\,1\textit{s} 					& 2.34(2) \\
      \cline{3-4}
      & 					& N\,1\textit{s} -- aminic 			& 2.28(5)  \\
      \cline{3-4}
      & 					& N\,1\textit{s} -- iminic 			& 1.97(8)  \\
      \hline
      \multirow{4}{*}{CuTPP~\cite{Burker_2014_JPhysChemC}} 	&\multirow{2}{*}{0.8\,ML, 294\,K} & C 1\textit{s} 	& 2.38(2) \\
      \cline{3-4}
      & 					& N\,1\textit{s}  					& 2.25(2)  \\
      \cline{2-4}
      & \multirow{2}{*}{0.8\,ML, 146\,K} 	& C 1\textit{s}						& 2.33(2) \\
      \cline{3-4}
      & 					& N 1\textit{s}  					& 2.25(2) \\
      \hline
      \multirow{4}{*}{CoP~\cite{Schwarz2018JPhysChemC}} 	&\multirow{3}{*}{} 	& C\,1\textit{s} -- C--C	 & 2.44(9) \\
      \cline{3-4}
      & 					& C\,1\textit{s} -- C--N 				& 2.37(5)  \\
      \cline{3-4}
      & 					& N\,1\textit{s} 					& 2.33(6)  \\
      \cline{3-4}
      &					& Co\,2\textit{p}					& 2.25(4) \\
      % \cline{3-4}
      % &	(1-11) reflection		& Co\,2\textit{p}	 				& -- \\
      \hline
      \multicolumn{4}{c}{\rule[-3mm]{0mm}{8mm}\textbf{Other compounds on Cu(111)}} \\
      \hline
      \multirow{4}{*}{Azo\-benzene~\cite{Willenbockel_2015_ChemCommun}} &\multirow{3}{*}{subML, 60\,K} & C\,1\textit{s} -- C--C 	& 2.36(2)   \\
      \cline{3-4}
      &  					& C\,1\textit{s} -- C--N				& 2.23(6)  \\
      \cline{3-4}
      &  					& N\,1\textit{s} 					& 2.02(2)  \\
      \cline{3-4}
      &  ML, 60\,K				& N\,1\textit{s} 					&   - \\
      \hline
      \multirow{2}{*}{COHON~\cite{Chen_2019_JPhysCondensMatter}} & \multirow{2}{*}{0.4\,ML} 	& C\,1\textit{s}			& 2.46(4) \\
      \cline{3-4}
      & 					& O\,1\textit{s}					& 2.15(3) \\
      \hline
      \multirow{3}{*}{F$_4$TCNQ~\cite{Romaner_2007_PhysRevLett}} & & C\,1\textit{s}			& - \\
      \cline{3-4}
      & 					& N\,1\textit{s}					& 2.7(1)  \\
      \cline{3-4}
      & 					& F\,1\textit{s} 					& 3.3(1)   \\
      % \hline
      % Furan~\cite{Mulligan_2003_Surf.Sci.} & & O\,1\textit{s} & 2.39(4) \\
      % \hline
      % THF~\cite{Mulligan_2003_Surf.Sci.} & & O\,1\textit{s} & 3.08(4) \\
      % \hline
      % \multirow{6}{*}{h-BN} 	& \multirow{2}{*}{\cite{Brulke_2017_JPhysChemC}, 300\,K} & N\,1\textit{s}		 & 3.22(3)  \\
      % \cline{3-4}
      % &  					& B\,1\textit{s} 					& 3.25(2)  \\
      % \cline{2-4}
      % & \multirow{4}{*}{\cite{Schwarz_2017_ACSNano}, 300\,K} & N\,1\textit{s} main 			& 3.37(4)  \\
      % \cline{3-4}
      % &  					& N\,1\textit{s} -- defective				& 3.30(6)  \\
      % \cline{3-4}
      % &					& B\,1\textit{s} -- main				& 3.39(4)  \\
      % \cline{3-4}
      % &  					& B\,1\textit{s} -- defective				& 3.26(9)  \\
      \hline
      \multirow{2}{*}{\shortstack[l]{Phenyl \\ nitrene~\cite{Willenbockel_2015_ChemCommun}}} &\multirow{2}{*}{60\,K} & C\,1\textit{s} 	& 4.25(4)  \\
      \cline{3-4}
      &  					& N\,1\textit{s} 					& 1.17(4)  \\
      \hline
      \multirow{2}{*}{PYT~\cite{Chen_2019_JPhysCondensMatter}} & \multirow{2}{*}{0.9\,ML} 	& C\,1\textit{s} 				& 2.31(3)  \\
      \cline{3-4}
      & 					& O\,1\textit{s} 					& 2.04(6)  \\
      \hline
      Naphtalene~\cite{Klein_2019_PhysRevX} & ML, 150\,K & C\,1\textit{s} 	& 3.04(3)  \\
      \hline
      Azulene~\cite{Klein_2019_PhysRevX} & ML. 150\,K & C\,1\textit{s} 	        & 2.30(3)  \\
      \hline
      \multicolumn{4}{c}{\rule[-3mm]{0mm}{8mm}\textbf{Pentacene (PEN) derivatives on Ag(111)}} \\
      \hline
      % \multirow{6}{*}{F$_4$PEN~\cite{}} & \multirow{3}{*}{0.50\,ML}& C\,1\textit{s} -- total				& 2.97(5) \\
      % \cline{3-4}
      % &					& \mbox{C\,1\textit{s} -- PEN backbone}			& 3.00(3) \\
      % \cline{3-4}
      % &					& F\,1\textit{s} 					& 3.05(4) \\
      % \cline{2-4}
      % & \multirow{3}{*}{0.75\,ML}		& C\,1\textit{s} -- total				& 2.91(5) \\
      % \cline{3-4}
      % & 	 				& \mbox{C\,1\textit{s} -- PEN backbone}			& 2.97(4) \\
      % \cline{3-4}
      % & 	 				& F\,1\textit{s} 					& 2.93(5) \\
      % \hline
      \multirow{2}{*}{P2O~\cite{Heimel_2013_NatureChem}} & 	& C\,1\textit{s} 					& 3.32 \\
      \cline{3-4}
      & 					& O\,1\textit{s} 					& 3.35 \\
      \hline
      \multirow{2}{*}{P4O~\cite{Heimel_2013_NatureChem}} & 	& C\,1\textit{s} 					& 2.69 \\
      \cline{3-4}
      & 					& O\,1\textit{s} 					& 2.43 \\
      \hline	
      \multirow{4}{*}{PEN~\cite{Duhm_2013_ACSApplMaterInterfaces}} & 0.50\,ML, 295\,K & C\,1\textit{s}		 & 2.99(1) \\
      \cline{2-4}
      & 0.50\,ML, 145\,K 			& C\,1\textit{s} 					& 3.04(1) \\
      \cline{2-4}
      & 0.75\,ML, 295\,K 			& C\,1\textit{s} 					& 3.13(1) \\
      \cline{2-4}
      & 0.75\,ML, 145\,K 			& C\,1\textit{s} 					& 3.13(1) \\
      \hline
      \multirow{2}{*}{PFP~\cite{Duhm_2010_PhysRevB}} 	&  	& C\,1\textit{s} 					& 3.16(6) \\
      \cline{3-4}
      &					& F\,1\textit{s} 					& 3.16(6) \\
      % \cline{2-4}
      % & \multirow{2}{*}{\cite{Goiri_2014_PhysRevLett} Heteromol. layer 1:1 PFP--CuPc} & C\,1\textit{s} -- C--F 	& 3.48(5) \\
      % \cline{3-4}
      % &					& F\,1\textit{s}					& 3.52(5)	 \\
      \hline
\newpage
      \multicolumn{4}{c}{\rule[-3mm]{0mm}{8mm}\textbf{Perylene derivatives on Ag(111)}} \\
      \hline
      DIP~\cite{Burker_2013_PhysRevB} & 0.5\,ML 		& C\,1\textit{s} 					& 3.01(4) \\
      \hline
      \multirow{8}{*}{NTCDA} &\multirow{2}{*}{\cite{Stanzel_2004_SurfSci}relaxed ML}& O\,1\textit{s}		 & 3.02(2) \\
      \cline{3-4}
      &   					& O\,KLL 						& 3.03 	 \\
      \cline{2-4}
      & \multirow{2}{*}{\cite{Stanzel_2004_SurfSci}compressed ML}& O\,1\textit{s}			& 3.12(3)  \\
      \cline{3-4}
      &  					& O\,KLL 						& 3.06 	\\
      \cline{2-4}
      & \multirow{4}{*}{\cite{Stadler_2007_NewJPhys}} & C\,1\textit{s} 				& 2.997(16)   \\
      \cline{3-4}
      &  					& \mbox{O\,1\textit{s} -- total/O\,KLL}		& 2.872(14) \\
      \cline{3-4}
      &  					& O\,1\textit{s} -- carb. 				& 2.747(25) \\
      \cline{3-4}
      &  					& O\,1\textit{s} -- anh. 				& 3.004(15) \\
      % \cline{2-4}
      % & \multirow{5}{*}{\cite{Stadtmuller_2014_PhysRevB} Heteromol. layer (CuPc-rich) NTCDA--CuPc} & C\,1\textit{s} -- total	& 3.02(2) \\
      % \cline{3-4}
      % &					& C\,1\textit{s} -- NTCDA 				& 3.02(1) \\
      % \cline{3-4}
      % &					& O\,1\textit{s} -- total				& 2.83(2) \\
      % \cline{3-4}
      % &					& O\,1\textit{s} -- carb.				& 2.78(2) \\
      % \cline{3-4}
      % &					& O\,1\textit{s} -- anh.				& 2.98(3) \\
      \hline	
      \multirow{14}{*}{PTCDA} & \multirow{3}{*}{~\cite{Henze_2007_SurfSci, Gerlach_2007_PhysRevB, Hauschild_2005_PhysRevLett}}& C\,1\textit{s}	& 2.86 \\
      \cline{3-4}
      & 					& O\,1\textit{s} -- carb. 				& 2.68 	 \\
      \cline{3-4}
      & 					& O\,1\textit{s} -- anh.				& 2.97 	 \\
      \cline{2-4}
      & \multirow{3}{*}{\cite{Kilian_2008_PhysRevLett} LT} 	& C\,1\textit{s} 			& 2.80(2) \\
      \cline{3-4}
      & 					& O\,1\textit{s} -- carb. 				& 2.49(4)  \\
      \cline{3-4}
      & 					& O\,1\textit{s} -- anh.				& 2.83(4)  \\
      \cline{2-4}
      & \multirow{4}{*}{\cite{Hauschild_2010_PhysRevB} 300\,K} & C\,1\textit{s} 			& 2.86(1) \\
      \cline{3-4}
      & 					& O\,1\textit{s} -- total	 			& 2.86(2) \\
      \cline{3-4}
      & 					& O\,1\textit{s} -- carb. 				& 2.66(3) \\
      \cline{3-4}
      & 					& O\,1\textit{s} -- anh.				& 2.98(8) \\
      \cline{2-4}
      & \multirow{4}{*}{\cite{Hauschild_2010_PhysRevB} 100\,K}  & C\,1\textit{s} 			& 2.81(2) \\
      \cline{3-4}
      & 					& O\,1\textit{s} -- total	 			& 2.67(3)  \\
      \cline{3-4}
      & 					& O\,1\textit{s} -- carb. 				& 2.50(4) \\
      \cline{3-4}
      & 					& O\,1\textit{s} -- anh.				& 2.83(4) \\
      \hline
      \multirow{14}{*}{PTCDA}& \multirow{8}{*}{\shortstack[l]{\cite{Stadtmuller_2016_PhysRevB} Surf. Pb$_1$Ag$_2$ \\alloy}} & C\,1\textit{s} site A		& 3.63(4) \\
      \cline{3-4}
      & 					& C\,1\textit{s} site B					& 3.80(4) \\
      \cline{3-4}
      & 					& \mbox{O\,1\textit{s} -- carb. site A}			& 3.33(2)  \\
      \cline{3-4}
      & 					& \mbox{O\,1\textit{s} -- carb. site B}			& 3.61(2)  \\
      \cline{3-4}
      & 					& O\,1\textit{s} -- anh. site A				& 3.52(9) \\
      \cline{3-4}
      & 					& O\,1\textit{s} -- anh. site B				& 3.74(5)  \\
      \cline{3-4}
      & 					& Pb\,4\textit{f} -- bare				& 0.41(2)  \\
      \cline{3-4}
      & 					& Pb\,4\textit{f} -- PTCDA				& 0.49(1)  \\
      \cline{2-4}
      & \multirow{6}{*}{\shortstack[l]{\cite{Baby_2017_ACSNano}K-doped ML \\ (K$_2$PTCDA)}} & C\,1\textit{s} perylene core	& 3.12(2) \\
      \cline{3-4}
      & 					& C\,1\textit{s} C=O					& 3.26(7) \\
      \cline{3-4}
      & 					& O\,1\textit{s} -- carb.				& 3.36(7)  	\\
      \cline{3-4}
      & 					& O\,1\textit{s} -- anh.	 			& 3.36(5) 	\\
      \cline{3-4}
      & 					& K\,2\textit{p}--PTCDA 				& 3.23(3)	\\
      \cline{3-4}
      & 					& K\,2\textit{p}--Ag 					& 3.26(30) 	\\
      \hline		
      \multirow{6}{*}{PTCDI~\cite{Franco-Canellas_2017_PhysRevMaterials}} & \multirow{2}{*}{} & \mbox{C\,1\textit{s} -- perylene core}		& 2.86(2) \\
      \cline{3-4}
      & 					& C\,1\textit{s} -- C=O					& 2.73(3)  \\
      \cline{3-4}
      & 					& C\,1\textit{s} -- C--H				& 2.87(2) \\
      \cline{3-4}
      & 					& C\,1\textit{s} -- C--C+C--C--O			& 2.85(2) \\
      \cline{3-4}
      &					& N\,1\textit{s} 					& 2.58(3) \\
      \cline{3-4}	
      &					& O\,1\textit{s} 					& 2.60(3) \\
      \hline
      \multirow{2}{*}{TAT~\cite{Yang_2016_PhysRevB}} & \multirow{2}{*}{} & C\,1\textit{s} 				 & 2.99(5) \\
      \cline{3-4}
      &					& N\,1\textit{s} 					& 2.88(10) \\
      \hline
      \multicolumn{4}{c}{\rule[-3mm]{0mm}{8mm}\textbf{Phthalocyanine (Pc) derivatives on Ag(111)}} \\
      \hline
      \multirow{21}{*}{CuPc} &\multirow{3}{*}{\cite{Kroger_2010_NewJPhys} \shortstack[l]{0.5 ML, \\ 300 K}} & C\,1\textit{s}		& 3.049(5) \\
      \cline{3-4}
      & 					& N\,1\textit{s} 					& 3.00(4) \\
      \cline{3-4}
      & 					& Cu\,2\textit{p}$_{3/2}$ 				& 2.98(4) \\
      \cline{2-4}
      & \multirow{3}{*}{\cite{Kroger_2010_NewJPhys} \shortstack[l]{0.85 ML, \\ 300 K}} 	& C\,1\textit{s}		& 2.993(3) \\
      \cline{3-4}
      & 					& N\,1\textit{s} 					& 3.03(4) \\
      \cline{3-4}
      & 					& Cu\,2\textit{p}$_{3/2}$ 				& 2.90(4) \\
      \cline{2-4}	
      & \multirow{3}{*}{\cite{Kroger_2010_NewJPhys} \shortstack[l]{1.00 ML, \\ 300 K} } 	& C\,1\textit{s}		& 3.089(3) \\
      \cline{3-4}
      & 					& N\,1\textit{s} 					& 3.04(4) \\
      \cline{3-4}
      & 					& Cu\,2\textit{p}$_{3/2}$ 				& 2.97(4) \\
      \cline{2-4}
      & \multirow{3}{*}{\cite{Kroger_2010_NewJPhys} \shortstack[l]{0.5 ML, \\ 153 K} } 	& C\,1\textit{s}		& 2.999(4) \\
      \cline{3-4}
      & 					& N\,1\textit{s} 					& 2.94(4) \\
      \cline{3-4}
      & 					& Cu\,2\textit{p}$_{3/2}$ 				& 2.89(4) \\
      \cline{2-4}	
      & \multirow{3}{*}{\cite{Kroger_2010_NewJPhys} \shortstack[l]{0.85 ML, \\ 140 K} } 	& C\,1\textit{s}		& 3.010(2) \\
      \cline{3-4}
      & 					& N\,1\textit{s}					& 3.01(4) \\
      \cline{3-4}	
      & 					& Cu\,2\textit{p}$_{3/2}$ 				& 2.94(4) \\
      \cline{2-4}
\newpage
      & \multirow{3}{*}{\cite{Kroger_2010_NewJPhys} \shortstack[l]{1.00 ML, \\ 140 K} } 	& C\,1\textit{s}		& 3.077(2) \\
      \cline{3-4}
      & 					& N\,1\textit{s} 					& 3.07(4) \\
      \cline{3-4}
      & 					& Cu\,2\textit{p}$_{3/2}$ 				& 3.02(4) \\
      \cline{2-4}
      & \multirow{3}{*}{\cite{Kleimann_2014_JPhysChemC} \shortstack[l]{1.00 ML, \\ RT} } & C\,1\textit{s} 		& 3.02(5) \\
      \cline{3-4}
      & 					& N\,1\textit{s} 					& 3.00(5) \\
      \cline{3-4}
      & 					& Cu\,2\textit{p}$_{3/2}$ 				& 3.09(5) \\
      \hline
      & \multirow{5}{*}{\shortstack[l]{\cite{Stadtmuller_2016_PhysRevB} Surface \\Pb$_1$Ag$_2$ alloy}} & C\,1\textit{s} & 3.77(2) \\
      \cline{3-4}
      & 					& N\,1\textit{s}					& 3.68(2) \\
      \cline{3-4}
      & 					& Cu\,2\textit{p}$_{3/2}$	 			& 3.62(2) \\
      \cline{3-4}
      & 					& Pb\,4\textit{f} -- bare				& 0.44(1)  \\
      \cline{3-4}
      & 					& Pb\,4\textit{f} -- CuPc	 			& 0.44(1)  \\
      \hline
      \multirow{2}{*}{F$_{16}$CuPc} &\multirow{2}{*}{\cite{Gerlach_2005_PhysRevB}} & C\,1\textit{s} 			 & 3.25 \\
      \cline{3-4}
      & 					& F\,1\textit{s} 					& 3.45 \\
      % \cline{2-4}
      % & \multirow{3}{*}{\cite{Kleimann_2014_JPhysChemC} Bilayer: CuPc bottom/F$_{16}$CuPc top, 50\,K} & C\,1\textit{s}\footnote{From vector analysis.\label{argand2}} 										& 6.11(5) \\
      % \cline{3-4}
      % & 					& N\,1\textit{s}\textit{s}\footnoteref{argand2}		& 6.16(2) \\
      % \cline{3-4}
      % & 					& F\,1 							& 6.16(2) \\
      \hline
      \multirow{12}{*}{FePc~\cite{Deimel_2016_ChemSci}} & \multirow{4}{*}{No dosing} & C\,1\textit{s} -- C--C 	& 2.92(5) \\
      \cline{3-4}
      & 					& C\,1\textit{s} -- C--N 				& 2.83(5) \\
      \cline{3-4}
      & 					& N\,1\textit{s} 					& 2.71(7) \\
      \cline{3-4}
      & 					& Fe\,2\textit{p}$_{3/2}$ 				& 2.61(1) \\
      \cline{2-4}
      & \multirow{4}{*}{NH$_3$ dosing}	& C\,1\textit{s} -- C--C 				& 2.98(3) \\
      \cline{3-4}
      & 					& C\,1\textit{s} -- C--N 				& 2.90(3) \\
      \cline{3-4}
      & 					& N\,1\textit{s} 					& 2.84(2) \\
      \cline{3-4}
      & 					& Fe\,2\textit{p}$_{3/2}$ 				& 2.80(7) \\
      \cline{2-4}
      & \multirow{4}{*}{H$_2$O dosing} 	& C\,1\textit{s} -- C--C 				& 2.90(1) \\
      \cline{3-4}
      & 					& C\,1\textit{s} -- C--N 				& 2.84(2) \\
      \cline{3-4}
      & 					& N\,1\textit{s} 					& 2.79(6) \\
      \cline{3-4}
      & 					& Fe\,2\textit{p}$_{3/2}$ 				& 2.68(4) \\
      \hline
      \multirow{12}{*}{H$_2$Pc~\cite{Kroger_2012_PhysRevB}} &\multirow{2}{*}{0.7 ML, 300\,K} & C\,1\textit{s} 		& 3.04(7) \\
      \cline{3-4}
      & 					& N\,1\textit{s} 					& 2.81(7) \\
      \cline{2-4}
      &\multirow{2}{*}{0.8\,ML, 300\,K} 	& C\,1\textit{s} 					& 3.07(7) \\
      \cline{3-4}
      & 					& N\,1\textit{s} 					& 2.82(7) \\
      \cline{2-4}
      & \multirow{2}{*}{0.93\,ML, 300\,K} 	& C\,1\textit{s} 					& 3.07(7) \\
      \cline{3-4}
      & 					& N\,1\textit{s} 					& 2.90(7) \\
      \cline{2-4}
      & \multirow{2}{*}{0.7\,ML, 183\,K} 	& C\,1\textit{s} 					& 3.06(7) \\
      \cline{3-4}
      & 					& N\,1\textit{s} 					& 2.82(7) \\
      \cline{2-4}
      & \multirow{2}{*}{0.8\,ML, 183\,K}	& C\,1\textit{s} 					& 3.03(7) \\
      \cline{3-4}
      & 					& N\,1\textit{s} 					& 2.93(7) \\
      \cline{2-4}
      & \multirow{2}{*}{0.93\,ML, 183\,K} 	& C\,1\textit{s} 					& 3.08(7) \\
      \cline{3-4}
      & 					& N\,1\textit{s}					& 3.02(7) \\
      \hline
      \multirow{6}{*}{SnPc~\cite{Stadler_2006_PhysRevB}} &\multirow{3}{*}{1\,ML, 300\,K} & C\,1\textit{s}		& 3.16(3) \\
      \cline{3-4}
      & 					& N\,1\textit{s} 					& 3.24(6) \\
      \cline{3-4}
      & 					& Sn\,3\textit{d} 					& 2.46(3) \\
      \cline{2-4}
      & \multirow{3}{*}{0.87\,ML,\,150 K} 	& C\,1\textit{s} 					& 2.93(6) \\
      \cline{3-4}
      & 					& N\,1\textit{s} 					& 3.12(7) \\
      \cline{3-4}
      & Sn down/up 				& Sn\,3\textit{d}  					& 2.59/4.01 \\
      % \cline{2-4}
      % & \multirow{4}{*}{\cite{Stadtmuller_2014_PhysRevB} Heteromol. layer 1:2 PTCDA--SnPc} & C\,1\textit{s} -- total		& 3.09(2) \\
      % \cline{3-4}
      % &					& C\,1\textit{s} -- SnPc 				& 3.13(2) \\
      % \cline{3-4}
      % &					& N\,1\textit{s}					& 3.19(2) \\
      % \cline{3-4}
      % &					& Sn\,2\textit{p}$_{3/2}$				& 2.29(2) \\
      \hline
      \multirow{4}{*}{TiOPc~\cite{Kroger_2016_NewJPhys}} &\multirow{4}{*}{0.95\,ML, all up} & C\,1\textit{s} 		& 3.01(3) \\
      \cline{3-4}
      & 					& N\,1\textit{s} 					& 2.88(6) \\
      \cline{3-4}
      & 					& O\,1\textit{s} 					& 5.38(10) \\
      \cline{3-4}
      & 					& Ti\,2\textit{p}$_{3/2}$ 				& 3.70(8) \\
      \cline{2-4}
      % & \multirow{4}{*}{2\,ML, up (first)/down (second)} & C\,1\textit{s} 				& 3.01(3)/6.30(8) \\
      % \cline{3-4}
      % & 					& N\,1\textit{s} 					& 2.88(6)/6.00(11) \\
      % \cline{3-4}
      % & 					& O\,1\textit{s} 					& 5.38(10)/3.60(18) \\
      % \cline{3-4}
      % & 					& Ti\,2\textit{p}$_{3/2}$ 				& 3.70(8)/5.30(26) \\
      \hline
      \multicolumn{4}{c}{\rule[-3mm]{0mm}{8mm}\textbf{Other compounds on Ag(111)}} \\
      \hline
      \multirow{6}{*}{Azobenzene} &\multirow{2}{*}{\cite{Mercurio_2010_PhysRevLett}} & C\,1\textit{s}			 & -    \\
      \cline{3-4}
      &  					& N\,1\textit{s} 					& 3.07(2)  \\
      \cline{2-4}
      & \multirow{2}{*}{\cite{Mercurio_2013_PhysRevBb} 210\,K} & C\,1\textit{s} 			& -   \\
      \cline{3-4}
      &  					& N\,1\textit{s} 					& 2.97(5)  \\
      \cline{2-4}
      & \multirow{2}{*}{\cite{Mercurio_2014_FrontPhysics} 210\,K} & C\,1\textit{s} 			& 2.99(5)   \\
      \cline{3-4}
      &  					& N\,1\textit{s} 					& 2.97(5)  \\
      \hline
      Benzene~\cite{Liu_2015_PhysRevLett} & 80\,K		& C\,1\textit{s}					& 3.04(2) \\
      \hline
      EC4T~\cite{Kilian_2002_PhysRevB} & 			& S\,1\textit{s}					& 3.15(5) \\
      \hline
      \multirow{3}{*}{NO$_2$PYT~\cite{Hofmann_2017_JPhysChemC}} 	& \multirow{3}{*}{0.6 ML} & C\,1\textit{s}		& 2.82(2) \\
      \cline{3-4}
      & 					& O\,1\textit{s} -- carb. 				& 2.23(3) \\
      \cline{3-4}
      & 					& O\,1\textit{s} -- nitr.				& 2.61(7) \\
      \hline
      \multirow{2}{*}{PYT~\cite{Hofmann_2017_JPhysChemC}} & \multirow{2}{*}{0.6 ML} & C 1\textit{s} 			 & 2.46(3) \\
      \cline{3-4}
      & 					& O\,1\textit{s} 					& 2.31(4) \\
      \hline
      TBA~\cite{McNellis_2010_ChemPhysLett} & 		& N\,1\textit{s} 					& 3.21(5)  \\
      \hline
\newpage
      \multicolumn{4}{c}{\rule[-3mm]{0mm}{8mm}\textbf{Other compounds on Ag(111) (cont.)}} \\
      \hline
      \multirow{10}{*}{TCNQ}& \multirow{4}{*}{\cite{Blowey_2017_FaradayDiscuss}} & C\,1\textit{s} -- C--H		& 2.86(2) \\
      \cline{3-4}
      & 					& C\,1\textit{s} -- C--C				& 2.78(2) \\
      \cline{3-4}
      & 					& C\,1\textit{s} -- C--N	 			& 2.76(2) 	\\
      \cline{3-4}
      & 					& N\,1\textit{s}					& 2.75(3)  	\\
      \cline{2-4}		
      & \multirow{6}{*}{\shortstack[l]{\cite{Blowey_2017_FaradayDiscuss} K-doped ML \\ (K-TCNQ)}}& C\,1\textit{s} -- C--H	& 2.85(1) \\
      \cline{3-4}
      & 					& C\,1\textit{s} -- C--C				& 2.79(1) \\
      \cline{3-4}
      & 					& C\,1\textit{s} -- C--N	 			& 2.79(1) 	\\
      \cline{3-4}
      & 					& N\,1\textit{s}					& 2.79(5)  	\\
      \cline{3-4}
      & 					& K\,2\textit{p} -- bare 				& 2.81(1)	\\
      \cline{3-4}
      & 					& K\,2\textit{p} -- TCNQ 				& 3.56(3) 	\\
      \hline
      \multicolumn{4}{c}{\rule[-3mm]{0mm}{8mm}\textbf{Pentacene (PEN) derivatives on Au(111)}} \\
      \hline
      P4O~\cite{Heimel_2013_NatureChem} 	& 1.0\,ML 	& C\,1\textit{s} 					& 3.27 \\
      \hline
      \multicolumn{4}{c}{\rule[-3mm]{0mm}{8mm}\textbf{Perylene derivatives on Au(111)}} \\
      \hline
      DIP~\cite{Burker_2013_PhysRevB} 	& 0.8\,ML 	& C\,1\textit{s} 					& 3.10(3) \\
      \hline
      Perylene~\cite{Franco-Canellas_2017_PhysRevMaterials} & & C\,1\textit{s} 					& 3.01(6) \\
      \hline
      PTCDA~\cite{Henze_2007_SurfSci} 	& 		& C\,1\textit{s} 					& 3.27(2) \\
      \hline
      \multirow{2}{*}{PTCDI~\cite{Franco-Canellas_2017_PhysRevMaterials}} & \multirow{2}{*}{} & C\,1\textit{s} 	& 3.29(2) \\
      \cline{3-4}
      & 					& N\,1\textit{s} 					& 3.26(3) \\
      \hline
      \multirow{2}{*}{TAT~\cite{Yang_2016_PhysRevB}} & \multirow{2}{*}{} & C\,1\textit{s} 				 & 3.06(7) \\
      \cline{3-4}
      & 					& N\,1\textit{s} 					& 3.04(2) \\
      \hline
      \multicolumn{4}{c}{\rule[-3mm]{0mm}{8mm}\textbf{Phthalocyanine (Pc) derivatives on Au(111)}} \\
      \hline
      \multirow{12}{*}{CuPc~\cite{Kroger_2011_PhysRevB}} &\multirow{3}{*}{0.7\,ML, 300\,K} & C\,1\textit{s}		& 3.31(7) \\
      \cline{3-4}
      & 					& N\,1\textit{s} 					& 3.26(7) \\
      \cline{3-4}
      &					& Cu\,2\textit{p}$_{3/2}$ 				& 3.20(7) \\
      \cline{2-4}
      & \multirow{3}{*}{1.0\,ML, 300\,K} 	& C\,1\textit{s} 					& 3.31(7) \\
      \cline{3-4}
      & 					& N\,1\textit{s} 					& 3.26(7) \\
      \cline{3-4}
      & 					& Cu\,2\textit{p}$_{3/2}$ 				& 3.25(7) \\
      \cline{2-4}
      & \multirow{3}{*}{0.7\,ML, 133\,K} 	& C\,1\textit{s}  					& 3.37(7) \\
      \cline{3-4}
      & 					& N\,1\textit{s} 					& 3.25(7) \\
      \cline{3-4}
      & 					& Cu\,2\textit{p}$_{3/2}$ 				& 3.25(7) \\
      \cline{2-4}
      & \multirow{3}{*}{1.0\,ML, 133\,K} 	& C\,1\textit{s}					& 3.28(7) \\
      \cline{3-4}
      & 					& N\,1\textit{s} 					& 3.27(7) \\
      \cline{3-4}
      & 					& Cu\,2\textit{p}$_{3/2}$ 				&  3.29(7) \\
      \hline
      \multirow{2}{*}{F$_{16}$CuPc~\cite{Oteyza_2010_JChemPhys}} & \multirow{2}{*}{} & C\,1\textit{s}			 & 3.25 \\
      \cline{3-4}
      & 					& F\,1\textit{s} 					& 3.26 \\
      \hline
      % \multicolumn{4}{c}{\rule[-3mm]{0mm}{8mm}\textbf{Other compounds on Au(111)}} \\
      % \hline
      % \multirow{2}{*}{HBC} & 0.3\,ML, 294\,K  		& C\,1\textit{s} 					& 3.10(2) \\
      % \cline{2-4}
      % & 0.3\,ML, 117\,K 			& C\,1\textit{s} 					& 3.09(2) \\
      % \hline
%
      \multicolumn{4}{c}{\rule[-3mm]{0mm}{8mm}\textbf{Various COMs on other surfaces}} \\
      \hline
      \multirow{5}{*}{\shortstack[l]{MnPc \\on Cu(001)}~\cite{Javaid_2013_PhysRevB}} & \multirow{5}{*}{(002) reflection}& C\,1\textit{s} total & 2.403(28) \\
      \cline{3-4}
      & 					& C\,1\textit{s} -- C--C				& 2.412(28)\\
      \cline{3-4}
      &					& C\,1\textit{s} -- C--N				& 2.376(26) \\
      \cline{3-4}
      & 					& N\,1\textit{s} 					& 2.312(19) \\
      \cline{3-4}
      &					& Mn\,2\textit{p}$_{3/2}$ 				& 2.24(45) \\
      % \cline{3-4}
      % & 					& Cu\,2\textit{p}$_{3/2}$ 				& 3.631(9) \\
      \hline
      \multirow{8}{*}{\shortstack[l]{PTCDA \\on Cu(100)~\cite{Weiss_2017_PhysRevB}}} & \multirow{4}{*}{(200) reflection}& \mbox{C\,1\textit{s} -- perylene core}& 2.44(2) \\
      \cline{3-4}
      & 				& C\,1\textit{s} -- C=O					& 2.53(2)\\
      \cline{3-4}
      &					& O\,1\textit{s} -- carb.				& 2.47(5) \\
      \cline{3-4}
      &					& O\,1\textit{s} -- anh.				& 2.76(2) \\
      \cline{2-4}
      & \multirow{4}{*}{(111) reflection}& \mbox{C\,1\textit{s} -- perylene core} 			& - \\
      \cline{3-4}
      & 				& C\,1\textit{s} -- C=O					& - \\
      \cline{3-4}
      &					& O\,1\textit{s} -- carb.				& - \\
      \cline{3-4}
      &					& O\,1\textit{s} -- anh.				& - \\
      \hline
      \multirow{4}{*}{\shortstack[l]{CoTPP \\on Ag(100)~\cite{Wechsler_2017_JPhysChemC}}} & \multirow{4}{*}{(200) reflection}& C\,1\textit{s} -- core 	& 3.10 \\
      \cline{3-4}
      &					& \mbox{C\,1\textit{s} -- phenyl rings}			& 2.45 \\
      \cline{3-4}
      & 				& N\,1\textit{s} 					& 3.10 \\
      \cline{3-4}
      &					& Co\,2\textit{p}$_{3/2}$ 				& 3.00 \\
      \hline
\newpage
      \multirow{12}{*}{\shortstack[l]{PTCDA \\ on Ag(100)~\cite{Bauer_2012_PhysRevB}}} & \multirow{6}{*}{\shortstack[l]{0.8\,ML, \\ (200) reflection}}& C\,1\textit{s} -- total	& 2.81(2) \\
      \cline{3-4}
      & 				& \mbox{C\,1\textit{s} -- perylene core}		& 2.84(2)\\
      \cline{3-4}
      & 				& C\,1\textit{s} -- C=O					& 2.73(1)\\
      \cline{3-4}
      & 				& O\,1\textit{s} -- total				& 2.64(2)\\
      \cline{3-4}	
      &					& O\,1\textit{s} -- carb.				& 2.53(2) \\
      \cline{3-4}
      &					& O\,1\textit{s} -- anh.				& 2.78(2) \\
      \cline{2-4}
      & \multirow{4}{*}{\shortstack[l]{0.8\,ML, \\(111) reflection}}& C\,1\textit{s} -- total  				& - \\
      \cline{3-4}
      & 				& \mbox{C\,1\textit{s} -- perylene core}		& - \\
      \cline{3-4}
      & 				& C\,1\textit{s} -- C=O					& - \\
      \cline{3-4}
      &					& O\,1\textit{s} -- total				& - \\
      \cline{3-4}
%      &					& O\,1\textit{s} -- carb.				& - \\
%      \cline{3-4}
%      &					& O\,1\textit{s} -- anh.				& - \\
      \hline
%
% add NTCDA / Ag(100) 
%
      \multirow{3}{*}{\shortstack[l]{NTCDA \\ on Ag(100)~\cite{Felter_2019_Nanoscale}}} & \multirow{3}{*}{\shortstack[l]{0.x\,ML, \\ (200) reflection}}& C\,1\textit{s} -- total	& 2.384(9) \\
      \cline{3-4}
      &					& O\,1\textit{s} -- carb.				& 2.25(2) \\
      \cline{3-4}
      &					& O\,1\textit{s} -- anh.				& 2.42(4) \\
      \cline{2-4}
      \hline
      \multirow{19}{*}{\shortstack[l]{PTCDA \\on Ag(110)}} & \multirow{6}{*}{\shortstack[l]{\cite{Bauer_2012_PhysRevB} 0.9\,ML, \\(220) reflection}}& C\,1\textit{s} -- total 	& 2.56(1) \\
      \cline{3-4}
      & 				& \mbox{C\,1\textit{s} -- perylene core}		& 2.58(1) \\
      \cline{3-4}
      & 				& C\,1\textit{s} -- C=O					& 2.45(11) \\
      \cline{3-4}
      &					& O\,1\textit{s} -- total				& 2.33(3) \\
      \cline{3-4}
      &					& O\,1\textit{s} -- carb.				& 2.30(4) \\
      \cline{3-4}
      &					& O\,1\textit{s} -- anh.				& 2.38(3) \\
      \cline{2-4}	
      & \multirow{6}{*}{\shortstack[l]{\cite{Mercurio_2013_PhysRevB} 0.89\,ML, \\(220) reflection}}& C\,1\textit{s} total	& 2.56(2) \\
      \cline{3-4}
      & 				& \mbox{C\,1\textit{s} -- perylene core}		& 2.59(1) \\
      \cline{3-4}
      & 				& C\,1\textit{s} -- C=O					& 2.45(11) \\
      \cline{3-4}
      &					& O\,1\textit{s} -- total				& 2.36(5) \\
      \cline{3-4}
      &					& O\,1\textit{s} -- carb.				& 2.32(5) \\
      \cline{3-4}
      &					& O\,1\textit{s} -- anh.				& 2.41(6) \\
      \cline{2-4}
      & \multirow{6}{*}{\shortstack[l]{\cite{Mercurio_2013_PhysRevBa} K-doped \\(K-PTCDA), \\(220) reflection}}& C\,1\textit{s} -- total 		& 2.66(3) \\
      \cline{3-4}
      & 				& \mbox{C\,1\textit{s} -- perylene core}		& 2.64(3) \\
      \cline{3-4}                       & 				& C\,1\textit{s} -- C=O					& 2.73(6) \\
      \cline{3-4}
      &					& O\,1\textit{s} -- total				& 2.70(10) \\
      \cline{3-4}
      &					& O\,1\textit{s} -- carb.				& 2.63(10) \\
      \cline{3-4}
      &					& O\,1\textit{s} -- anh.				& 2.76(11) \\
      \cline{3-4}
      &					& K\,1\textit{s}					& 1.44(6) \\	
      \hline
      \multirow{4}{*}{\shortstack[l]{Graphene (Gr)\\on Ir(111)~\cite{Runte_2014_PhysRevB}}} & 0.22\,ML   & C\,1\textit{s} 		& - \\
      \cline{3-4}
      & 0.39\,ML				& C\,1\textit{s}					& - \\
      \cline{3-4}
      & 0.63\,ML				& C\,1\textit{s} 					& - \\
      \cline{3-4}
      & 1.00\,ML				& C\,1\textit{s} 					& - \\
      \hline
      \multirow{9}{*}{\shortstack[l]{Graphene  \\on SiC(0001)~\cite{Emery_2013_PhysRevLett}}} & \multirow{3}{*}{\shortstack[l]{UHV-grown \\0.5\,ML}}  & \mbox{C\,1\textit{s} -- second layer} 	& - \\
      \cline{3-4}
      & 					& \mbox{C\,1\textit{s} -- first layer (1)}		& 2.3(2) \\
      \cline{3-4}
      & 					& \mbox{C\,1\textit{s} -- first layer (2)}		& 2.0(1) \\
      \cline{2-4}
      & \multirow{3}{*}{\shortstack[l]{UHV-grown \\1.3\,ML}}  	& \mbox{C\,1\textit{s} -- second layer}	& - \\
      \cline{3-4}
      & 					& \mbox{C\,1\textit{s} -- first layer (1)}		& 2.4(1) \\
      \cline{3-4}
      &					& \mbox{C\,1\textit{s} -- first layer (2)}			& 2.1(1) \\
      \cline{2-4}
      & \multirow{3}{*}{\shortstack[l]{Ar-grown \\  1.7\,ML}}  	& \mbox{C\,1\textit{s} -- second layer}	& - \\
      \cline{3-4}
      & 					& \mbox{C\,1\textit{s} -- first layer (1)}		& 2.52(13)\\
      \cline{3-4}
      & 					& \mbox{C\,1\textit{s} -- first layer (2)}		& 2.12(7) \\
      \hline
      \multirow{1}{*}{\shortstack[l]{Graphene on 6H-SiC(0001)~\cite{Sforzini_2015_PhysRevLett}}} &  & C\,1\textit{s} -- graphene	& 4.272(60)  \\
      \hline
      \multirow{11}{*}{\shortstack[l]{Graphene \\on SiC(0001)~\cite{Sforzini_2016_PhysRevLett}}} & \multirow{2}{*}{\shortstack[l]{Gr buffer layer \\(BL) bare}} & C\,1\textit{s} -- BL bare & 2.30(1)\\
      \cline{3-4}
      & 					& \mbox{C\,1\textit{s} -- BL bare doped}		& 2.28(2) \\
      \cline{2-4}
      \multirow{1}{*}{} 	& \multirow{5}{*}{\shortstack[l]{Epitaxial ML\\ on a BL}} & C\,1\textit{s} -- BL 				& 2.37(2)\\
      \cline{3-4}
      & 					& C\,1\textit{s} -- BL doped	 			& 2.37(2) \\
      \cline{3-4}
      & 					& C\,1\textit{s} -- graphene	 			& 5.67(1) \\
      \cline{3-4}
      & 					& \mbox{C\,1\textit{s} -- graphene doped}		& 5.65(1) \\
      \cline{3-4}
      & 					& \mbox{N\,1\textit{s} -- graphene doped}		& 5.72(5) \\
      \cline{2-4}
      & \multirow{4}{*}{ML on 6H-termin.} 	& C\,1\textit{s} -- graphene				& 4.28(1) \\
      \cline{3-4}
      & 					& \mbox{C\,1\textit{s} -- graphene doped}		& 4.56(1) \\
      \cline{3-4}
      &					& \mbox{N\,1\textit{s} -- graphene doped}			& 4.49(5) \\
      \cline{3-4}
      &					& N\,1\textit{s} -- interstitial			& 2.71(4) \\
      \hline
      \multirow{6}{*}{h-BN on Ir(111)~\cite{Hagen_2016_ACSNano}} &   & N\,1\textit{s} -- total			 & - \\
      \cline{3-4}
      & 					& \mbox{N\,1\textit{s} -- strongly bound}			& 2.22(2) \\
      \cline{3-4}
      & 					& \mbox{N\,1\textit{s} -- weakly bound}			& 3.72(2) \\
      \cline{3-4}
      & 					& B\,1\textit{s} -- total				& - \\
      \cline{3-4}
      & 					& \mbox{B\,1\textit{s} -- strongly bound}		& 2.17(2) \\
      \cline{3-4}
      & 					& \mbox{B\,1\textit{s} -- weakly bound}			& 3.70(2) \\
      \hline
\end{longtable}
\endgroup

%\newpage

%%% Local Variables: 
%%% mode: pdflatex
%%% TeX-master: "manuscript_review_monolayers_MASTER"
%%% End: 

\section{Case studies} \label{sec:case_studies}

\subsection{Overview and compilation of adsorption distances} \label{table}
We organize the case studies according to the strength of the interaction with the substrate.  In the
following sections, we present some well-studied systems that illustrate the current understanding of the
different adsorption regimes, i.e.\ (vdW-dominated) physisorption, clear chemisorption and different cases in
between. Most of the systems have been studied with both XSW and UPS.

For a general overview, we also refer to Table~\ref{tab:xsw-data} which provides a comprehensive list with
adsorption distances of COMs on metals obtained by XSW measurements. For reasons of space only the molecule,
the substrate and the adsorption distance for the different elements (if applicable) are reported.

\subsection{Weakly interacting systems}
Weakly interacting systems, which are dominated by dispersion forces and lack stronger covalent interactions,
represent the limiting case of physisorptive bonding.  Because they may be considered as test ground for DFT
calculations with van der Waals (vdW) corrections, precise XSW results have become particularly important for
evaluating the accuracy of these methods.  A prototypical system falling in this category would be simply
benzene on Ag(111), which was also pursued in Refs.~\onlinecite{Liu_2015_PhysRevLett,
  Maass_2018_JChemPhys}. With regard to possible applications in organic electronics, though, larger acenes or
similar systems have greater practical relevance due to their more suitable energy-levels and smaller
HOMO-LUMO gap, as well as greater thermal stability, since benzene desorbs already at 300\,K.

As model system we may consider diindenoperylene (DIP, see Figure~\ref{fig:molecules}), a $\pi$-conjugated
organic semiconductor with excellent optoelectronic device performance, which has been studied over the
last decade both in thin films~\cite{Durr_2003_PhysRevLett, Kowarik_2006_PhysRevLett,
  Heinemeyer_2010_PhysRevLett, Banerjee_2013_PhysRevLett, Kurrle_2008_ApplPhysLett, Wagner_2012_JApplPhys,
  Simbrunner_2019_JApplCryst} and in monolayers on noble metal surfaces~\cite{Huang_2011_PhysChemChemPhys,
  Yonezawa_2016_ApplPhysExpress, Hosokai_2017_OrgElec, Oteyza_2008_JPhysChemC,
  Oteyza_2009_PhysChemChemPhys, Burker_2013_PhysRevB, Aldahhak_2015_PhysChemChemPhys}. With respect to its
chemical structure, DIP is a relatively simple, planar hydrocarbon without heteroatoms. In contrast to the
intensely studied PTCDA~\cite{Weiss_2015_NatCommun, Puschnig_2017_JPhysChemLett,
  Weinhardt_2016_JPhysChemC, Ziroff_2010_PhysRevLett, Romaner_2009_NewJPhys, Tautz_2007_ProgSurfSci,
  Zou_2006_SurfSci, Stallberg_2019_PhysRevB, Nicoara_2019_PhysStatusSolidiB}, i.e.\ a molecule with the
same perylene core but with four carbonyl groups, the specific DIP--substrate interaction is not
complicated by polar side groups -- see Figure~\ref{fig:perylene_derivatives} which illustrates the
significant influence of funtional groups on the bonding distances for these molecules.  Moreover, the
influence of intermolecular (lateral) interactions is expected to be smaller than for PTCDA.

Generally, the reliable prediction of the equilibrium structure and energetics of hybrid inorganic/organic
systems from first principles represents a significant challenge for theoretical methods due to the interplay
of, generally, covalent interactions, electron transfer processes, Pauli repulsion, and vdW
interactions. Recent years have seen substantial efforts to incorporate vdW interactions into density
functional theory (DFT) calculations in order to determine the structure and stability of $\pi$-conjugated
organic molecules on metal surfaces~\cite{Liu_2014_AccChemRes, Berland_2015_RepProgPhys,
  Yanagisawa_2015_JElectronSpectroscRelatPhenom, Maurer_2016_ProgSurfSci, Hermann_2017_ChemRev,
  Morbec_2017_JChemPhys}. This is particularly important for systems with significant vdW contributions to the
overall bonding (i.e.\ in the absence of covalent interactions, etc.)  such as most $\pi$-conjugated molecules
on weakly interacting substrates.

\begin{figure} \centering
  \includegraphics[width=\columnwidth]{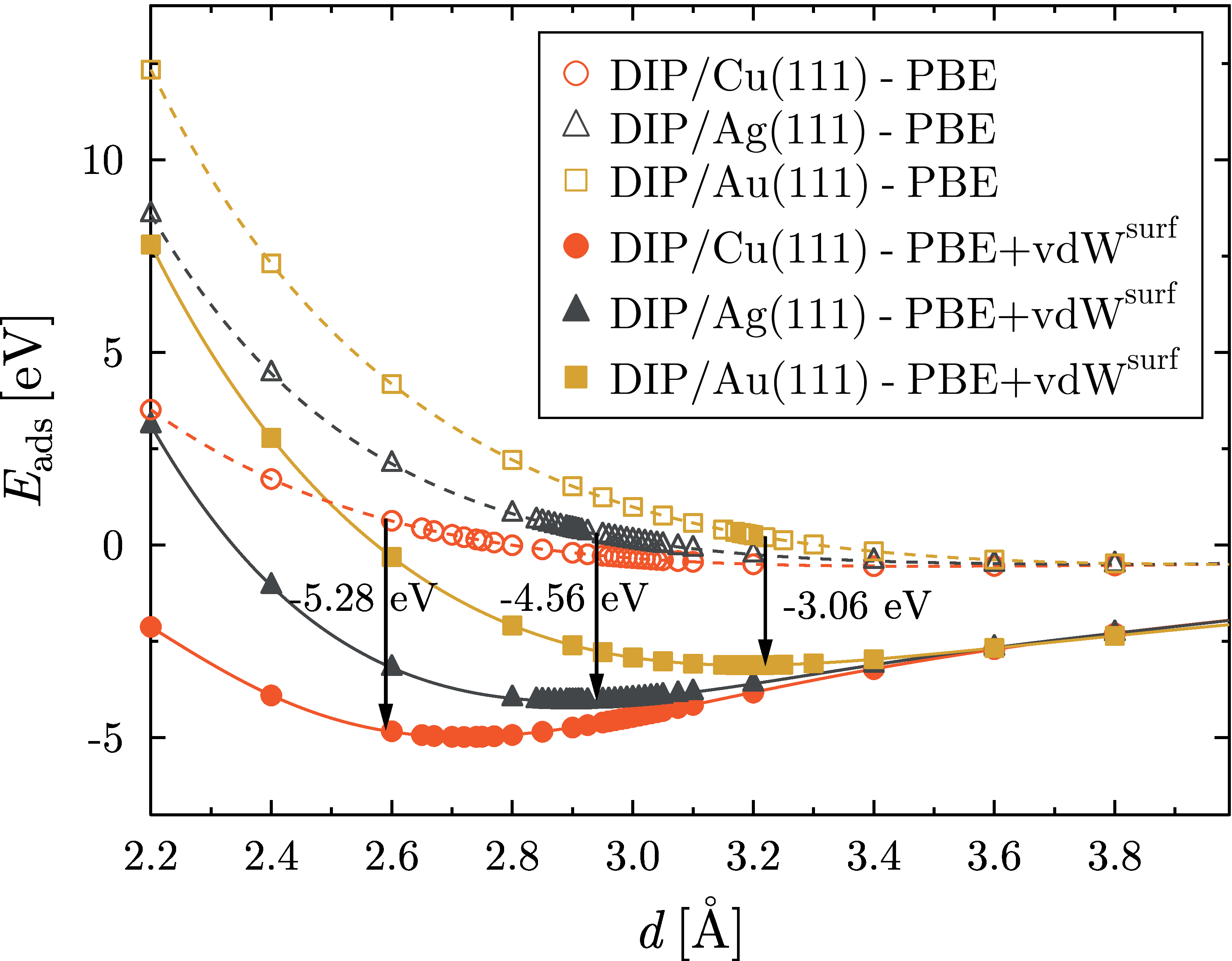}
  \caption{Comparison of DFT calculations performed for DIP adsorbed on the three (111)-surfaces of Cu, Ag
    and Au using the Perdew, Burke und Ernzerhof (PBE) exchange-correlation functional with and without vdW
    corrections. Note that in some cases, i.e.\ if no vdW interactions are included, the outcome would be no
    binding at all. Taken from~Ref.~\onlinecite{Burker_2013_PhysRevB} }
  \label{fig:vdw}
\end{figure}

The effect of dispersion forces is nicely demonstrated in Figure~\ref{fig:vdw} by comparing DFT results
obtained with the PBE exchange-correlation functional with and without including vdW interactions. 
It was found that dispersion corrected DFT calculations applied to DIP on three different noble metal
surfaces yield vertical bonding distances that agree very well with the experimental data.  The XSW
results averaged over all carbon species of DIP, i.e.\ $d_H = 2.51 \pm 0.03$\,\AA{} for Cu(111), $3.01 \pm
0.04$\,\AA{} for Ag(111) and, taking the reconstruction of the gold surface into account, $3.10 \pm
0.03$\,\AA{} for Au(111), differ less than 0.12\,\AA{} from the minima of the calculated adsorption
energies $E_\mathrm{ads}(z)$ as they are marked by arrows in Fig.~\ref{fig:vdw}.  As expected, those
energies follow the trend $|E_\mathrm{ads}(\mathrm{Cu})| > | E_\mathrm{ads}(\mathrm{Ag})| > |
E_\mathrm{ads}(\mathrm{Au})|$ (i.e., with the strongest interaction for the most reactive substrate, which
matches the discussion in Sec.~\ref{ssec:concepts_substrate}). Importantly, the rather shallow and broad
minima of $E_\mathrm{ads}(d)$, which correspond to the equilibrium distances, form only if the vdW
corrections are included, otherwise there is no stable adsorption at all. Moreover, Figure~\ref{fig:vdw}
shows that on Cu(111) the Pauli repulsion sets in rather weakly (a less steep $E_\mathrm{ads}(d)$ for
small distances) compared to Ag(111) and Au(111), which is due to significant interaction between DIP and
Cu(111), i.e.\ contributions beyond the vdW attraction.

In order to understand the contribution of the vdW interactions in more detail, first one has to consider the
impact of the specific symmetry on the vdW interaction (which for individual atoms goes as (distance)$^{-6}$):
Integrating the vdW energy of a single atom over the semi-infinite substrate yields the atom-surface vdW
energy as $C_3(z - z_0$)$^{-3}$, where $C_{3}$ determines the interaction strength between atom and
surface~\cite{Tkatchenko_2009_PhysRevLett,Grimme_2006_JComputChem}, $z$ corresponds to the distance of the
atom to the uppermost surface layer, and $z_0$ indicates the position of the surface image plane. Naively, one
might attempt to determine the $C_3$ coefficients for the different surfaces from all the two-body atom-atom
vdW energies and thereby neglect any interactions of the substrate atoms with each other. However, it can be
shown that the dielectric function, i.e.\ the collective electronic response of the underlying solid, has a
strong influence on the interaction strength~\cite{Tkatchenko_2009_PhysRevLett,Grimme_2006_JComputChem}.

Using the Lifshitz-Zaremba-Kohn (LZK) expression for calculating the $C_3$ coefficients, one obtains (in units
of Hartree$\cdot$Bohr$^3$) 0.35 for Cu, 0.35 for Ag and 0.33 for Au, which leads to essentially the same
interaction energy at large distances for DIP on Cu(111), Ag(111), and Au(111)
(Figure~\ref{fig:vdw}). However, at shorter molecule-surface distances, which include the equilibrium
distance, the adsorption energy is determined by an interplay between the vdW attraction and the Pauli
repulsion with a possible covalent component. The Pauli repulsion follows roughly the trend of decreasing vdW
radii, with a faster onset in terms of the molecule-surface distance for Au (the largest vdW radius), and then
decreases for Ag and Cu. Therefore, for Au the balance between vdW attraction and the Pauli repulsion is
obtained further away from the substrate (i.e.\ at larger adsorption distances) than for Cu, which in turn
makes the adsorption energies lower for Au than for Cu, in contrast to the possible naive expectation of Au
with its higher electron density and polarizability leading to stronger interactions than Cu.

\begin{figure}
  \centering
  \includegraphics[width=\columnwidth]{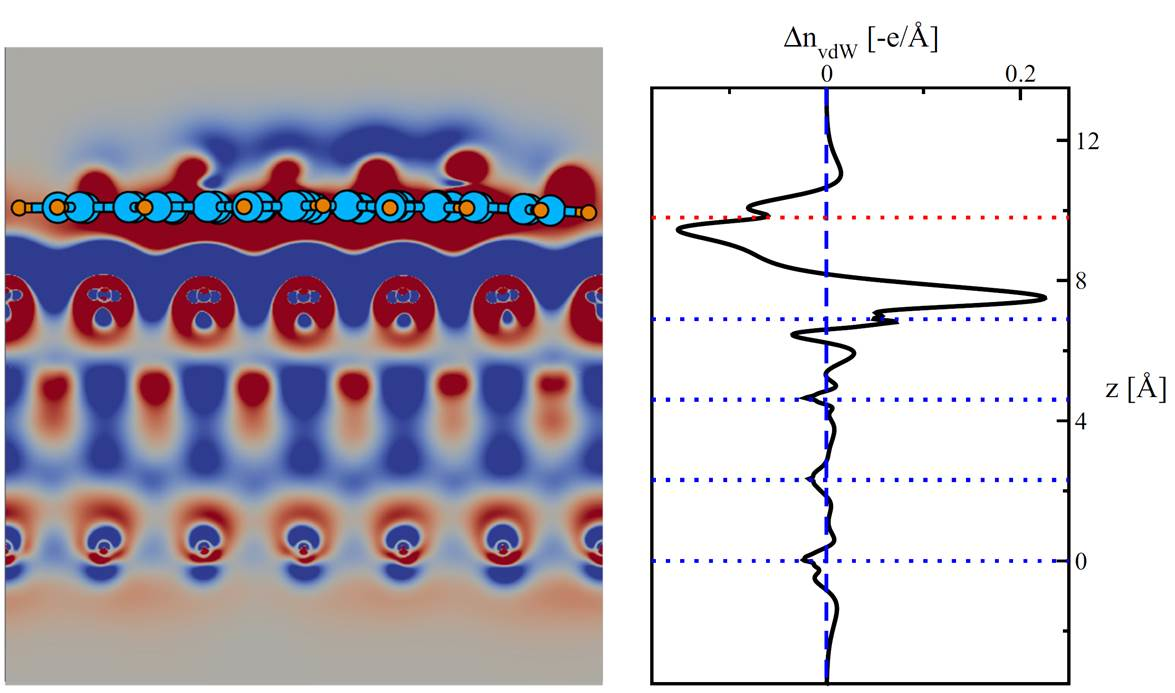}
  \caption{Left panel: vdW effect on the electron density distribution $\Delta n(\mathbf{r})_\mathrm{vdW}$
    upon adsorption of DIP on Ag(111), accumulation is in blue, depletion is in red. Right panel: The integral
    of $\Delta n(\mathbf{r})_\mathrm{vdW}$ plotted as a function of $z$, the axis perpendicular to the
    surface. A dipole-like density redistribution emerges at the interface. Taken from
    Ref.~\onlinecite{Ferri_2017_PhysRevMaterials}.}
  \label{fig:DIP_Ag111_vdW}
\end{figure}

Generally, we note that semi-local DFT calculations, i.e.\ the different versions of the generalized
gradient approximation (GGA), might not provide very accurate
energy-levels~\cite{Maurer_2016_ProgSurfSci}. More advanced methods, however, are computationally
prohibitively expensive because of the large number of atoms within the unit cell of larger molecules on
surfaces~\cite{Klimevs_2012_JChemPhys}.  For weakly interacting systems, one may not expect major changes
of the electronic structure. Nevertheless, even for purely vdW-driven systems there will be at least
variations of the molecular energy-levels and the vacuum level. As discussed in Sec.~\ref{ssec:energetics},
the push-back effect~\cite{Wandelt_1997_ApplSurfSci, Smoluchowski_1941_PhysRev, Witte_2005_ApplPhysLett,
  Rusu_2010_PhysRevB, Egger_2015_NanoLett, Toyoda_2010_JChemPhys} decreases the interface dipole of the
clean metal surface. However, recently it was shown~\cite{Ferri_2017_PhysRevMaterials} that vdW
interactions can also cause significant charge rearrangements in the vicinity of the adsorbed COM, as shown
for DIP on Ag(111) in Figure~\ref{fig:DIP_Ag111_vdW}.
Using DIP on noble metals as a model system, we believe that the limiting case of weakly interacting systems
-- although indeed not as simple as at first assumed -- is essentially understood. The key for this are
state-of-the-art vdW-corrected DFT calculations in combination with very precise experimental data from XSW
and other techniques, which agree within less than $\sim$0.1\,\AA{} for the adsorption distance.

This also applies to other examples in this category of weakly interacting systems. Benzene is probably a
prototypical case, which, however, due to its smallness and thus high vapor pressure at room temperature, can
only be studied at lower temperatures. In a DFT benchmark study Liu et al.\ combined XSW and TPD using benzene
on Ag(111) as model system to test the accuracy of the performed calculations~\cite{Liu_2015_PhysRevLett,
  Maass_2018_JChemPhys}. They found that the adsorption distance of $d_\mathrm{Ads} = 3.04 \pm 0.02$\,\AA{}
and the adsorption energy of $0.68 \pm 0.05$\,eV, which were measured for this clearly physisorptive system,
are in excellent agreement with their DFT calculations.

Again, we note that there is a gradual transition from truly weakly interacting systems such as DIP or
benzene on Au(111) to those on Ag(111) or Cu(111), which are more reactive due to the increased orbital
overlap of molecular states at smaller adsorption distances (see also Sec.\ref{ssec:concepts_substrate}).

\subsection{Strongly interacting systems} \label{sec:CTC}
In the other limiting case, the coupling and the interaction between molecule and substrate is so strong that
there is, \textit{inter alia}, significant charge donation and/or back donation, significant shifts of
energy-levels and presumably an associated significant distortion of the molecule (but, notably, not yet a
chemical reaction). In Sec.~\ref{sec:overview} we used P4O on Ag(111) as example for a strongly coupled
system. The schematic energy-level diagram (Figure~\ref{fig:ELA_concept}) shows the charge transfer from the
substrate into the former LUMO of P4O in the monolayer. This CT goes along with strong chemical shifts of the
core-levels and a bending of the P4O oxygen atoms below the plane of the carbon backbone
(Figure~\ref{fig:dH_PxO_Ag111})~\cite{Heimel_2013_NatureChem}. Overall, P4O re-hybridizes in the contact
layer to Ag(111) (possible resonance structures shown in the bottom of Figure~\ref{fig:ELA_concept}). P4O
exhibits a similar chemisorptive behavior on Cu(111), but physisorbs on
Au(111)~\cite{Chen_2019_JPhysCondensMatter, Heimel_2013_NatureChem}. In contrast, F$_4$TCNQ chemisorbs on
virtually \emph{all} clean metal surfaces~\cite{Romaner_2007_PhysRevLett, Koch_2005_PhysRevLett,
  Braun_2008_JChemPhys, Rangger_2009_PhysRevB, Duhm_2009_ApplPhysLett, Yamane_2017_JPhysChemLett,
  Borghetti_2017_JPhysChemC, Duhm_2011_OrgElec, Faraggi_2012_JPhysChemC, Gerbert_2018_JPhysChemC} showing a
qualitatively similar behavior. These systems shall hence be discussed as model systems for strongly coupled
organic-metal interfaces.

\begin{figure}
  \centering
  \includegraphics[width=\columnwidth]{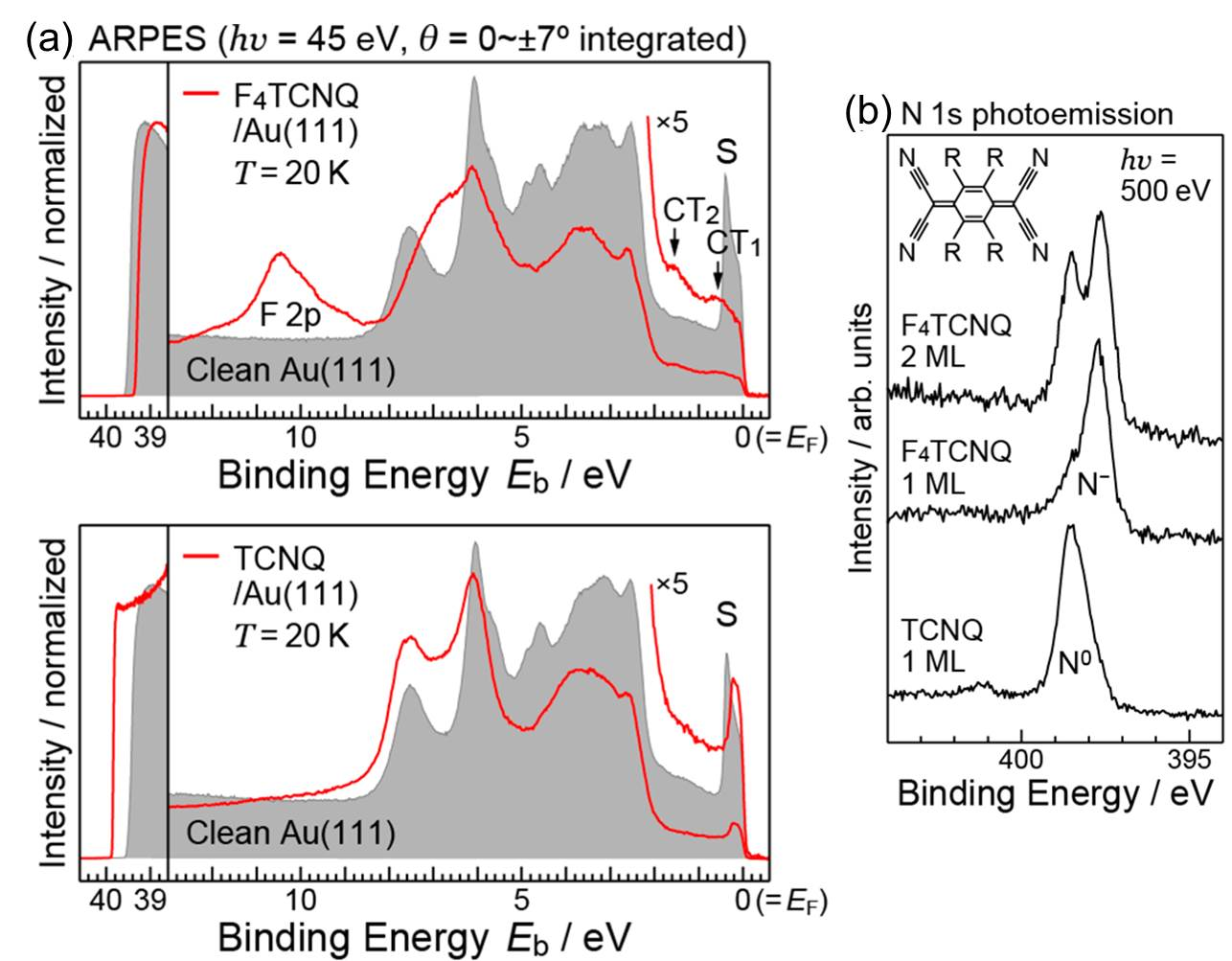}
  \caption{(a) UP spectra of the clean Au(111) surface (gray shadows) and monolayer (F$_4$)TCNQ films on
    Au(111) (red curves). For F$_4$TCNQ two charge transfer states (CT$_1$ and CT$_2$) close to the substrate
    Fermi level $E_F$ are apparent. The shift of the secondary-electron cutoff (left panel) to lower binding
    energy upon F$_4$TCNQ deposition evidences a work function increase. (b) N\,1s core-level: No charge
    transfer takes place into the TCNQ monolayer and all molecules are neutral (N$^0$). In contrast, in the
    F$_4$TCNQ monolayer almost all molecules are charged (N$^-$). Taken from
    Ref.~\onlinecite{Yamane_2017_JPhysChemLett}.}
  \label{fig:F4TCNQ_Au111}
\end{figure}

The EA of F$_4$TCNQ in multilayer thin films on Au is 5.08\,eV~\cite{Kanai_2009_ApplPhysA} to
5.25\,eV~\cite{Gao_2001_ApplPhysLett} and thus larger than the work functions of most clean noble
metals~\cite{Derry_2015_JVacSciTechnolA}. Therefore, one can expect a charge transfer into the LUMO of
F$_4$TCNQ in monolayers on such substrates to increase the effective work function and maintain thermodynamic
equilibrium. In fact, monolayers of F$_4$TCNQ increase the work functions of
Au(111)~\cite{Yamane_2017_JPhysChemLett}, Ag(111)~\cite{Duhm_2011_OrgElec, Rangger_2009_PhysRevB} and
Cu(111)~\cite{Romaner_2007_PhysRevLett}. As an example Figure~\ref{fig:F4TCNQ_Au111}a shows UP spectra of
F$_4$TCNQ on Au(111) and compares them with spectra of the unfluorinated parent molecule, TCNQ, which
interacts only weakly with Au(111)~\cite{Yamane_2017_JPhysChemLett}. The increase in the work function
(evidenced by the SECO shift) is accompanied by two charge transfer peaks (CT$_1$ and CT$_2$) close to
$E_F$. Likewise, the N\,1s core-level shows strong chemical shifts from mono- to multilayer F$_4$TCNQ
coverage (Figure~\ref{fig:F4TCNQ_Au111}b), which are caused by the charge transfer into F$_4$TCNQ.

\begin{figure}
  \centering
  \includegraphics[width=\columnwidth]{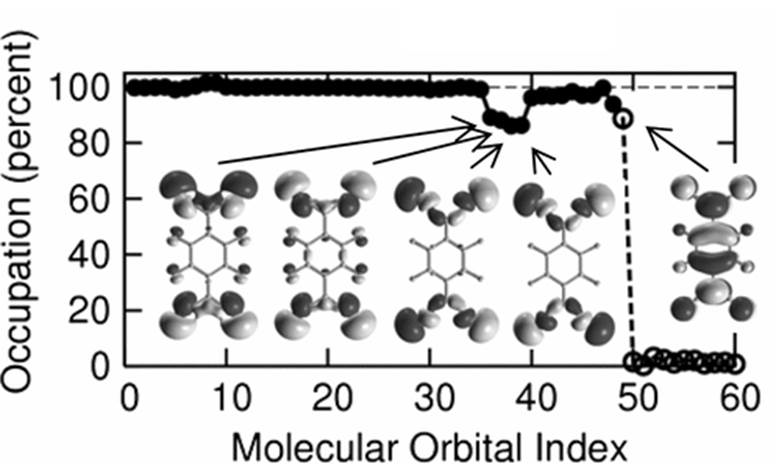}
  \caption{Calculated occupation (in percent) of the lowest 60 molecular orbitals of F$_4$TCNQ in a monolayer
    on Cu(111). The full (open) circles and solid (dashed) lines correspond to the orbitals which are occupied
    (unoccupied) in the isolated molecule. Taken from Ref.~\onlinecite{Romaner_2007_PhysRevLett}.}
  \label{fig:F4TCNQ}
\end{figure}

\begin{figure}
  \centering
  \includegraphics[width=\columnwidth]{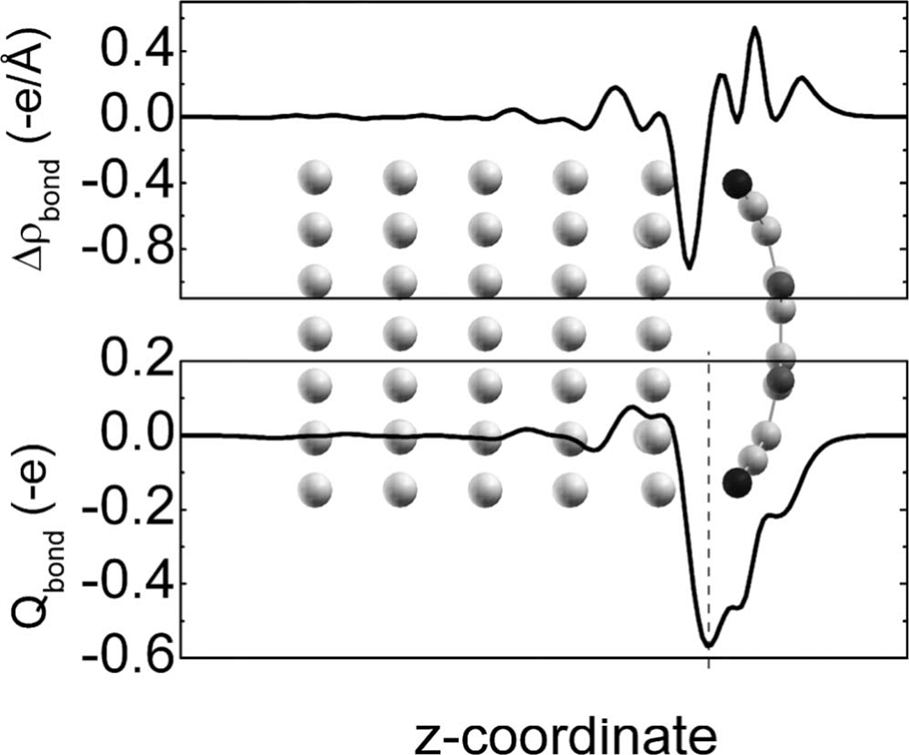}
  \caption{Top: charge-density rearrangement, $\Delta \rho_\mathrm{bond}$, upon adsorption of a densely packed
    F$_4$TCNQ monolayer on a Ag(111) surface integrated over the $x$-$y$ plane within the unit cell; Bottom:
    resulting total charge transferred, $Q_\mathrm{bond}$. The vertical line denotes the maximum value of
    $Q_\mathrm{bond}$. The structure of the combined system is shown in the background as a guide to the
    eye. $e$ corresponds to the (positive) elementary charge and positive $\Delta \rho_\mathrm{bond}$ values
    correspond thus to electron accumulation and negative values to electron depletion. Taken from
    Ref.~\onlinecite{Rangger_2009_PhysRevB}.}
  \label{fig:F4TCNQ_Ag111}
\end{figure}

The work function increase on Au(111) and all other metal surfaces is smaller than 1\,eV
\cite{Duhm_2011_OrgElec, Rangger_2009_PhysRevB, Yamane_2017_JPhysChemLett, Romaner_2007_PhysRevLett}. However,
a complete filling of the LUMO of the molecules in the monolayer would lead approximately to a work function
increase of around 5\,eV~\cite{Romaner_2007_PhysRevLett}. The calculated occupation of the lowest 60 molecular
orbitals (MOs) for a F$_4$TCNQ monolayer on Cu(111) as displayed in Figure~\ref{fig:F4TCNQ} indicates that the
charge donation to the LUMO is, in fact, accompanied by a back-donation to deeper lying MOs. Especially, the
HOMO-9 to HOMO-12 levels are involved and each is only 80\% to 90\% occupied after adsorption. They correspond
to $\sigma$-orbitals localized on the four nitrile groups of the molecule, which participate most prominently
in the chemical bonding with the substrate. Summing over all MOs gives a net negative charge of $\sim$0.6\,$e$
per F$_4$TCNQ molecule, which is significantly less than 2\,$e$ corresponding to a complete filling of the
F$_4$TCNQ LUMO. In addition, as discussed below in more detail, adsorption induced conformation changes --
shown in Figure~\ref{fig:F4TCNQ_Ag111} for F$_4$TCNQ on Ag(111) -- lead to additional interface dipole
moments.

The significant molecular charging causes aromatization of the central quinone ring and makes the molecule
structurally flexible. This allows the molecule to bend and hybridize with the substrate through the lone
electron pairs of nitrogen~\cite{Borghetti_2017_JPhysChemC}. On Cu(111), the fluorine atoms are found
$\sim$0.6\,\AA{} above the nitrogen atoms as determined by XSW~\cite{Romaner_2007_PhysRevLett}. This is a
consequence of the carbon atoms carrying the nitrile groups re-hybridizing from $sp^2$ toward $sp^3$ upon
contact formation~\cite{Romaner_2007_PhysRevLett}. Additionally, the strong molecule-metal interaction
leads to marked changes in bond lengths within F$_4$TCNQ. In the gas phase the molecule adopts a fully
planar, quinoid-like geometry~\cite{Koshino_2004_JElectronSpectroscRelatPhenom}. Adsorption on the Cu(111)
surface results in a nearly aromatic ring~\cite{Romaner_2007_PhysRevLett}. In the context of strong
chemisorption and hybridization with metal surfaces also surface adatoms have been discussed, this applies
especially to F$_4$TCNQ on Au(111)~\cite{Yamane_2017_JPhysChemLett, Faraggi_2012_JPhysChemC} and for TCNQ
on Ag(111)~\cite{Blowey_2018_Nanoscale}.

For strongly coupled systems, fractional charge transfer including donation and back-donation is usually
observed. The impact of such charge rearrangements on the VL shall be discussed with the example of F$_4$TCNQ
on Ag(111)~\cite{Rangger_2009_PhysRevB}. For a better understanding of vacuum level shifts caused by
organic/inorganic contact formation, the calculated total change of the VL, i.e., the interface dipole $\Delta
\mathrm{VL}$, is often decomposed into two contributions~\cite{Romaner_2009_NewJPhys, Hofmann_2008_JPhysChemC,
  Romaner_2007_PhysRevLett, Rangger_2009_PhysRevB, Hofmann_2017_JPhysChemC}: $\left. i \right)$ the molecular
contribution $\Delta \mathrm{VL}_\mathrm{mol}$ related to the surface-normal component of the molecular dipole
(permanent or adsorption induced) and $\left. ii \right)$ the contribution due to the interfacial charge
rearrangement (including charge transfer from or to the metal), the so-called bond dipole $\Delta
\mathrm{VL}_\mathrm{bond}$, i.e.:
\begin{equation}\label{Delta_phi}
  \Delta \mathrm{VL} = \Delta \mathrm{VL}_\mathrm{mol} + \Delta \mathrm{VL}_\mathrm{bond}.
\end{equation}

An (infinitely) extended dipole layer results in a shift of the vacuum level by
\begin{equation}\label{eq:helmholtz}
  \Delta \mathrm{VL} = \frac{1}{\varepsilon_0 A} \mu_\bot  ,
\end{equation}
where $\mu _{\bot}$ refers to the surface-normal component of the dipole moment per molecule in the monolayer
(including depolarization effects~\cite{Fukagawa_2011_PhysRevB, Natan_2007_AdvMater}), $\varepsilon_0$ to the
vacuum permittivity and $A$ to the area per molecule. In the gas phase F$_4$TCNQ is planar, thus, all
contributions to $\Delta \mathrm{VL}_\mathrm{mol}$ are due to adsorption-induced conformation changes. In the
distorted conformation of the monolayer (Figure~\ref{fig:F4TCNQ_Ag111}), each individual F$_4$TCNQ molecule
possesses a dipole moment of $-2.69$\,D. In effect, it points away from the metal surface and would result in
a work function decrease by $-0.85$\,eV. In addition to the distortion, however, also $\Delta
\mathrm{VL}_\mathrm{bond}$ due to adsorption-induced charge rearrangements, $\Delta \rho_\mathrm{bond}$, has
to be taken into account. $\Delta \rho_\mathrm{bond}$ is calculated as the difference of the total electron
density of the combined metal-organic interface $\rho^\mathrm{sys}$ and the non-interacting densities of metal
$\rho^\mathrm{metal}$ and monolayer $\rho^\mathrm{monolayer}$:
\begin{equation}
  \Delta \rho _\mathrm{bond} = \rho^\mathrm{sys} - (\rho^\mathrm{metal} + \rho^\mathrm{monolayer}).
\end{equation}

From $\Delta \rho_\mathrm{bond}$, $\Delta \mathrm{VL}_\mathrm{bond}$ is then obtained by solving the Poisson
equation. For F$_4$TCNQ, $\Delta \mathrm{VL}_\mathrm{bond}$ amounts to +1.70\,eV. The net effect, i.e., the
sum of $\Delta \mathrm{VL}_\mathrm{bond}$ and $\Delta \mathrm{VL}_\mathrm{mol}$, is a work function increase
by +0.85\,eV, which fits very well with the experimental value of 0.65\,eV~\cite{Rangger_2009_PhysRevB}. For
molecules that undergo charge-transfer reactions with the surface, $\Delta \mathrm{VL}_\mathrm{mol}$ and
$\Delta \mathrm{VL}_\mathrm{bond}$ are \emph{not} independent. Rather, any error in the description of the
bending is made up for by a change in charge transfer, making $\Delta \mathrm{VL}$, which is the experimental
observable, a very robust quantity~\cite{Hofmann_2010_NanoLett, Hofmann_2017_JPhysChemC}.

Figure~\ref{fig:F4TCNQ_Ag111} shows the $x$-$y$-plane integrated charge density rearrangements $\Delta
\rho_\mathrm{bond}(z)$ of a monolayer F$_4$TCNQ on Ag(111). The pronounced electron depletion directly above
the top metal layer is attributed to push-back. The largest electron accumulation can be found in the
$\pi$-electron region of F$_4$TCNQ. The dip in $\Delta \rho_\mathrm{bond}(z)$ in the region of the CN groups
is consistent with the decreased $\sigma$-electron density in that part of the molecule
(\textit{cf.}~Figure~\ref{fig:F4TCNQ})~\cite{Rangger_2009_PhysRevB}.

In general, such strongly coupled systems can be used for work function engineering and consequently for
energy-level tuning of subsequently deposited organic layers. This has been first demonstrated for the
molecular acceptor TCAQ, which lowers the hole injection barrier into 6T layers on Au and
Ag~\cite{Koch_2005_ApplPhysLett}. Other examples of strongly coupled electron accepting molecules on metal
surfaces include PEN \cite{Koch_2008_JAmChemSoc,Lu_2016_JPhysCondensMatter},
DIP~\cite{Hosokai_2017_OrgElec, Yonezawa_2016_ApplPhysExpress}, PTCDA~\cite{Weiss_2015_NatCommun,
  Puschnig_2017_JPhysChemLett, Weinhardt_2016_JPhysChemC, Ziroff_2010_PhysRevLett,
  Romaner_2009_NewJPhys,Khoshkhoo_2017_OrgElec, Baby_2017_ACSNano, Hofmann_2013_NewJPhys,
  Gerlach_2007_PhysRevB,Hauschild_2005_PhysRevLett, Willenbockel_2014_PhysChemChemPhys,
  Tautz_2007_ProgSurfSci, Kawabe_2008_OrgElec, Zou_2006_SurfSci, Duhm_2008_OrgElec},
PTCDI~\cite{Franco-Canellas_2017_PhysRevMaterials}, TAT~\cite{Yang_2016_PhysRevB},
FAQ~\cite{Duhm_2006_JPhysChemB}, HATCN~\cite{Broker_2010_PhysRevLett, Amsalem_2011_JApplPhys}, Pcs
\cite{Peisert_2015_JElectronSpectroscRelatPhenom, Maughan_2017_PhysRevB, Stadtmuller_2011_PhysRevB,
  Kroger_2010_NewJPhys,Ruocco_2008_JPhysChemC, Thussing_2016_JPhysChemC}, and TCNQ
\cite{DellaPia_2016_Nanoscale, Tseng_2010_NatChem,Stradi_2016_RSCAdv, Lindquist_1988_JPhysChem}. Most of
the above mentioned COMs can serve different purposes in addition to being a suitable ELA modifier
\cite{Tang_1986_ApplPhysLett, Fung_2016_AdvMater,Liao_2008_AdvMater, Zhang_2015_ACSApplMaterInterfaces,
  Jung_2018_OrgElec, Lussem_2016_ChemRev}. For example, F$_4$TCNQ is also a popular molecular dopant
\cite{Mendez_2013_AngewChemIntEd, Jung_2018_OrgElec, Jacobs_2017_AdvMater, Lussem_2016_ChemRev,
  Mendez_2015_NatComms, Salzmann_2015_JElectronSpectroscRelatPhenom,
  Salzmann_2016_AccChemRes}. Furthermore, work function modification is not restricted to metal
surfaces~\cite{Wang_2015_AdvMaterInterfaces, Zhang_2016_AdvMater, Schultz_2016_PhysRevB,
  Zu_2017_ACSApplMaterInterfaces, Abellan_2016_AngewChemIntEd, Hoffmann-Vogel_2018_RepProgPhys,
  Winkler_2019_ElectronStruct, Schamoni_2019_MaterResExpress, Wang_2019_AdvElectronMater} and electron
donating COMs can also \emph{lower} effective work functions \cite{Timpel_2018_AdvFunctMat,
  Hofmann_2008_JPhysChemC, Broker_2008_ApplPhysLett, Lindell_2008_ApplPhysLett, Akaike_2016_AdvFunctMater}
by the reversed process as electron accepting molecules, i.e., by an electron transfer from the adsorbate
to the substrate. Strongly interacting organic-metal systems have thus a significant relevance for
applications.

\subsection{Intermediate cases}
Between the two extreme scenarios discussed above, there exist plenty of systems whose phenomenology can
neither be described by only considering ``chemical'' interactions nor vdW attraction alone. Most
interestingly, these cases may tend towards one or the other side depending on the particular characteristics
of the system. The following consists mostly of prototypical systems of the kind ``molecule A on substrate
B'', which are taken as a reference to discuss the interfacial changes that occur when A or B are (slightly)
modified. In addition, particular cases that exemplify the possibilities of surface/interface tuning are also
outlined. We note that some systems may fit in two or more subsections.

\begin{figure*}
  \centering
  \includegraphics[width=0.9\textwidth]{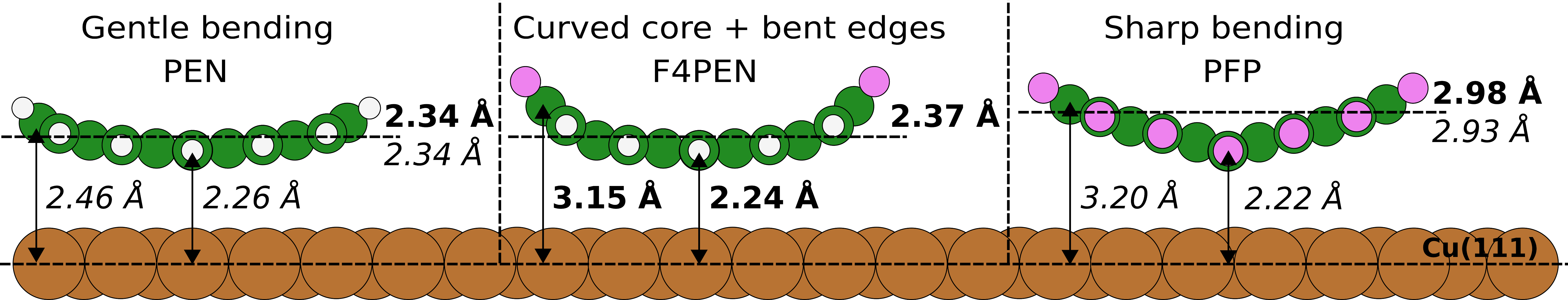}
  \caption{Adsorption geometry of PEN, F$_4$PEN and PFP on Cu(111) that combines experimental data obtained
    with XSW (in bold) and state-of-the-art DFT calculations with vdW corrections (in italics). The
    calculations are obtained from Ref.~\onlinecite{Shi_2012_JPhysChemC}, the measured adsorption distances
    for PEN and PFP from Ref.~\onlinecite{Koch_2008_JAmChemSoc} and F$_4$PEN from
    Ref.~\onlinecite{Franco-Canellas_2018_PhysRevMaterials}. Note that only values for different carbon
    positions are included. In addition, the average adsorption distance of the fluorine atoms in F$_4$PEN
    is 3.40\,{\AA}~\cite{Franco-Canellas_2018_PhysRevMaterials} and in PFP
    3.08\,{\AA}~\cite{Koch_2008_JAmChemSoc}. Figure adapted from
    Ref.~\onlinecite{Franco-Canellas_2018_PhysRevMaterials} with permission.}
  \label{fig:PEN_PFP_F4PEN_Cu111}
\end{figure*}

\subsubsection{Fluorination}
Among the many options to functionalize a COM through specific chemical
modifications~\cite{Babudri_2007_ChemCommun}, fluorination, i.e.\ the substitution of peripheral hydrogen
atoms by fluorine, is one of the most widely employed. In the gas phase and thin films, it increases the
resistance to oxidation, changes the electron affinity, modifies the intrinsic molecular dipole as well as
the optical properties~\cite{Milian-Medina_2017_JPhysChemLett}. At the interface, fluorination modifies the
ELA and the nature of the substrate-molecule and molecule-molecule interactions. One of the most illustrative
examples when discussing the effects of fluorination is the case of PEN and its perfluorinated derivative
PFP~\cite{Sakamoto_2004_JAmChemSoc} adsorbed on copper surfaces~\cite{Lukas_2004_ChemPhysChem,
  Ferretti_2007_PhysRevLett, Gross_2009_Science, Koch_2008_JAmChemSoc,
  Yamane_2009_JElectronSpectroscRelatPhenom, Oteyza_2010_ChemPhysLett, Glowatzki_2012_JPhysChemC,
  Schmidt_2012_JPhysChemC,Schuler_2013_PhysRevLett}. Even on the moderately reactive (111) noble metal
surfaces, PEN molecules can experience strong interactions.

The hybridization of the molecular states with the surface atoms~\cite{Koch_2008_JAmChemSoc,
  Glowatzki_2012_JPhysChemC, Lu_2016_JPhysCondensMatter} renders a completely filled LUMO on Cu(111) well
below the Fermi level~\cite{Lu_2016_JPhysCondensMatter} and a remarkably short adsorption distance
(Figure~\ref{fig:PEN_PFP_F4PEN_Cu111})~\cite{Koch_2008_JAmChemSoc}. In contrast, PFP shows no LUMO filling,
no sign of hybridization is seen in XPS~\cite{Glowatzki_2012_JPhysChemC} and the average adsorption
distance of carbon is $\sim$0.6\,\AA{} higher than PEN, with the fluorine atoms further up (see
Figure~\ref{fig:PEN_PFP_F4PEN_Cu111}), on average, by $\sim$0.1\,{\AA} \cite{Koch_2008_JAmChemSoc}. DFT
calculations with vdW-corrections~\cite{Shi_2012_JPhysChemC} yielded adsorption geometries with average
adsorption distances in perfect agreement with experiments and gave a more precise description of the
actual arrangement: PEN would adsorb forming a small canoe-like shape with the short molecular edges
slightly above the average carbon distance, whereas PFP would adsorb in a strong V-shape with the central
carbon atoms, being the most reactive in acenes~\cite{}, very close to the surface and the short edges
$\sim$1\,\AA{} above. These dramatic changes in the interface properties upon fluorination are due to the
strong Pauli and steric repulsion (fluorine is larger than hydrogen) between the fluorine atoms and the
substrate that weakens the electronic coupling. In this direction, a recent PES-XSW combined study on the
partially fluorinated PEN derivative F$_4$PEN~\cite{Franco-Canellas_2018_PhysRevMaterials}, with fluorine
atoms only at the short molecular edges, adsorbed on Cu(111) revealed that the selective fluorination of
PEN only yields a local conformational change. Despite the increased adsorption distance of the fluorine
and carbon atoms nearby (see Figure~\ref{fig:PEN_PFP_F4PEN_Cu111}) the structural, electronic and chemical
properties of the PEN backbone remain unaffected because the strong interaction of the core with the copper
atoms prevails~\cite{Franco-Canellas_2018_PhysRevMaterials}.

On the less interacting silver surface, PEN has been shown to have different growth phases that depend on
the temperature as well as on the coverage~\cite{Dougherty2008JPCC, Duhm_2013_ACSApplMaterInterfaces}. Such
behavior has been defined as ``soft chemisorption''~\cite{Duhm_2013_ACSApplMaterInterfaces} since from TPD
a remarkable thermal stability is found and NEXAFS shows a significant modification of the PEN orbitals in
the monolayer~\cite{Kafer_2007_ChemPhysLett}, but there is no trace of LUMO
filling~\cite{Lu_2016_JPhysCondensMatter} and the molecules form a disordered liquid-like phase at
RT. Indeed, different studies have reported the disorder present in the first PEN layer in contact with
Ag(111)~\cite{Eremtchenko_2005_PhysRevB, Kafer_2007_ChemPhysLett, Dougherty2008JPCC,
  Lu_2016_JPhysCondensMatter} with the interesting and controversial~\cite{Kafer_2007_ChemPhysLett} fact
that an \emph{ordered second layer} may grow on top~\cite{Eremtchenko_2005_PhysRevB,
  Dougherty2008JPCC}. Upon cooling, ordered areas are found in STM~\cite{Dougherty2008JPCC} but no clear
diffraction pattern is observable in LEED~\cite{Lu_2016_JPhysCondensMatter}. Notably, cooling as well as
increasing the coverage modify the adsorption distances, with the remarkable displacement of $+0.14$\,\AA{}
in adsorption distance upon coverage increase from 0.5 to 0.75\,ML at
RT~\cite{Duhm_2013_ACSApplMaterInterfaces} as a consequence of the shifting balance between intermolecular
and substrate-molecule interactions (see Sec.~\ref{ssec:concepts_intermolecular}). In this situation,
fluorination of PEN has a similar effect as on copper, namely, the molecule-substrate interaction
decreases. As reported by G{\"o}tzen et al.~\cite{Gotzen_2011_Langmuir} the TPD spectrum of PFP, compared
to that of PEN, does not show a monolayer feature, which indicates that the bonding strength for the latter
is higher. This appears in line with the increased adsorption distance of PFP compared to PEN for a similar
coverage (2.98\,{\AA}~\cite{Duhm_2013_ACSApplMaterInterfaces} vs.\ 3.16\,{\AA}~\cite{Duhm_2010_PhysRevB})
and the absence of CT to the LUMO~\cite{Kera_2018_JPhysSocJpn, Duhm_2010_PhysRevB}. Similar to PEN on
Ag(111), temperature, coverage and even the preparation method seem to impact the supramolecular
arrangement of PFP: monolayers prepared via desorption of a multilayer appear as ordered patches that leave
substrate regions uncovered~\cite{Marks_2012_JPhysChemC} at $T<$130\,K, then become disordered and
homogeneously distributed all over the surface at $T>$160\,K. On the contrary, (sub)monolayers prepared via
direct evaporation of the desired coverage adopt ordered arrangements at LT (dislocation network) and RT
(Moir{\'e} pattern)~\cite{Goiri_2012_JPhysChemLett}.

On gold, both PEN and PFP, show a clear physisorptive behavior with no evidence of LUMO filling, nor
hybridization of the molecular orbitals~\cite{Koch_2007_AdvMater, Lu_2016_JPhysCondensMatter,
  Lo_2013_JpnJApplPhys}. Quite interestingly, despite the \textit{a priori} higher ionization energy of
PFP, an almost identical HIB was measured for both molecules on gold, which comes along with a larger (by
0.45\,eV) VL shift for PEN~\cite{Koch_2007_AdvMater}. The authors argued that the weaker pushback effect
and the unexpected ELA should be explained by a much larger adsorption distance of PFP compared to
PEN~\cite{Koch_2007_AdvMater}. Recently, direct XSW measurements have confirmed
this~\cite{unpublished_1}. Another interesting finding, which indicates a considerable interaction even
within the physisorptive regime, was reported by Lo et al.~\cite{Lo_2013_JpnJApplPhys}: Using STM it was
found that PEN may change the surface reconstruction of Au(111) and thereby suggesting a stronger
interaction with the substrate than PFP. Yet, the PEN molecules appear to be mobile while PFP forms
assemblies that are stabilized by intermolecular interactions~\cite{Lo_2013_JpnJApplPhys}.

For the sake of completeness, we shall mention that the influence of fluorination on the metal-organic
interface has been studied also for phthalocyanines~\cite{Peisert_2003_JApplPhys, Peisert_2002_SurfSci,
  Schwieger_2004_ChemPhysLett, Oteyza_2010_JChemPhys, Schwarze_2016_Science},
rubrene~\cite{Anger_2015_JPhysChemC} and thiophenes~\cite{Reisz_2017_JMaterRes}. Of course not only (111)
surfaces, but also several others have been studied, e.g., PFP on
Ag(110)~\cite{Navarro-Quezada_2018_JPhysChemC} or F$_4$PEN on Au(100)\cite{Savu_2015_ACSApplMaterInterfaces,
  Savu_2015_JPhysChemC}.

As concluding remark, it is worth noting that within the monolayer regime the combination of fluorinated and
non-fluorinated compounds has been shown to be an effective way to tune the work function of a metal
substrate~\cite{El-Sayed_2013_ACSNano} and, in the thin-film regime, the ionization energy as
well~\cite{Salzmann_2008_JAmChemSoc, Schwarze_2016_Science, Li_2018_ACSApplMaterInterfaces}. In both cases,
this method provides a suitable pathway to systematically modify the interface properties and adapt them to
the particular device requirements.

\subsubsection{Core substitutions of phthalocyanines} \label{ssec:Pcs}
\begin{figure}
  \centering
  \includegraphics[width=\columnwidth]{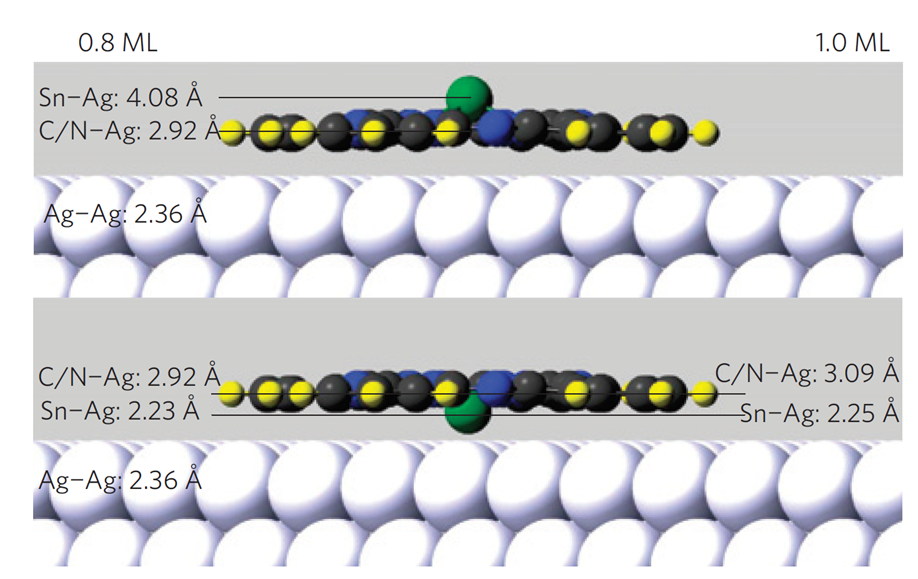}
  \caption{Side views of a molecule in the Sn-up (upper part of the figure) and Sn-down position (lower part)
    on Ag(111).  For a coverage of 0.8~monolayers (left side) Sn-up and Sn-down oriented molecules are present
    on the surface and the vertical bonding distances (as measured by XSW) are given. For a monolayer coverage
    (right side) only Sn-down oriented molecules are present on the surface. Taken from
    Ref.~\onlinecite{Stadler_2009_NaturePhysics} with permission.}
  \label{fig:SnPc_Ag111}
\end{figure}

All COMs discussed so far are intrinsically non-polar and therefore do not offer the possibility to tune the
ELA by a permanent molecular dipole moment, whose magnitude and orientation may influence the vacuum level
and thus the ELA~\cite{Maughan_2017_PhysRevB, Lerch_2017_JPhysChemC, Taucher_2016_JPhysChemC,
  Natan_2007_AdvMater, Fukagawa_2011_PhysRevB, Gerlach_2011_PhysRevLett, Huang_2013_PhysRevB,
  Lin_2017_PhysRevB}. An important class of polar COMs are particular porphyrins and phthalocyanines, as they
offer numerous possibilities of functionalization through substitution of the central metal atom and
insertion of further heteroatoms \cite{Peisert_2002_JApplPhys, Yamane_2010_PhysRevLett,
  Peisert_2015_JElectronSpectroscRelatPhenom, Gottfried_2015_SurfSciRep, Maughan_2017_PhysRevB,
  Schwieger_2002_PhysRevB, Lerch_2017_JPhysChemC, Fukagawa_2011_PhysRevB, Gerlach_2011_PhysRevLett,
  Petraki_2010_JPhysChemC, Huang_2013_PhysRevB, Lin_2017_PhysRevB, Woolley_2007_SurfSci,
  Stadler_2009_NaturePhysics, Stadler_2006_PhysRevB, Petraki_2014_BeilsteinJNanotechnol,
  Auwaerter_2015_NatChem, Hipps_2018_Langmuir, Zhao_2018_JPhysChemC, Singh_2019_MaterResExpress}. Since only
a few adsorption distances have been measured for porphyrins by XSW~\cite{Burker_2014_JPhysChemC,
  Schwarz2018JPhysChemC}, we focus on Pcs. We shall discuss SnPc~\cite{Woolley_2007_SurfSci,
  Stadler_2009_NaturePhysics, Stadler_2006_PhysRevB, Wang_2009_JAmChemSoc, Baran_2012_JPhysChemC,
  Greif_2013_PhysRevB, Schwarz_2015_JPhysChemC, Kashimoto_2018_JPhysChemC} as example for a Pc for which the
central atom is simply too big to fit into the aromatic macrocycle and which is thus polar in the gas
phase. Thus, it can adsorb in two different flat-lying geometries (Figure~\ref{fig:SnPc_Ag111}), i.e.\ with
the Sn atom either below (Sn-down) or above the molecular plane (Sn-up).
For a submonolayer coverage on Ag(111) both orientations were found and the adsorption distances have been
measured by XSW~\cite{Woolley_2007_SurfSci, Stadler_2009_NaturePhysics, Stadler_2006_PhysRevB}. For a full
monolayer coverage substrate mediated intermolecular interactions lead to a reorientation of the Sn-up
molecules and only Sn-down can be found on the surface~\cite{Stadler_2009_NaturePhysics}. In this case the
tin atom plays a crucial role in the coupling with the substrate leading to pronounced charge
rearrangements, which are negligible for the Sn-up orientation~\cite{Baran_2012_JPhysChemC,
  Greif_2013_PhysRevB}. Overall, for such systems the orientation has a significant impact on the
intermolecular interaction. However, the molecular dipole moments are rather weak and hence the impact on
the vacuum level marginal.

\begin{figure*}
  \centering
  \includegraphics[width=0.85\textwidth]{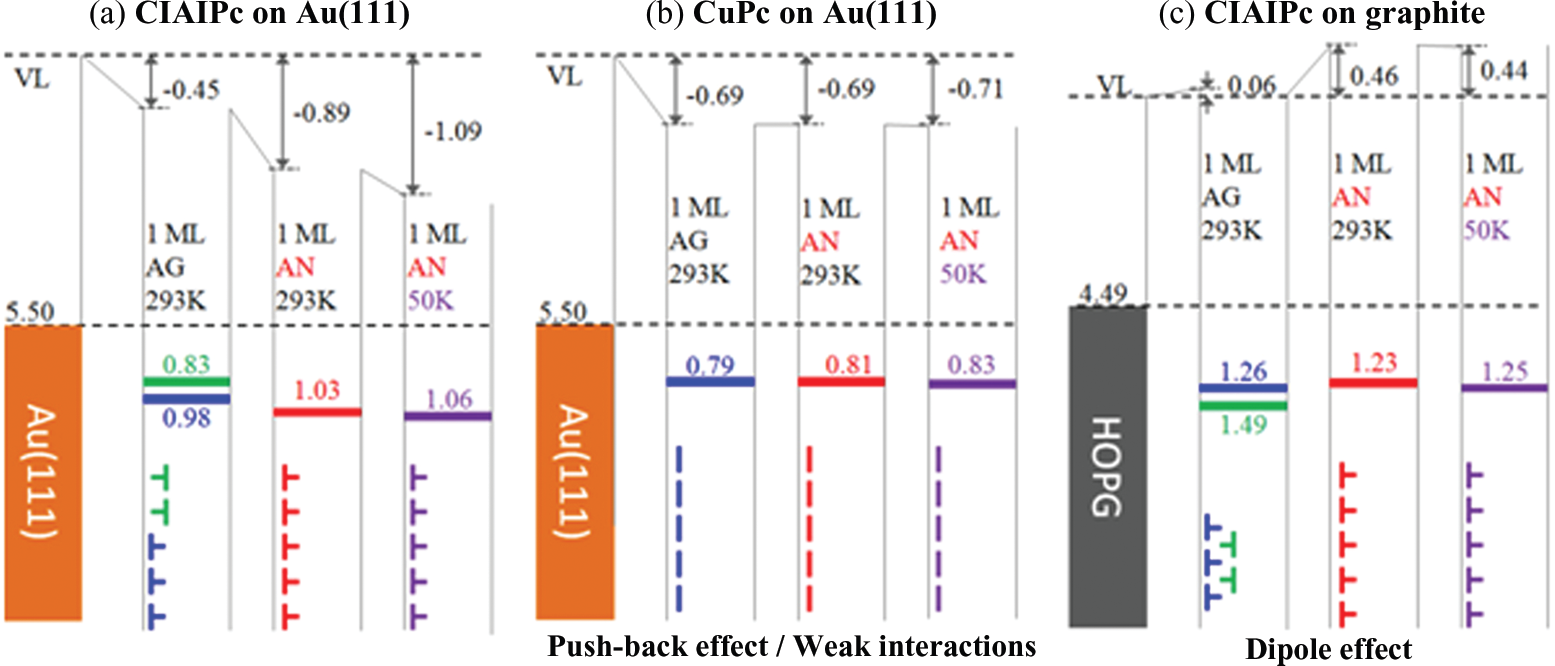}
  \caption{Energy-level diagrams for monolayers (ML) of (a) ClAlPc/Au(111), (b) CuPc/Au(111), and (c)
    ClAlPc/HOPG. For each system, UPS measurements of the as-grown (AG) film at room temperature (RT) and
    of the annealed (AN) film at RT and low temperature (LT) are shown. Upon annealing and cooling, vacuum
    level (VL) shifts and changes of HOMO states are observed. Here, $\perp$ corresponds to the Cl-up and
    the reversed symbol to the Cl-down orientation. Taken from Ref.~\onlinecite{Huang_2013_PhysRevB} with
    permission.}
  \label{fig:ClAlPc_compare}
\end{figure*}

For ``umbrella-shaped'' Pc molecules with an additional heteroatom attached to the central metal ion the
situation is different. For example, the dipole moment of ClAlPc in the gas phase is
1.87\,D~\cite{Kera_2004_SurfSci} and, according to equation~(\ref{eq:helmholtz}), the collective impact of
these dipoles on the vacuum level ($\Delta \mathrm{VL}_\mathrm{mol}$) for an aligned monolayer should yield a
$\Delta \mathrm{VL}$ value of several hundred meV. For as-deposited ClAlPc on Au(111) a mixed Cl-up/down
orientation has been observed and the resulting $\Delta \mathrm{VL} = -0.45$\,eV
(Figure~\ref{fig:ClAlPc_compare}a) has been mainly ascribed to the push-back
effect~\cite{Huang_2013_PhysRevB}. Strikingly, aligning the molecules to a Cl-up orientation by annealing
leads to a further VL decrease ($\Delta \mathrm{VL} = -0.89$\,eV) and the total $\Delta \mathrm{VL}$ is
larger than that of planar CuPc on the same substrate (Figure~\ref{fig:ClAlPc_compare}b). Apparently, the
permanent dipole moment of ClAlPc is decreasing the vacuum level -- whereas in fact a Cl-up orientation
should lead to an VL increase (by $+0.47$\,eV), which was indeed observed for ClAlPc on HOPG
(Figure~\ref{fig:ClAlPc_compare}c). On inert HOPG the dipole moment of the adsorbate is thus not changed in
the contact layer to the substrate. On metal substrates, however, adsorption induced bond-length changes,
which can lead to a partial depolarization of the COM on the surface, are frequently
observed~\cite{Ilyas_2016_JPhysChemC, Niu_2013_JPhysChemC, Wruss_2016_JPhysChemC,
  Harivyasi_2018_JPhysChemC}. In addition, interfacial charge rearrangements due to strong interactions can
further impact the vacuum level, which has been suggested to be the reason for the unexpected ELA of ClAlPc
on Au(111)~\cite{Huang_2013_PhysRevB}. A similar behavior has been observed on Ag(111): Also on this
substrate annealing changes a mixed orientation of a ClAlPc monolayer to a predominant Cl-up arrangement and
concomitantly decreases VL~\cite{Lin_2017_PhysRevB}. Interestingly, for very low deposition rates of ClAlPc
on Ag(111) ($\sim$0.1\,\AA/min) a Cl-down orientation is favorable.

Unfortunately, the modulo-$d$ ambiguity of the XSW technique (Eq.~(\ref{eq:d_H})) can hinder a
straightforward assignment of adsorption distances and even the orientation (X-down or X-up). For example,
the DFT-modeled adsorption distances of GaClPc on Cu(111) in the Cl-up and Cl-down
orientation~\cite{Wruss_2016_JPhysChemC} do not match the experimental
values~\cite{Gerlach_2011_PhysRevLett}. It turns out that in the most likely adsorption geometry the Cl
atom is dissociated \cite{Wruss_2016_JPhysChemC}. Moreover, as mentioned above, the ``up'' and ``down''
orientation can also coexist, as also observed for VOPc on Cu(111)~\cite{Blowey_2019_JPhysChemC}. In these
cases, having a very well characterized system may help to address this issue~\cite{Kroger_2016_NewJPhys}.

\begin{figure}
  \centering
  \includegraphics[width=0.95\columnwidth]{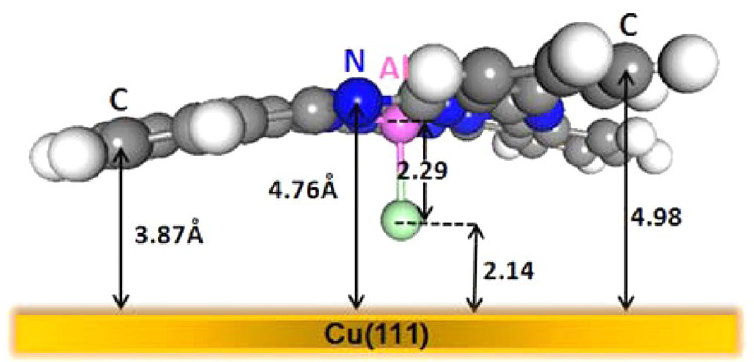}
  \caption{DFT-based adsorption geometry of ClAlPc on Cu(111). Taken from
    Ref.~\onlinecite{Niu_2013_JPhysChemC} with permission.}
  \label{fig:ClAlPc_Cu111}
\end{figure}

Another challenge for XSW measurements are the above mentioned pronounced distortions of the $\pi$-system
leading to significantly different adsorption distance of the carbon atoms. For example, for ClAlPc on
Cu(111)~\cite{Niu_2013_JPhysChemC} in the Cl-down orientation the DFT-modeled adsorption distances of
individual carbon atoms differ by up to 1.11\,\AA{} (Figure~\ref{fig:ClAlPc_Cu111}). The strong distortion is
a consequence of a charge transfer from the Cu(111) into ClAlPc, which is mainly localized on two of the four
ClAlPc lobes as shown by STM~\cite{Niu_2013_JPhysChemC}. This involves a symmetry reduction of ClAlPc from
4-fold in the gas phase [and the Cl-up orientation on Cu(111)] to 2-fold in the Cl-down orientation on
Cu(111). Similar symmetry reductions have been observed also for other Pcs on different substrates, e.g., for
CuPc on Cu(111)~\cite{Karacuban_2009_SurfSci} and on Ag(100) \cite{Mugarza_2010_PhysRevLett}, for FePc on
Cu(111)~\cite{Snezhkova_2016_JChemPhys} and for PtPc as well as PdPc on
Ag(111)~\cite{Sforzini_2017_PhysRevB}. In general, both orientations (Cl-up or Cl-down) have been observed
for vacuum-sublimed ClAlPc on the (111)-surfaces of noble metals~\cite{Huang_2013_PhysRevB,
  Lin_2017_PhysRevB, Niu_2013_JPhysChemC, Matencio_2018_JPhysChemC}. Moreover, the orientation can be changed
by, e.g., the deposition rate~\cite{Lin_2017_PhysRevB}, post-deposition annealing~\cite{Huang_2013_PhysRevB}
or by pulsing using an STM tip~\cite{Huang_2012_Small} and can thus act as molecular
switches~\cite{Huang_2012_Small, Song_2017_PhysChemChemPhys}.

\subsubsection{Functional groups}
Obviously, also COMs other than phthalocyanines can be functionalized. We already discussed the impact of
oxygen substitution on the coupling of pentacene with metal substrates
(Figures~\ref{fig:ELA_concept}~and~\ref{fig:dH_PxO_Ag111}). In that case, the oxygen atoms break the
conjugation of the pentacene backbone and the impact on the gas phase properties (namely, it increases IE and
EA) as well as on the coupling with metal substrates is severe~\cite{Heimel_2013_NatureChem}. Also nitrogen
substitution is frequently used to increase the EAs of COMs~\cite{Bunz_2015_AccChemRes}. One example is the
nitrogen-substituted terrylene analogue TAT~\cite{Fan_2012_OrgLett, Wise_2014_PhysChemChemPhys,
  Yang_2016_PhysRevB}. However, in this case, the nitrogen atoms are a central part of the TAT $\pi$-system
and, although a nitrogen-specific interaction in TAT monolayers on Ag(111) takes place, the vertical
adsorption distances are not substantially affected~\cite{Yang_2016_PhysRevB}. We also briefly discussed the
impact of functionalization of perylene (Figure~\ref{fig:perylene_derivatives}). Intriguingly, already the
substitution with indeno-groups, i.e., the change from perylene to DIP, changes the adsorption behavior
significantly~\cite{Franco-Canellas_2017_PhysRevMaterials, Burker_2013_PhysRevB}. In contrast, the
functionalization of perylene with oxygen (PTCDA) or with oxygen and nitrogen (PTCDI) does \emph{not} change
the averaged adsorption distances of the carbon atoms on the (111)-surfaces of noble metals
(Figure~\ref{fig:perylene_derivatives})~\cite{Henze_2007_SurfSci, Hauschild_2005_PhysRevLett,
  Franco-Canellas_2017_PhysRevMaterials, Gerlach_2007_PhysRevB}. However, the adsorption distances of the
atoms in the functional groups differ notably and the ELA is considerably different: The energy-levels of
PTCDA are Fermi level pinned on the (111)-surfaces of noble metals~\cite{Duhm_2008_OrgElec} and virtually all
substrates~\cite{Khoshkhoo_2017_OrgElec}, whereas for PTCDI the ELA is vacuum-level
controlled~\cite{Franco-Canellas_2017_PhysRevMaterials}. The coupling of PTCDA (and to minor extent also of
PTCDI) to metals has been subject to extensive research (see Refs.~\onlinecite{Hirose_1996_PhysRevB,
  Glockler_1998_SurfSci, Guillermet_2004_SurfSci, Hauschild_2005_PhysRevLett, Zou_2006_SurfSci,
  Bauer_2012_PhysRevB, Temirov_2006_Nature, Gerlach_2007_PhysRevB, Henze_2007_SurfSci, Duhm_2008_OrgElec,
  Mura_2009_JPhysChemC, Weiss_2010_PhysRevLett, Ziroff_2010_PhysRevLett, Yu_2012_NanoRes,
  Franco-Canellas_2017_PhysRevMaterials, Zaitsev_2018_JPhysCondensMatter, Yang_2018_ChemCommun} as well as
the review papers~\onlinecite{Stadtmuller_2015_JElectronSpectroscRelatPhenom, Tautz_2007_ProgSurfSci,
  Willenbockel_2014_PhysChemChemPhys}).

In the following, we will focus on molecular functionalizations that change the orientation of the COM on
metal surfaces. It is well known that the organic-inorganic ELA depends on the orientation of the COMs in the
molecular thin film~\cite{Chen_2011_AdvFunctMater, Akaike_2018_JpnJApplPhys, Duhm_2008_NatMater,
  Yamada_2018_PhysRevB}. For all cases discussed so far, the molecules have a lying-down orientation in the
contact layer to the metal, as such a face-on orientation maximizes the wave function overlap between
adsorbate and substrate. Monolayers of tilted or standing COMs on metal surfaces are rather exceptional and
in most cases the result of a transition from flat-lying in a loosely packed monolayer to a standing or
tilted orientation in a closely packed monolayer~\cite{Braatz_2016_SurfSci, Broker_2010_PhysRevLett,
  Hofmann_2017_JPhysChemC}. In some other rare cases the molecular surface unit cell includes two molecules
with one of them lying flat and the other one being tilted~\cite{Muellegger_2003_ApplPhysLett,
  Huempfner_2016_JChemPhys, Wang_2017_PhysStatusSolidiRRL}. On other surfaces, e.g., on metal oxides,
standing orientations of (polar) COMs in monolayers are successfully used for ELA engineering
\cite{Winkler_2019_ElectronStruct, Pang_2013_ChemRev, Diebold_2003_SurfSciRep}. The question arises how this
can be achieved for organic-metal interfaces, i.e., what are the driving forces for a molecular semiconductor
to adopt a tilted or standing orientation on a clean metal surface?

One of the first experimental demonstration of a COM with a standing orientation on a clean metal surface has
been made for the electron accepting COM HATCN on Ag(111)~\cite{Broker_2010_PhysRevLett}. In a combined UPS,
TPD, RAIRS, DFT and Kelvin probe study Br\"{o}ker et al.\ showed that, up to a threshold coverage, HATCN
adopts a lying-down orientation on Ag(111). Increasing the coverage leads to an orientational transition to
standing molecules, i.e., HATCN forms a transient monolayer on Ag(111). In the standing monolayer $\Delta
\mathrm{VL}$ is almost 1\,eV~\cite{Broker_2010_PhysRevLett} and thus considerably larger than for monolayers
of lying-down electron accepting molecules on the same surface such as F$_4$TCNQ~\cite{Duhm_2011_OrgElec,
  Rangger_2009_PhysRevB}, PTCDA~\cite{Kawabe_2008_OrgElec} or FAQ~\cite{Duhm_2006_JPhysChemB}. For HATCN
specific interactions of the peripheral molecular cyano groups with the metal are believed to be one of the
driving forces for an orientational transition, since for the edge-on conformation the CT becomes more
localized on the C$\equiv$N docking groups. In contrast, for the face-on conformation the whole molecule is
involved in the interaction with the substrate, including the $\sigma$-electrons of the C$\equiv$N groups as
well as the entire $\pi$-system~\cite{Broker_2010_PhysRevLett}.

\begin{figure*}
  \centering
  \includegraphics[width=0.95\textwidth]{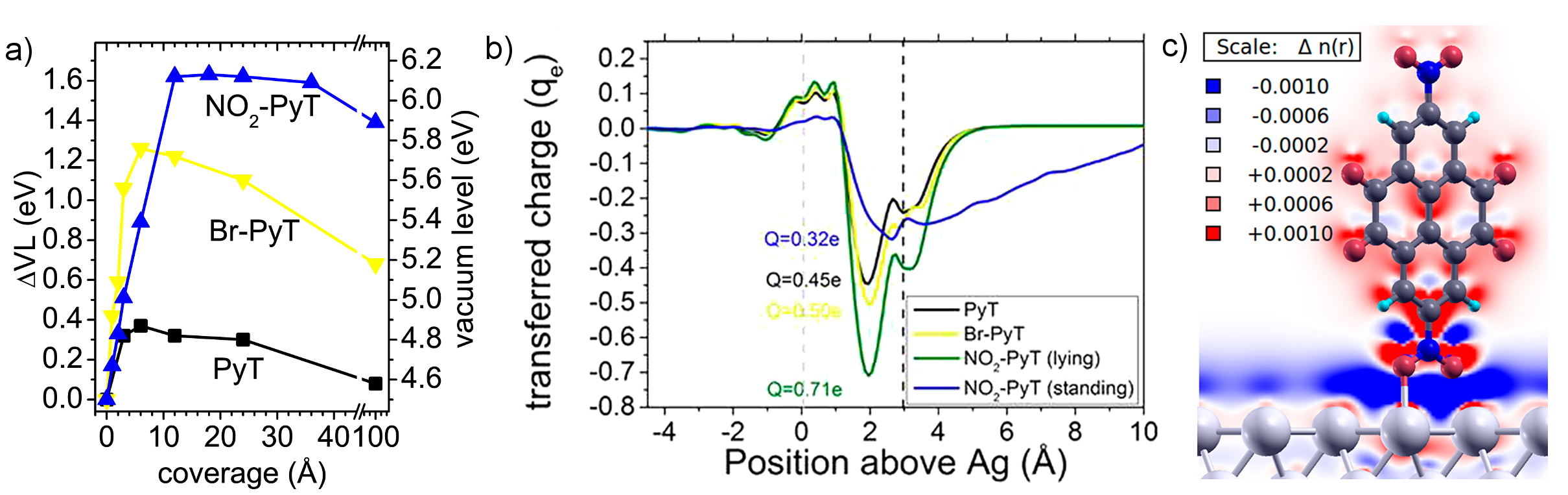}
  \caption{(a) Experimentally determined vacuum level changes ($\Delta$VL) upon stepwise deposition of
    pyrene-tetraone (PyT) and its derivatives on Ag(111). (b) Calculated cumulative charge transfer. The
    averaged position of the carbon atoms (for lying molecules) and the topmost Ag plane are indicated by
    vertical dashed lines. (c) Adsorption induced charge rearrangements derived by DFT calculations for the
    adsorption of an upright standing NO$_2$-PyT monolayer, averaged in the direction perpendicular to the
    paper plane. Br-PyT, which also shows a rather large increase of the substrate work function, is included
    for the sake of completeness. It could not be unambiguously shown whether the molecules adsorb intact or
    whether Br atoms detach during the adsorption process. Adapted from
    Ref.~\onlinecite{Hofmann_2017_JPhysChemC} with permission.}
  \label{fig:PyT_Ag111}
\end{figure*}

While HATCN has been an early example showing a large $\Delta \mathrm{VL}$ for a monolayer of edge-on COMs, a
more recent example is dinitropyrene-tetraone (NO$_2$-PyT) on Ag(111), which also exhibits a transient
monolayer structure and where $\Delta \mathrm{VL}$ for the edge-on orientation even amounts to $\sim$1.6\,eV
(Figure~\ref{fig:PyT_Ag111}a)~\cite{Hofmann_2017_JPhysChemC}. The unsubstituted parent molecule
pyrene-tetraone (PyT), which is flat-lying for all coverages, only increases the work function of Ag(111) by
$\sim$0.3\,eV. For submonolayer coverages the valence electronic structure of NO$_2$-PyT and PyT on Ag(111)
as measured by UPS is nearly identical: The former LUMO is partially filled due to a CT from the substrate
and the work function increases by $\sim$0.3\,eV. Also, the adsorption distances of the oxygen atoms in the
carbonyl groups (around 2.30\,\AA) as measured by XSW are rather similar. However, the averaged adsorption
distances of the carbon atoms are strikingly different: 2.46\,\AA{} for PyT and 2.83\,\AA{} for
NO$_2$-PyT~\cite{Hofmann_2017_JPhysChemC}. This was attributed to the bulky NO$_2$ groups ``pushing away''
the carbon skeleton from the substrate. In fact, the oxygen atoms in these groups have adsorption distances
of 2.75\,\AA. The valence electron structure and the measured adsorption distances could be reproduced quite
well by DFT modelling with vdW corrections.~\cite{Hofmann_2017_JPhysChemC}.

While these calculations were confirming the $\Delta \mathrm{VL}$ data for PyT, the results for NO$_2$-PyT
were at variance.  This can be attributed to an orientational transition of NO$_2$-PyT to a standing
monolayer. As XSW is intrinsically limited to lying (sub)monolayers, one has to rely on DFT for the adsorption
geometry. Indeed, DFT modelling of a standing monolayer of NO$_2$-PyT on Ag(111) yields almost the same $\Delta
\mathrm{VL}$ as the measurements~\cite{Hofmann_2017_JPhysChemC}. Figure~\ref{fig:PyT_Ag111}b compares the
charge rearrangements upon adsorption. For PyT and lying NO$_2$-PyT they are qualitatively similar and,
moreover, they also fit with the charge rearrangements of F$_4$TCNQ upon adsorption on the same substrate
(Figure~\ref{fig:F4TCNQ_Ag111}). In all cases, the minimum of charge density rearrangements (i.e., the maximum
in electron density accumulation) can be found between the metal surface and the molecular $\pi$-system. For
standing NO$_2$-PyT the minimum is located at the NO$_2$ docking groups
(Figure~\ref{fig:PyT_Ag111}c). Moreover, the electron accumulation extends more than 10\,\AA{} above the
surface and thus much further than for lying NO$_2$-PyT. Notably, the averaged charge transfer per molecule is
smaller for standing (0.32\,$e$) than for lying (0.71\,$e$) NO$_2$-PyT. The dipole moment, however, is
increased due the larger charge separation, which causes in turn a pronounced work function increase. Finally,
one finds that the electron affinity measured for standing NO$_2$-PyT molecules is significantly increased
because of electrostatic effects.

\subsubsection{Surface modification and decoupling}
Another area, in which the connection between electronic and geometric structure becomes evident, are efforts
towards decoupling adsorbates from the metal substrate, which are often related to effects of charge transfer
or exciton lifetimes.  In the spirit of pioneering studies, such as the decoupling of Xe from Pd(001) by the
adsorption of Kr monolayers~\cite{Kaindl_1980_PhysRevLett}, salt layers may be used to decouple COMs from
metal surfaces~\cite{Repp_2006_Science, Hollerer_2017_ACSNano, Wang_2014_AdvMater, Hofmann_2015_ACSNano,
  Imai-Imada_2018_PhysRevB}. Oxidation of Cu(100) via O$_2$-dosing decouples deposited PTCDA molecules from
the surface and hinders organic-metal charge transfer: The averaged adsorption distance of the PTCDA carbon
atoms on the oxygen-reconstructed ($\sqrt{2} \times \sqrt{2}$)\textit{R}45$^\circ$ Cu(100) surface is
3.27\,\AA~\cite{Yang_2018_ChemCommun} and thus much larger than that on pristine Cu(100)
(2.46\,\AA)~\cite{Weiss_2017_PhysRevB}. The PTCDA/Ag(111) model system has also been studied with respect to
doping by K~\cite{Zwick_2016_ACSNano, Baby_2017_ACSNano}. The experimental and theoretical results point
towards a reduced electronic coupling between the adsorbate and the substrate, which goes hand in hand with
an increasing adsorption distance of the PTCDA molecules caused by a bending of their carboxylic oxygen away
from the substrate and towards the potassium atoms~\cite{Baby_2017_ACSNano}. In principle, the organic-metal
interaction strength can also be decreased by molecular functionalization with bulky
side-groups~\cite{Jung_1997_Nature}. However, only for one system adsorption distances have been measured by
XSW and functionalizing azobenzene by alkyl groups only increases the averaged adsorption distance of the
carbon atoms on Ag(111) by 0.14\,\AA~\cite{McNellis_2010_ChemPhysLett} compared to the unsubstituted parent
molecule~\cite{Mercurio_2010_PhysRevLett}.

\subsection{Chemical reactions at interfaces}
Chemical reactions at surfaces~\cite{Lindner_2015_ChemPhysChem} involving COM molecules, a very
important topic in the context of catalysis and surface functionalization, have been addressed in
recent publications (see for instance Refs.~\onlinecite{Oteyza_2013_Science,
  Klappenberger_2014_ProgSurfSci, Liu_2014_AccChemRes, Zhang_2015_ChemSocRev, Li_2016_JAmChemSoc,
  Bjork_2016_JPhysCondMatter, Otero_2017_SurfSciRep, Shen_2017_NanoToday, Lackinger_2017_ChemCommun}).
Therefore, we shall only highlight some cases that involved a precise determination of the geometric
structure by XSW: $\left. i \right)$ on-surface formation of porous
systems~\cite{Matena_2014_PhysRevB}, $\left. ii \right)$ self-metalation reactions of
porphyrins~\cite{Burker_2014_JPhysChemC}, $\left. iii \right)$ the dissociation reaction of azobenzene
(AB)~\cite{Willenbockel_2015_ChemCommun} and $\left. iv \right)$ surface-mediated
\textit{trans}-effects~\cite{Deimel_2016_ChemSci} involving phthalocyanines.

\subsubsection{On-surface formation of porous systems}

\begin{figure}
  \centering
  \includegraphics[width=\columnwidth]{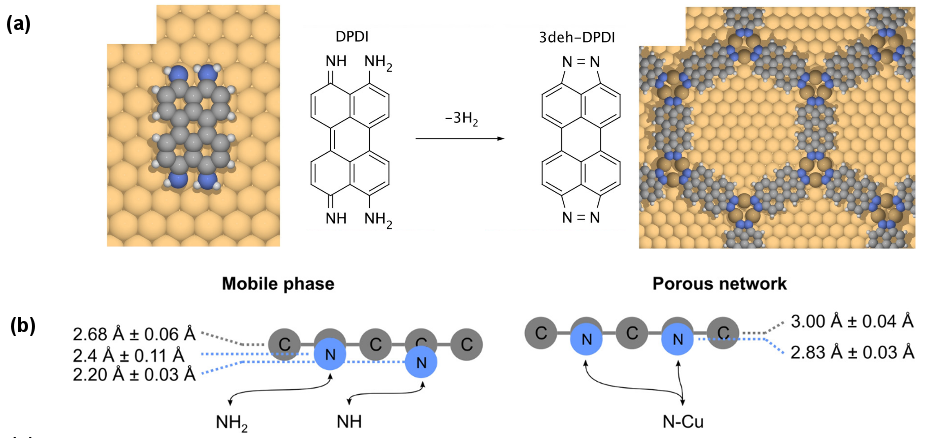}
  \caption{On-surface formation of a 2D porous network and the related chemical and structural molecular
    changes. (a) Schematics of the reaction leading to the network formation. The annealing at
    200$^\circ$C of a submonolayer coverage of DPDI induces the complete de-hydrogenation of the amine and
    imine groups, which are stabilized by the mediation of the Cu adatoms thus acting as the coordination
    centers for the network formation.  Figure adapted from Ref.~\onlinecite{Matena_2014_PhysRevB} with
    permission. }
  \label{fig:pore_formation}
\end{figure}

The perylene derivative 4,9-diaminoperylene-quinone-3,10-diimine (DPDI) has been shown to dehydrogenate
and become 3deh-DPDI after annealing of a submonolayer adsorbed on
Cu(111)~\cite{Shchyrba_2014_JAmChemSoc}. After loosing the hydrogen atoms, the two nitrogen atoms
coordinate to copper adatoms and form a highly ordered nanoporous network
(Figure~\ref{fig:pore_formation}a). Matena and coworkers studied the chemical and structural changes
induced by the formation of the network~\cite{Matena_2014_PhysRevB}. Initially, the core-level signature
of nitrogen is composed of two peaks separated by 1.8\,eV that belong to the amine (NH$_3$) and imide (NH)
groups, the latter being $\sim$0.2\,\AA{} closer to the surface, as obtained with XSW
(Figure~\ref{fig:pore_formation}b). Thus, the measurements show two different environments for the
nitrogen atoms, which, however, become chemically and structurally indistinguishable upon network
formation through the equal binding of the nitrogen ligands to the Cu adatoms. In addition, the XSW
measurements show an upward lifting of the molecule, i.e.\ $\sim$0.3\,\AA{} for the perylene core and
$\sim$0.4\,\AA{} for the nitrogen atoms compared to the less interacting NH and NH$_2$ groups
(Figure~\ref{fig:pore_formation}b). This was interpreted in terms of the interplay between
molecule-substrate vs.\ intermolecular interactions, which is clearly balanced towards the latter upon
network formation~\cite{Matena_2014_PhysRevB}. More precisely, the obtained adsorption distances with
respect to the surface correspond to a physisorptive scenario, implying that the molecule is decoupled
from the surface and the bonding occurs only through the copper adatoms. Interestingly, DFT calculations
of the network with and without the surface indicate that a planar geometry, with the copper adatoms at
the same plane as the molecules, is disrupted by the presence of the surface that pulls the copper adatoms
closer and thus bend the molecule. Consequently, the adatoms mediate the intermolecular interactions
acting as coordination centers but also influence the bonding between the molecules and the
substrate~\cite{Matena_2014_PhysRevB}.

\subsubsection{Self-metalation reactions of porphyrins}
\begin{figure}
  \centering
  \includegraphics[width=.9\columnwidth]{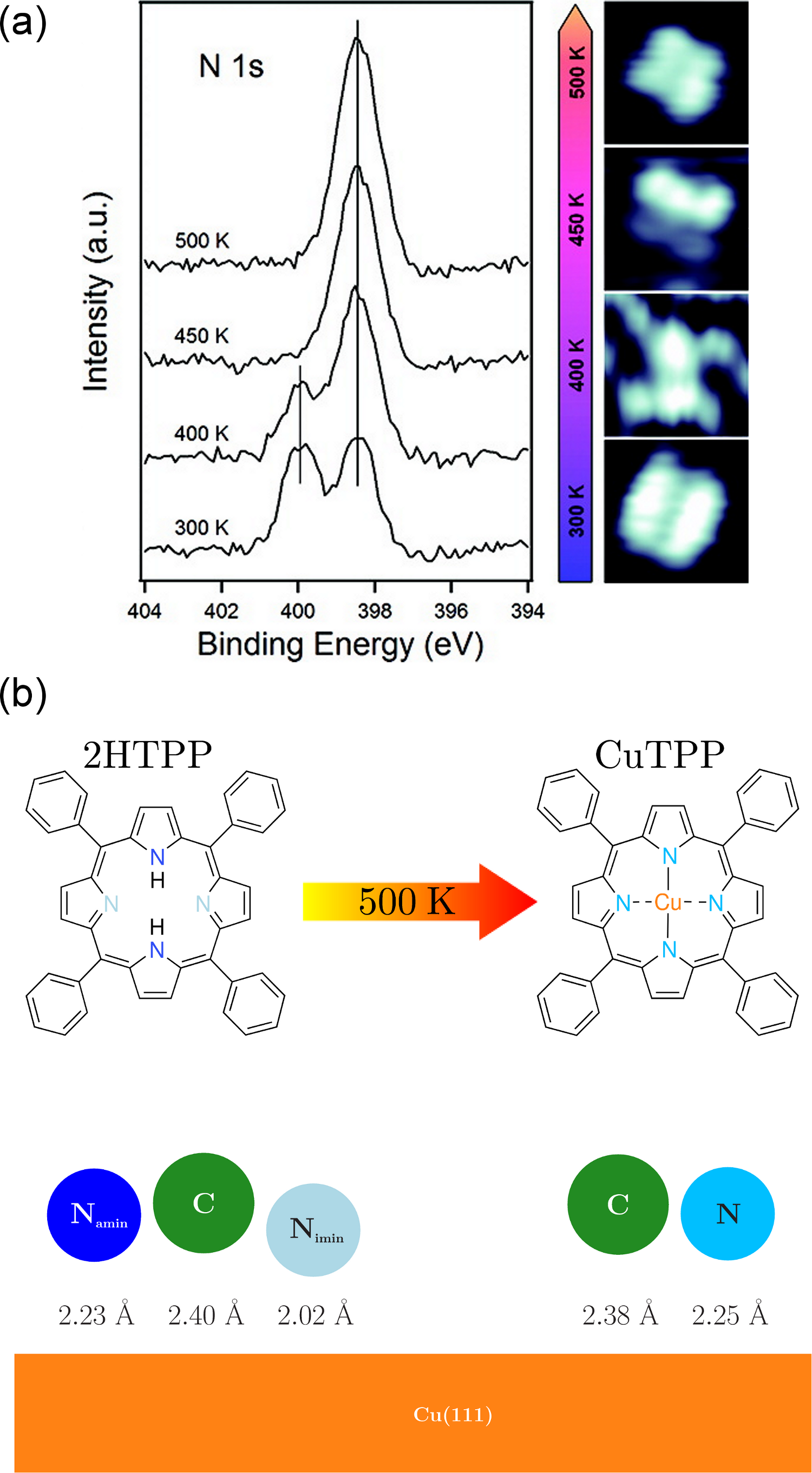}
  \caption{(a) Evolution of the nitrogen N\,1s core-level signal of 2HTPP as a function of the temperature
    and the corresponding STM images. Image taken from Ref.~\onlinecite{Xiao_2012_JPhysChemC} with
    permission. (b) Adsorption distances of the average carbon atoms and the nitrogen species (aminic in dark
    and iminic in light blue) before and after annealing at 500\,K. Image adapted from
    Ref.~\onlinecite{Burker_2014_JPhysChemC} with permission.}
  \label{fig:metalation}
\end{figure}

As introduced in Sec.~\ref{ssec:Pcs}, porphyrins as well as phthalocyanines can host a metal ion within the
molecular core (Figure~\ref{fig:molecules}). For metal-free molecules, these can also be incorporated through
various metalation reactions~\cite{Marbach_2015_AccChemRes, Gottfried_2015_SurfSciRep,
  Diller_2016_ChemSocRev}, whereby a H$_2$-molecule is released and the ion becomes coordinated to the
central nitrogen atoms. Similar to what has been discussed in the previous paragraph, the metalation reaction
can be followed by monitoring the change in the N\,1s core-level signal (see Figure~\ref{fig:metalation}a),
i.e.\ from two clearly distinguishable aminic (or pyrrolic, --NH--) and iminic (--N=) nitrogen species for
the free-base molecule towards one single species for the equally-coordinated nitrogen
atoms~\cite{Gonzalez-Moreno_2011_JPhysChemC, Diller_2012_JChemPhys,Xiao_2012_JPhysChemC,
  Burker_2014_JPhysChemC}.

A particular case of metalation is realized through the direct incorporation of surface atoms, the
so-called \emph{self-metalation} reaction~\cite{Gonzalez-Moreno_2011_JPhysChemC,
  Diller_2012_JChemPhys}. In this context, the thermally induced self-metalation of
2\textit{H}-tetraphenylporphyrin (2HTPP, Figure~\ref{fig:molecules}j) adsorbed on Cu(111) and the
subsequent formation of copper(II)-tetraphenylporphyrin (CuTPP) is one of the most thoroughly studied
reaction~\cite{Diller_2012_JChemPhys, Xiao_2012_JPhysChemC, Rockert_2014_ChemEurJ,Rockert_2014_JPhysChemC,
  Burker_2014_JPhysChemC}. For instance, in a temperature-dependent STM and XPS study of this
reaction~\cite{Xiao_2012_JPhysChemC} it was found that along with the self-metalation (starting at 400\,K)
the molecule undergoes a gradual hydrogen loss until a total de-hydrogenation occurs at 500\,K. As imaged
with STM (Figure~\ref{fig:metalation}a), 2HTPP molecules appear rather planar, with the phenyl groups
parallel to the surface~\cite{Xiao_2012_JPhysChemC} but the increasing loss of hydrogen reduces the steric
repulsion between phenyl rings and enables their rotation~\cite{Xiao_2012_JPhysChemC}. Interestingly, the
full de-hydrogenation again renders a flat molecule. Consequently, the adsorption geometry has possible
contributions from the metalation, which relaxes the strong interaction of the nitrogen atoms with the
substrate, and also from the rotation of the functional groups. In order to study the influence of the
metalation on the vertical adsorption distance, B{\"u}rker et al.~\cite{Burker_2014_JPhysChemC} followed
the changes in the adsorption upon self-metalation at 500\,K to avoid strong contributions of the rotating
phenyl groups to the conformational properties. Thus, for the free-base porphyrin the two inequivalent
nitrogen atoms have two distinct adsorption distances (see Figure~\ref{fig:metalation}b), i.e.\ the iminic
ones closer to the surface as a result of the stronger interaction with the copper
atoms~\cite{Diller_2012_JChemPhys}. Because both nitrogen species are located below the average carbon
adsorption distance, the macrocycle takes a saddle-like shape on the surface. Upon metalation at 500\,K,
the incorporation of the copper atoms lifts this distortion, since the preferential interaction of the
iminic nitrogens with the substrate is switched off (Figure~\ref{fig:metalation}b).  Interestingly, the
average carbon adsorption distance remains virtually unchanged, although the vertical order is increased
(as deduced from the higher coherent fractions).  The authors therefore conclude that the metalation of
2HTPP relaxes the molecular core without impacting the overall adsorption distance of the molecule and
rather possible rotations and/or bending of the phenyl groups determine the overall adsorption
geometry~\cite{Burker_2014_JPhysChemC}. Notably, recent DFT calculations of 2HTPP adsorbed on Cu(111) have
shown that an inverted macrocycle reproduces the experimental data better than a saddle-shape
geometry~\cite{Lepper_2017_ChemCommun}.

\begin{figure}
  \centering
  \includegraphics[width=\columnwidth]{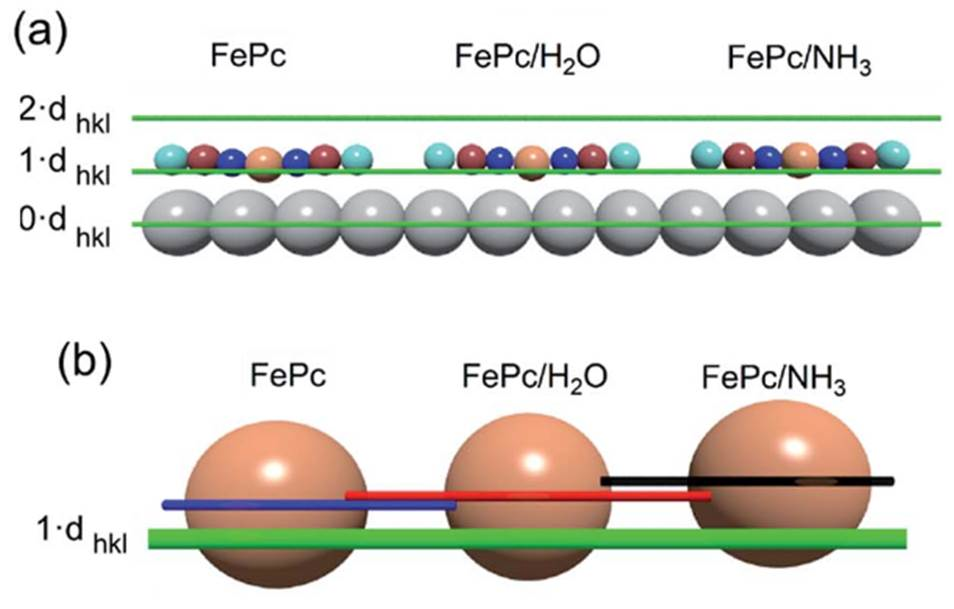}
  \caption{XSW measurements performed for FePc adsorbed on Ag(111) in the case of no-, H$_2$O- and
    NH$_3$-dosing. The \textit{trans}-effect increases in the order H$_2$O $<$ NH$_3$ $<$ Ag(111). (a)
    To-scale schematics of the XSW data (note that the graphic representation of the $n$ ambiguity of the
    $d_\mathrm{hkl} = d_0(n + P_\mathrm{hkl}$). (b) Detail of the change of the Fe adsorption distance for the
    different cases, which clearly describes the structural implications of the surface
    \textit{trans}-effect. Figure reprinted from Ref.~\onlinecite{Deimel_2016_ChemSci} with permission.}
  \label{fig:surface_trans_reaction}
\end{figure}

\subsubsection{The dissociation reaction of azobenzene}
As prototypical molecular switches azobenzene (AB) and its derivative tetrabutyl-AB (TBA) have been studied
with XSW on Cu(111)~\cite{Willenbockel_2015_ChemCommun} and on Ag(111)~\cite{Mercurio_2010_PhysRevLett,
  McNellis_2010_ChemPhysLett, Mercurio_2013_PhysRevBb, Mercurio_2014_FrontPhysics}. Willenbockel et
al.~\cite{Willenbockel_2015_ChemCommun} reported a coverage-dependent dissociation of AB on Cu(111), which
itself is not observed on the Ag(111) surface~\cite{Mercurio_2010_PhysRevLett, Mercurio_2013_PhysRevBb,
  Mercurio_2014_FrontPhysics}. The authors attribute the difference to the balance between molecule-molecule
and substrate-molecule interactions. More precisely, the stronger bond between the nitrogen atoms of the
\mbox{(--N=N--)} azo-bridge and the copper substrate forces AB to decompose into two phenyl nitrene molecules
to accommodate the increasing molecular packing. In contrast, the N--Ag bond is weaker and allows the molecule
to deform upon coverage increase. Through a sophisticated vector analysis of the XSW
data~\cite{Mercurio_2014_FrontPhysics} the authors obtained tilt and rotational angles for the phenyl rings on
both substrates. The derived rotation of those groups on silver is larger than on copper, which is considered
as indication for the increased flexibility of the N--Ag bond.

Interestingly, in another study it was found that the isomerization reaction of AB, which is essential for
the switching mechanism, can be quickly reversed by CT from the substrate to the molecule, thus preventing
the switching effect to be measured~\cite{Morgenstern_2011_ProgSurfSci}. This would explain why the
switching is observed on Au(111)~\cite{Comstock_2007_PhysRevLett} but not on
Ag(111)~\cite{Maurer_2012_AngewChemIntEd}.

\subsubsection{Surface-mediated trans-effects of MePc}
In inorganic chemistry, it is known that adding a new ligand to a metal ion influences the bond between
the ion and the other previously present ligands.  One can distinguish two cases: the ligands are opposed
(trans) to each other or the ligands are at the same side (cis). The new coordination affects the
ground-state properties, the length as well as the thermodynamic and vibrational properties of the other
bonds. In an analogous situation, it was seen that one can reproduce the \textit{trans}-effect with
metal-ion-containing molecules adsorbed on a metal surface, where the substrate acts as one of the
ligands. This is known as surface \textit{trans}-effect. For the particular case of of metal
phthalocyanines (MePc) adsorbed on Ag(111), a study with complementary PES, STM and DFT determined that Co
and Fe ions are forming bonds with the substrate, but not
Zn~\cite{Hieringer_2011_JAmChemSoc}. Interestingly, upon dosing of nitric oxide (NO), the changes in the
electronic characteristics indicate that the ion-to-substrate bond is weakened or entirely suppressed and
DFT calculations show that the Ag--Me bond length is increased~\cite{Hieringer_2011_JAmChemSoc}.

In this context, it was recently found that the coordination of ligands with different reactive character to
the Fe ion of FePc adsorbed on Ag(111) indeed changes the Ag-to-Fe adsorption
distance~\cite{Deimel_2016_ChemSci}. More precisely, H$_2$O and ammonia (NH$_3$) were dosed, which renders an
increasing \textit{trans}-effect in the order H$_2$O $<$ NH$_3$ $<$ Ag(111), thus one expects that the Fe atom
should show a larger adsorption distance when the Ag(111) surface is \textit{trans} to the ammonia than to
water. As shown in Figure~\ref{fig:surface_trans_reaction}, the XSW results confirmed this scheme as the Fe
atom is rather shifted away from the surface when NH$_3$ is dosed compared to H$_2$O.  This behavior was also
reproduced with vdW-corrected DFT.

\subsection{Heterostructures}
\begin{figure}
  \centering
  \includegraphics[width=\columnwidth]{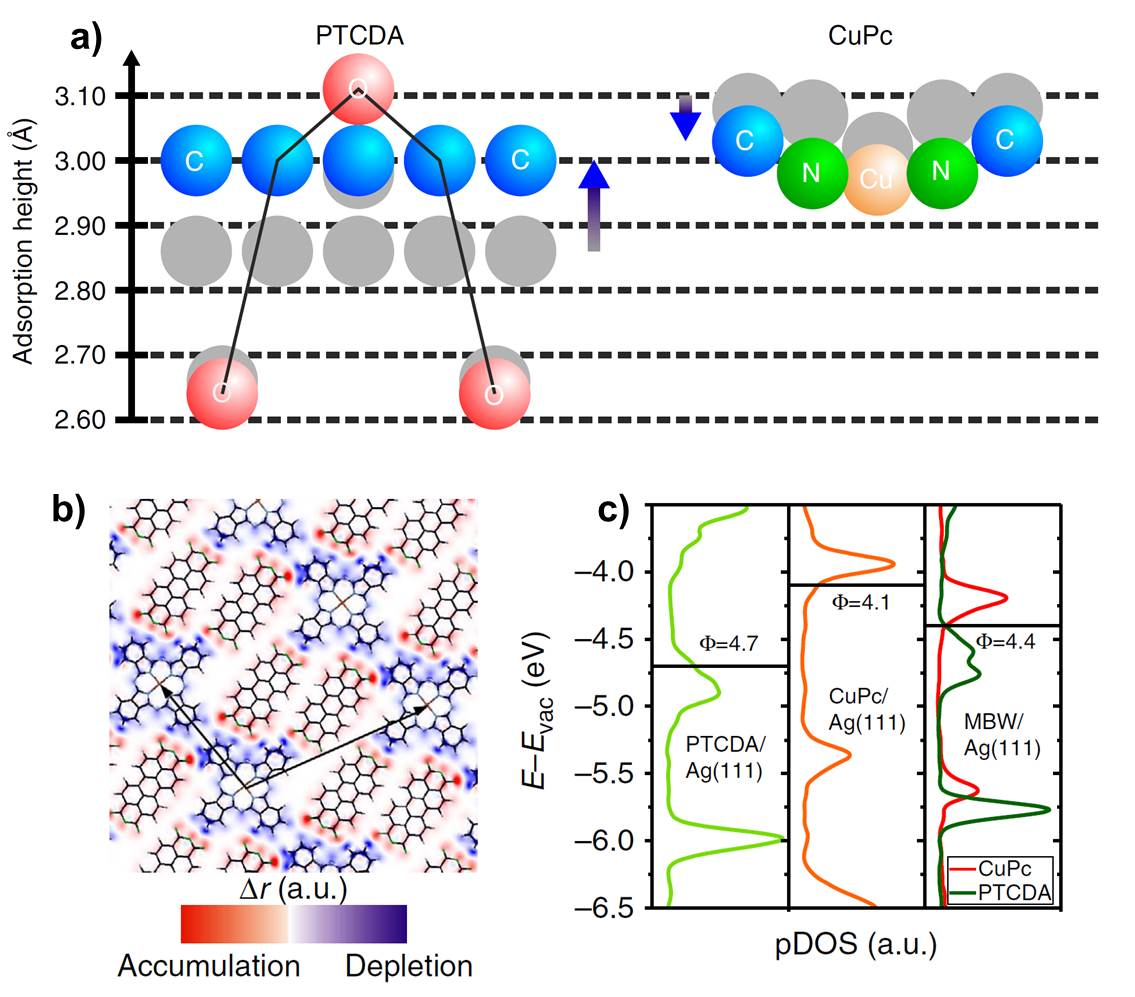}
  \caption{Bimolecular mixed layer of PTCDA and CuPc on Ag(111). (a) Vertical adsorption geometry as revealed
    by XSW. The adsorption heights of all involved atomic species are illustrated for the bimolecular
    monolayer (colored spheres) and the homomolecular monolayer (grey spheres). (b) Charge density difference
    plot showing depletion (blue) and accumulation (red) of electronic charge in a plane parallel to the
    surface (in a height of maximum DOS of the LUMO orbitals). (c) Projected DOS of the $\pi$-orbitals of
    PTCDA and CuPc in the homomolecular PTCDA/Ag(111) (left), the homomolecular CuPc/Ag(111) (middle) and the
    bimolecular layer (right). Energies are aligned with the vacuum level, the Fermi energies are indicated by
    black lines revealing the work functions.  Taken from Ref.~\onlinecite{Stadtmuller_2014_NatComms} with
    permission.}
  \label{fig:PTCDA_CuPc_Ag111}
\end{figure}

Organic heterostructures in the monolayer or bilayer regime on clean metals~\cite{Huang_2011_JPhysDApplPhys,
  Wakayama_2016_JpnJApplPhys, Stadtmuller_2015_JElectronSpectroscRelatPhenom, Goiri_2016_AdvMater,
  Bouju_ChemRev_2017} may be considered as model systems for organic-organic interfaces, i.e.\ the essential
component for most electronic devices applications.  Most studies in the area are dealing with bimolecular
mixed layers~\cite{Henneke_2017_NatMater, Zhong_2014_ACSNano, El-Sayed_2013_ACSNano,
  Stadtmuller_2014_NatComms}, whereas only a few studies focus on bilayers~\cite{Borghetti_2016_JPhysChemC,
  Stadtmuller_2014_JPhysChemC, Thussing_2017_JPhysChemC, Egger_2013_JPhysChemC, Wang_2018_JPhysChemC,
  Wang_2020_ACSApplMaterInterfaces}. For both types of heterostructures CuPc and PTCDA on Ag(111) have become
popular~\cite{Stadtmuller_2015_NewJPhys, Gruenewald_2015_PhysRevB, Stadtmuller_2014_JPhysChemC,
  Thussing_2017_JPhysChemC, Egger_2013_JPhysChemC, Stadtmuller_2014_NatComms, Henneke_2017_NatMater,
  Straaten_2018_JPhysChemC, Lerch_2018_JPhysCondensMatter}. Figure~\ref{fig:PTCDA_CuPc_Ag111}a shows the
experimentally determined adsorption geometries of PTCDA and CuPc in their respective monolayers on Ag(111)
and in the bimolecular mixed layer~\cite{Stadtmuller_2014_NatComms}. Strikingly, PTCDA is lifted up in the
bimolecular system, whereas CuPc is pushed down. Naively, this could lead to the notion that the coupling of
PTCDA with Ag(111) decreases and that of CuPc with Ag(111) increases. However, the situation is more complex
and Stadtm{\"u}ller et al.~\cite{Stadtmuller_2014_NatComms} showed by means of STM, STS, orbital tomography
and DFT modelling that the adsorption height changes are driven by intermolecular interactions, which are
increased by the equalization of adsorption heights. As illustrated in Figure~\ref{fig:PTCDA_CuPc_Ag111}b,
which highlights the charge rearrangement between PTCDA and CuPc, the acceptor character of PTCDA and the
donor character of CuPc are increased in the bimolecular system. Consequently, the LUMOs of PTCDA and CuPc
move away from the common Fermi level in opposite directions (Figure~\ref{fig:PTCDA_CuPc_Ag111}c). Overall,
this example shows how observables such as vertical adsorption distances, frontier orbital binding energies
and charge transfer are linked and influence each other.

\begin{figure}
  \centering
  \includegraphics[width=\columnwidth]{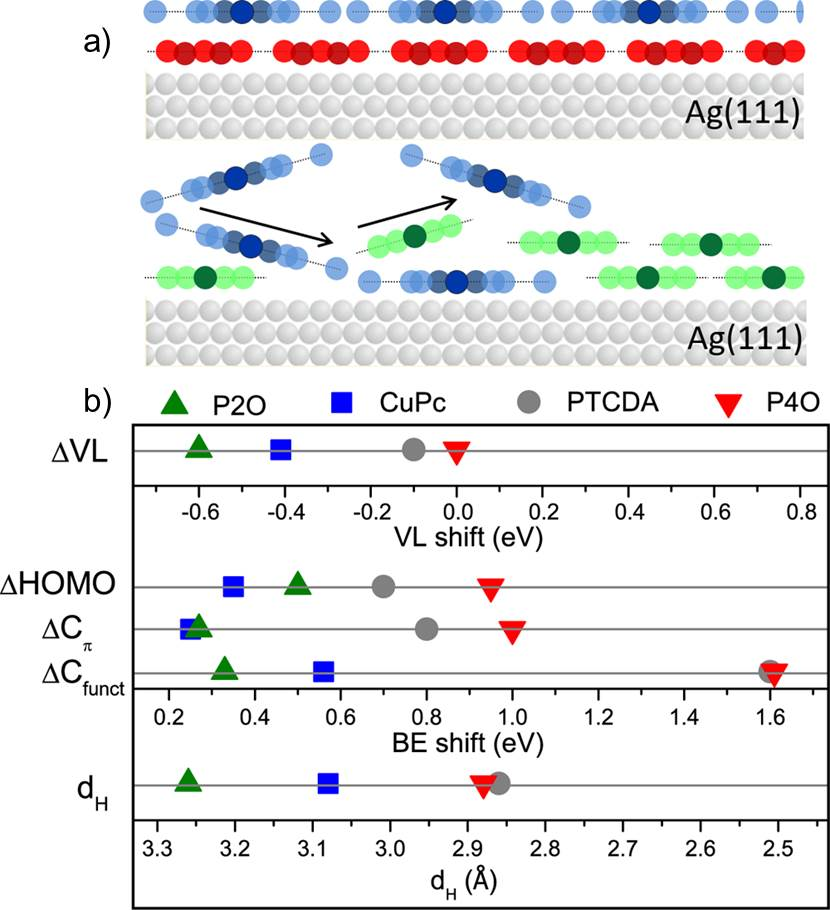}
  \caption{(a) Bilayer formation (top) vs.\ molecular exchange (bottom). In both cases, CuPc (blue) has been
    vacuum-sublimed on a closed monolayer of P4O (red) or P2O (green) on Ag(111). (b) Vacuum level shift
    ($\Delta$VL) between clean Ag(111) and a monolayer of the respective COM. Binding-energy shift between
    monolayer and multilayer of the HOMO-maximum ($\Delta$HOMO) and the C\,1s peak of the molecular backbone
    ($\Delta$C$_{\pi}$) and the functional group ($\Delta$C$_\mathrm{funct}$) of the respective COM on
    Ag(111). Averaged bonding distance (d$_\mathrm{H}$) of the carbon atoms in the molecular core in
    sub-monolayers on Ag(111). Taken from Ref.~\onlinecite{Wang_2018_JPhysChemC} with permission.}
  \label{fig:coupling_indicators}
\end{figure}

For bilayer systems, the most fundamental question is whether the deposition sequence reflects the actual
arrangement in the heterostructure. At room temperature, this is the case for CuPc deposited on a closed
monolayer of PTCDA on Ag(111)~\cite{Thussing_2017_JPhysChemC}. However, for the reverse deposition sequence,
i.e., PTCDA on CuPc, molecular exchange occurs and PTCDA replaces CuPc in the contact layer to
Ag(111)~\cite{Stadtmuller_2014_JPhysChemC, Thussing_2017_JPhysChemC}. One could expect that this is related to
the different interaction strength of the adsorbates with the substrate, which is weaker for CuPc than for
PTCDA. This assumption has been tested by using P2O and P4O monolayers on Ag(111), which have been introduced
as reference systems for physisorption and chemisorption (Figures~\ref{fig:ELA_concept}
and~\ref{fig:dH_PxO_Ag111}). Indeed, subsequently deposited CuPc molecules can replace P2O in the contact
layer to Ag(111), while they remain on top of P4O on Ag(111)
(Figure~\ref{fig:coupling_indicators}a)~\cite{Wang_2018_JPhysChemC}. The different behavior of the
CuPc/PxO/Ag(111) bilayer systems allows, thus, to conclude that the interaction of CuPc with Ag(111) is beyond
physisorption (although still relatively ``weak''~\cite{Kroger_2010_NewJPhys}).

As mentioned in Sec.~\ref{ssec:energetics}, rigid shifts of valence and core-levels observed for monolayer
and multilayer coverage can serve as indicator for organic-metal interaction strength. These shifts are
shown in Figure~\ref{fig:coupling_indicators}b for P2O, P4O, CuPc and PTCDA on Ag(111). With the exception
of the shift between HOMO position in the monolayer and the multilayer ($\Delta$HOMO) all indicators show
that the interaction strength with Ag(111) increases in the order P2O--CuPc--PTCDA--P4O. The largest shifts
have been found for the core-levels of carbon atoms in functional groups ($\Delta$C$_\mathrm{funct}$),
which might be the best indicator for the interaction strength. Notably, all the data are taken from
measurements of monomolecular systems~\cite{Gerlach_2007_PhysRevB, Zou_2006_SurfSci, Duhm_2008_OrgElec,
  Haeming_2010_PhysRevB, Wang_2018_JPhysChemC, Kroger_2010_NewJPhys}, but still allow to predict the
sequential arrangement in heterostructures. However, we are aware that also other factors such as the
particular molecular weight or shape also impact possible molecular exchange
processes~\cite{Jacobs_2017_AdvMater, McEwan_2018_AdvMaterInterfaces}.

Figure~\ref{fig:coupling_indicators}b also includes vacuum level shifts between the clean Ag(111) surface and
the respective monolayer and the vertical adsorption heights. As discussed throughout this review, several
often competing factors impact dipoles at organic-metal interfaces. The ``correct'' order of the $\Delta$VLs
might thus be merely coincidental. Vertical adsorption distances are a better indicator (for a detailed
discussion see Sec.~\ref{sec:summary}). However, they have the disadvantage of requiring measurements at
highly specialized beamlines at synchrotron radiation facilities, whereas the photoelectron spectroscopy
based indicators can be accessed with standard lab equipment.

\section{Summary and conclusions} \label{sec:summary}
As discussed in this review, the contact formation of specific adsorbate-substrate systems is by now
reasonably well understood. At the same time, numerous studies addressing the relation of structural and
electronic properties at organic-metal interfaces~\cite{Gerlach_2013_book,
  Stadtmuller_2015_JElectronSpectroscRelatPhenom, Willenbockel_2014_PhysChemChemPhys, Goiri_2016_AdvMater,
  Otero_2017_SurfSciRep, Kera_2018_JPhysSocJpn, Klein_2019_PhysRevX} have demonstrated that there are
actually no ``simple rules'' and that a prediction of the energy-level alignment requires significant
computational efforts.

\begin{figure}
  \centering
  \includegraphics[width=\columnwidth]{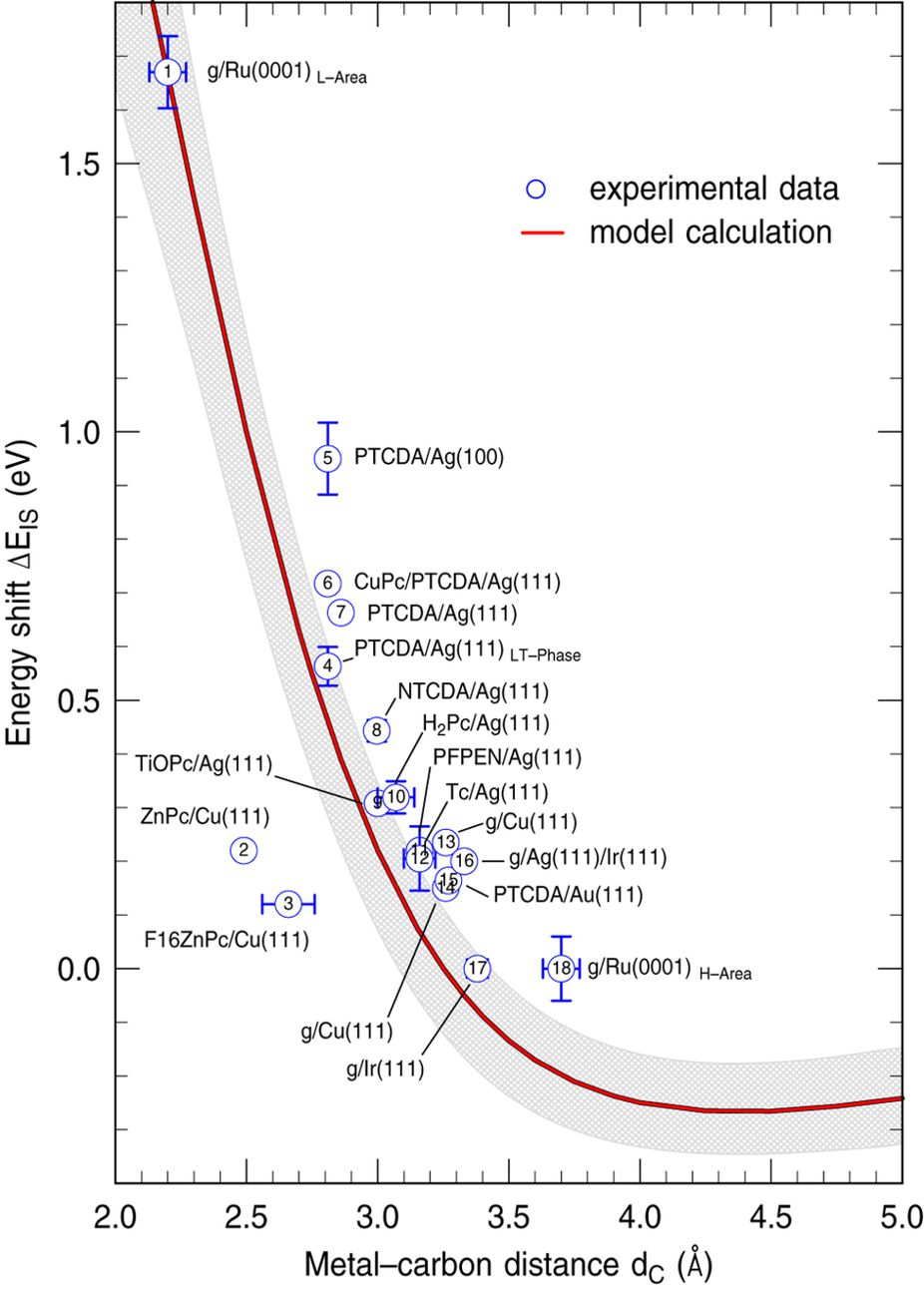}
  \caption{Energy shift $\Delta E_\mathrm{IS}$ of the interface state with respect to the Shockley surface
    state on the pristine metal as a function of the carbon-metal distance $d_\mathrm{C}$. The solid red line
    shows the calculated results for a carbon layer on Ag(111). Taken from
    Ref.~\onlinecite{Armbrust_2017_SciRep} with permission.}
  \label{fig:surface_state}
\end{figure}

Nevertheless, we can identify a few general trends that connect the adsorption geometry and the energy-level
alignment.  For example, a clear correlation was found for the shift of the Shockley surface state $\Delta
E_\mathrm{IS}$ on clean metals due to adsorption of a molecular monolayer
(\textit{cf.}~Figure~\ref{fig:surface_state}). Apparently, this shift is related to the organic-metal coupling
strength and can be explained using a relatively simple one-dimensional model
potential~\cite{Armbrust_2017_SciRep}. A closer inspection of Figure~\ref{fig:surface_state}, however, reveals
that most of the data points refer to Ag(111) surfaces and that the two outliers on the left of the calculated
model curve correspond to energy shifts on Cu(111) surfaces. This indicates that the situation is more
complicated and that in some cases effects beyond LUMO filling~\cite{Armbrust_2017_SciRep} play a role for the
surface state shift.

\begin{figure}
  \centering
  \includegraphics[width=\columnwidth]{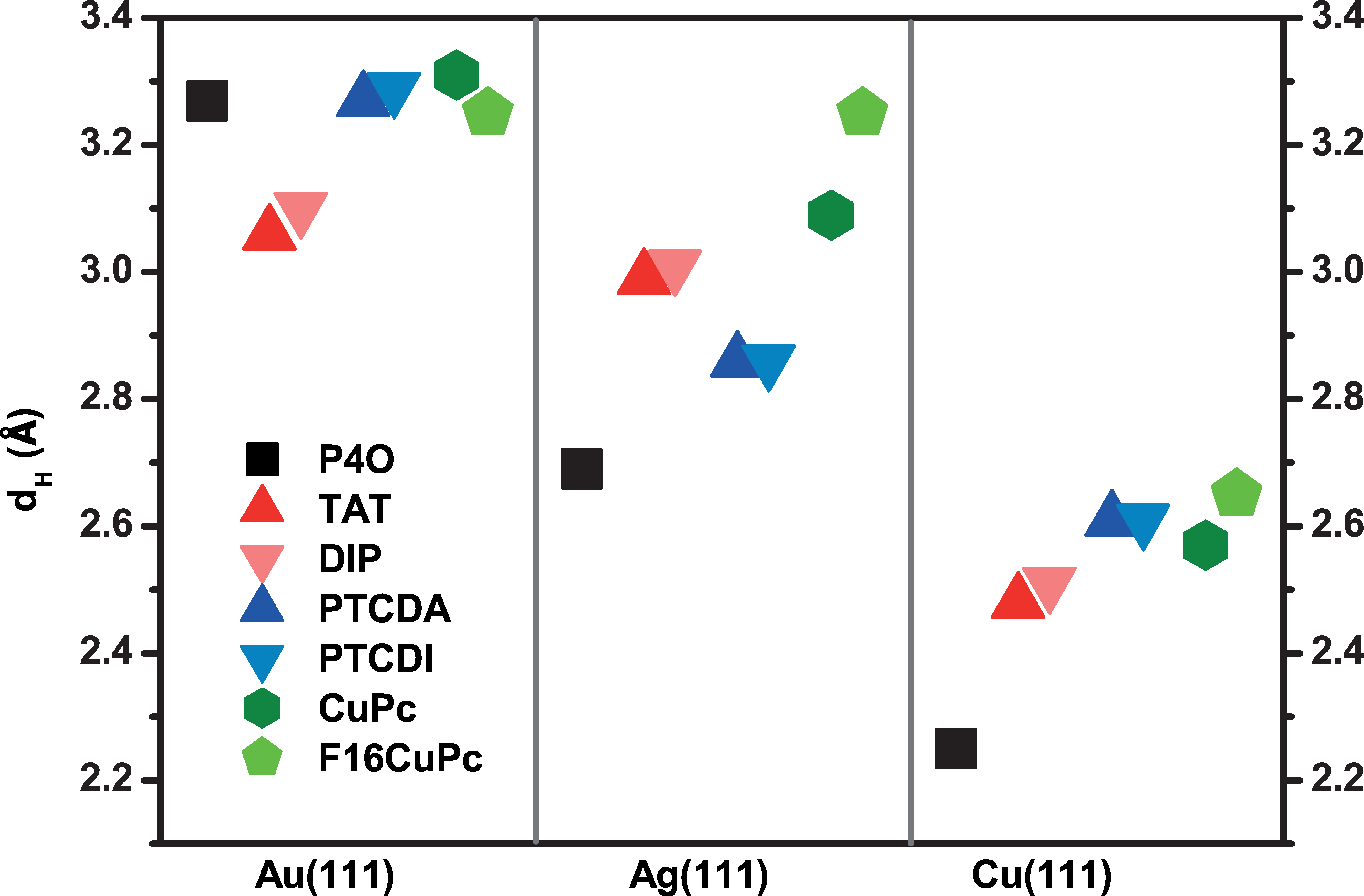}
  \caption{Compilation of vertical adsorption distances $d_H$ for carbon atoms within the molecular backbone
    of selected COMs. The plot includes those systems for which XSW results on all (111)-surfaces of the
    noble metals Au, Ag, and Cu are available. The data shown here are taken from
    Refs.~\onlinecite{Kleimann_2014_JPhysChemC, Kroger_2010_NewJPhys, Oteyza_2010_JChemPhys,
      Gerlach_2007_PhysRevB, Kroger_2011_PhysRevB, Gerlach_2005_PhysRevB, Hauschild_2005_PhysRevLett,
      Henze_2007_SurfSci, Franco-Canellas_2017_PhysRevMaterials, Burker_2013_PhysRevB, Yang_2016_PhysRevB,
      Heimel_2013_NatureChem}.}
  \label{fig:table_plot}
\end{figure}

Elaborating on this issue, Figure~\ref{fig:table_plot} shows adsorption distances $d_H$ of carbon atoms in
an aromatic environment of seven COMs, for which they have been determined on all three (111)-surfaces of
the noble metals. While the plot is certainly simplistic (possible coverage and/or temperature effects are
neglected) and the selection of molecules is to some degree arbitrary, it highlights some important
findings for organic-metal interfaces. Obviously, for all COMs the adsorption distances decrease in the
order Au--Ag--Cu (see also Sec.~\ref{ssec:concepts_substrate}). Moreover, on Au(111) and Cu(111) the
bonding distances exhibit a rather narrow distribution of only $\sim$0.25\,\AA{} (if the outlier P4O on
Cu(111) is excluded), whereas on Ag(111) the bonding distances span $\sim$0.60\,\AA. This difference can be
explained by noting that generally -- and in particular for the selected COMs in
Figure~\ref{fig:table_plot} -- the interaction with Au(111) is mostly physisorptive and that with Cu(111)
mostly chemisorptive, whereas COMs on Ag(111) may tend to either weak or strong coupling.

To illustrate this variability we may first consider two pairs of COMs, i.e.\ TAT/DIP (red symbols in
Figure~\ref{fig:table_plot}) and PTCDA/PTCDI (blue symbols in Figure~\ref{fig:table_plot}), which each have
nearly the same vertical adsorption distances on the different surfaces.  For TAT/DIP, the adsorption
distances on Au(111) and Ag(111) are very similar, but decrease significantly on Cu(111). For PTCDA/PTCDI, on
the other hand, a pronounced decrease occurs already when going from Au(111) to Ag(111). This behaviour shows
that for molecules with more reactive functional groups (see also Sec.~\ref{ssec:concepts_molecule})
chemisorption already sets in on Ag(111), as confirmed by the pronounced molecular distortion
(Figure~\ref{fig:perylene_derivatives}) and strong chemical core-level shifts (Figures~\ref{fig:PTCDI_XPS}
and~\ref{fig:coupling_indicators}) on this surface.  We note that having virtually the same vertical
adsorption distances, does not imply that the energy-level alignment is identical: PTCDA is Fermi-level
pinned on all the three surfaces~\cite{Duhm_2008_OrgElec} (and virtually all other
substrates~\cite{Khoshkhoo_2017_OrgElec}), whereas the ELA of PTCDI is vacuum level controlled on the
(111)-surfaces of the noble metals~\cite{Franco-Canellas_2017_PhysRevMaterials}.  The influence of
site-specific interactions is even more pronounced for P4O on these surfaces (black symbol in
Figure~\ref{fig:table_plot}), showing adsorption distances which differ by more than 1\,\AA{} due to the
re-hybridization of P4O on Ag(111) (Figure~\ref{fig:ELA_concept}) and Cu(111).  In fact, the vertical
adsorption distances of PEN and P4O on Cu(111) are rather similar (2.34\,\AA\ and 2.25\,\AA,
respectively)~\cite{Koch_2008_JAmChemSoc, Heimel_2013_NatureChem}.  While P4O is Fermi-level pinned, the ELA
of P2O and PEN are vacuum level controlled on these three surfaces~\cite{Chen_2019_JPhysCondensMatter}.

A common approach to reduce the organic-metal interaction strength is (per)fluorination of the
adsorbate~\cite{Gerlach_2005_PhysRevB, Oteyza_2010_JChemPhys,Kroger_2010_NewJPhys}. Comparing CuPc and
F$_{16}$CuPc (green symbols in Figure~\ref{fig:coupling_indicators}) shows that on Ag(111) this method is
indeed working and the repulsive interaction of the fluorine atoms prevent coupling beyond physisorption -- as
observed for CuPc/Ag(111). On Cu(111), however, where the adsorption distances of both phthalocyanine
molecules are rather similar, the attractive interaction between the substrate and the Pc core is already too
strong and the fluorine atoms cannot ``repel'' the entire molecule. Consequently, for
F$_{16}$CuPc/Cu(111)~\cite{Gerlach_2005_PhysRevB} (and other perfluorinated Pcs such as
F$_{16}$ZnPc/Cu(111)~\cite{Yamane_2010_PhysRevLett}) a significant molecular distortion with the fluorine
atoms above the carbon backbone is found.

Overall, these results demonstrate that the interplay between adsorption geometry and electronic structure is
complex and measuring the element-specific adsorption distances of $\pi$-conjugated molecules on metals is
essential for understanding the interface dipoles and thus the energy-level alignment.  Because of the
different driving forces for charge rearrangements upon contact formation, such as push-back effect or
chemical-bond formation, the bonding behavior cannot be characterized using few parameters like the metal
work function, the ionization energy and electron affinity of the organic thin film.  While XSW has become a
well-established high-precision technique in the field of metal-organic interfaces, it has not yet been used
extensively to study all relevant systems. However, we are confident that the results reviewed here and, most
importantly, new state-of-the-art facilities, such as beamline I09 at the Diamond Light Source will encourage
further systematic studies of such a vivid and interesting area of surface science.

\begin{acknowledgments}
  The authors thank the European Synchrotron Radiation Facility (ESRF) and the Diamond Light Source (DLS) for
  making their facilities available to us. We thank the different local contacts and beamline scientist that
  helped us during the numerous XSW experiments at ID32 (ESRF, until 2011) and I09 (DLS, since 2013), in
  particular J{\"o}rg Zegenhagen and Tien-Lin Lee.

  It is a pleasure to acknowledge interactions with a large number of colleagues in the field, including in
  alphabetical order J.\ Banerjee, C.\ B\"urker, B.\ Detlefs, D.\ A.\ Duncan, G.\ Heimel, O.\ T.\ Hofmann, T.\
  Hosokai, S.\ Kera, N.\ Koch, C.\ Kumpf, J.\ Niederhausen, I.\ Salzmann, A.\ Sch\"oll, M.\ Sokolowski, P.\
  K.\ Thakur, F.\ S.\ Tautz, A.\ Tkatchenko, N.\ Ueno, A.\ Vollmer, Q.\ Wang, E.\ Zojer and others, too many
  to name them all.

  Financial support from the Deutsche Forschungsgemeinschaft (DFG), the Soochow University-Western University
  Center for Synchrotron Radiation Research, the 111 Project of the Chinese State Administration of Foreign
  Experts Affairs and the Collaborative Innovation Center of Suzhou Nano Science \& Technology (NANO-CIC) is
  gratefully acknowledged.
\end{acknowledgments}%
%
%%%%%%%%%%%%%%%%%%%%%	
% bibliography file
%%%%%%%%%%%%%%%%%%%%%
%
%\newpage
%\bibliographystyle{my-iopart-num}
%\bibliography{references_review_monolayers_cleaned2}
%
%
\providecommand{\newblock}{}
\providecommand{\url}[1]{{\tt #1}}
\providecommand{\urlprefix}{}
\providecommand{\href}[2]{#2}

\end{document}